\documentclass[11pt,draftcls,onecolumn]{IEEEtran}
\IEEEoverridecommandlockouts
\usepackage{xcolor}
\usepackage{mathtools}
\ifCLASSINFOpdf
\else
\fi

\usepackage{cite}

\usepackage{color}
\usepackage{graphics}
\usepackage[dvips]{epsfig}
\usepackage{boxedminipage}
\usepackage{subcaption}

\usepackage{rotating}
\usepackage{amssymb,amsmath,amsthm}
\newtheorem{theorem}{Theorem}
\newtheorem{definition}{Definition}
\newtheorem{lemma}{Lemma}
\newtheorem{remark}{Remark}
\newtheorem{corollary}{Corollary}

\usepackage{tikz}
\usepackage{cite}

\newcommand{\indp}{\small\raisebox{-0.4mm}{\rotatebox{90}{$\models$}\,}}

\newcommand{\mmse}{\mathrm{mmse}}

\newcommand{\SNR}{\text{$\mathsf{SNR}$}}
\newcommand{\uSDR}{\text{$\mathsf{uSDR}$}}

\usepackage[]{algorithm2e}

\begin{document}
%
\title{Incremental Refinements and Multiple Descriptions with Feedback}

\author{Jan \O stergaard\thanks{This work was presented in part at the IEEE International Symposium on Information Theory, 2011 \cite{ostergaard:2011}, the IEEE Data Compression Conference, 2020 \cite{erez:2020}, and the IEEE Information Theory Workshop 2021~\cite{erez:2021}. The work of J.\ {\O}stergaard was partially supported by VILLUM FONDEN Young Investigator Programme grant agreement No. 19005. The work of R. Zamir was partially supported by 
the Israel Science Foundation grant \# 676/15.}, Uri Erez, and Ram Zamir\thanks{J. \O stergaard (jo@es.aau.dk) is with Aalborg University, Denmark.} \thanks{R. Zamir (zamir@eng.tau.ac.il) and U. Erez (uri@eng.tau.ac.il)  are with Tel Aviv University, Israel.}}

\maketitle

\begin{abstract}
It is well known that independent (separate) encoding of $K$ correlated sources may incur some rate loss compared to joint encoding, even if the decoding is done jointly.
This loss is particularly evident in the multiple descriptions problem, where it is the same source that is encoded in each description. 
We observe that under mild conditions about the source and distortion measure, the sum-rate of $K$ separately encoded individually good descriptions tends to the rate-distortion function of the joint decoder in the limit of vanishing small coding rates of the descriptions. Moreover, we then propose to successively encode the source into $K$ independent descriptions in each round in order to achieve a final distortion $D$ after $M$ rounds. 

We provide two examples -- a Gaussian source with mean-squared error and an exponential source with one-sided  error -- for which the excess rate vanishes in the limit as the number of rounds $M$ goes to infinity, for any fixed  $D$ and $K$.
This result has an interesting interpretation for a multi-round variant of the multiple descriptions problem, where after each round the encoder gets a (block) feedback regarding which of the descriptions arrived: In the limit as the number of rounds $M$ goes to infinity (i.e., many incremental rounds), 
the total rate of received descriptions approaches the rate-distortion function.
We provide theoretical and experimental evidence showing that this phenomenon is in fact more general than in the two examples above. 
\end{abstract}



%
\IEEEpeerreviewmaketitle



\section{Introduction}

The problem of lossy communication of  a source over a packet network has received considerable attention over the years, from various different angles. 
If the packet network is totally predictable in terms of thoughput and reliable  that packets are never dropped, the problem reduces to the standard rate distortion tradeoff. 
When the packet network does not satisfy the assumptions above, different formulations may be considered, depending on the constraints that are most important. 

Perhaps the simplest scenario is when there is  no  constraint on latency (and sparse utilization of the network is not of concern) and further there is a (perfect) feedback mechanism informing the sender whether a packet has been successfully received (ACK/NACK mechanism). In such a case, if the source is to be reconstructed at a given fidelity, one can simply divide the compressed source into many packets, sending each only after the previous packet has been successfully received (with enough re-transmissions). Note that the erasure pattern (packet drops), whether it is modeled statistically or not, has no bearing on such a transmission scheme, and simply translates to the total transmission time required (latency). We further note that one can avoid the need for 
supporting an ACK mechanism per packet by using a rateless code, in which case the only feedback needed is a termination flag to be fed back when sufficiently many packets have arrived for decoding (of the channel code) to be possible. 

Quite a different scenario arises when latency constraints are imposed. While in the case of an ergodic (e.g., i.i.d.) erasure process, one can still use a separation approach to compress the source to a fixed distortion and employ a fixed-rate code for the erasure channel (provided the latency constraint is not too strict), when ergodicity cannot be assumed, one cannot guarantee any pre-determined distortion level. 

The latter scenario therefore falls within the framework of joint source channel coding
and more specifically, for the considered case of an erasure channel, the scenario is the well known multiple description (MD) problem.
In contrast to the ergodic scenario, in the multiple description problem, the distortion level achieved is variable, depending on the erasure pattern experienced. In other words, in this scenario, the user experiences graceful degradation in accordance with the channel quality. 

Ideally, one would wish to achieve in the multiple description scenario the same level of distortion as dictated solely by the  source's rate distortion function (RDF), i.e., the same distortion one could achieve had the erasure pattern been known to the sender a priori. Unfortunately, this is known not to be possible except for under very special conditions, most notably, the case of successive refinement \cite{equitz:1991}. While  all sources are at least almost successively refinable at high resolution \cite{lastras:2001} (and the important case of a Gaussian source is strictly successively refinable), in order to employ successive refinement one must use the stop and wait (for ACK) approach discussed above, or prioritize the most important packet, which may limit its applicability. 
In general, and in particular in the symmetric case when all the packets have the same importance, the distortion guarantees
in MD coding are very far from the ``ideal" (known erasure
pattern) RDF.

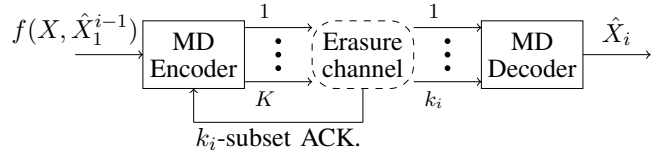
\begin{figure}
\centering
\scalebox{0.9}{
\begin{tikzpicture}

\draw [->] (1,2) -- (2,2)
	node[pos=0, above]{$f(X,\hat{X}_1^{i-1})$};
	
\draw 
   (2,1.5) rectangle (3.5,2.5);

\node[] at (2.75,2.2) {MD};
\node[] at (2.75,1.8) {Encoder};

\draw [->] (3.5,2.4) -- (4.5,2.4)
	node[pos=0.3, above]{\footnotesize $1$};

\filldraw (4, 2.2) circle (1pt);
\filldraw (4, 2) circle (1pt);
\filldraw (4, 1.8) circle (1pt);

\draw [->] (3.5,1.6) -- (4.5,1.6)
	node[pos=0.3, below]{\footnotesize $K$};

\draw [dashed, rounded corners=2ex]  
   (4.5,1.5) rectangle (6,2.5);

\node[] at (4.5+1.5/2,2.2) {Erasure};
\node[] at (4.5+1.5/2,1.8) {channel};

\draw [-] (4.5+1.5/2, 1.5) -- ((4.5+1.5/2, 1);
\draw [-] (2.75, 1) -- (4.5+1.5/2, 1);
\draw [->] (2.75, 1) -- (2.75, 1.5);

\node[] at (4,0.75) {$k_i$-subset ACK.};

\draw [->] (6,2.4) -- (7,2.4)
	node[pos=0.3, above]{\footnotesize $1$};

\filldraw (6.5, 2.2) circle (1pt);
\filldraw (6.5, 2) circle (1pt);
\filldraw (6.5, 1.8) circle (1pt);

\draw [->] (6,1.6) -- (7,1.6)
	node[pos=0.3, below]{\footnotesize $k_i$};

\draw 
   (7,1.5) rectangle (8.5,2.5);

\node[] at (7+1.5/2,2.2) {MD};
\node[] at (7+1.5/2,1.8) {Decoder};

\draw [->] (8.5,2) -- (9.5,2)
	node[pos=0.5, above]{$\hat{X}_i$};

\end{tikzpicture}
}
\caption{Incremental MDs with block acknowledgement. A subset of $k_i$ out of $K$ descriptions are received in round $i$ and the encoder is informed about this. In the first round, the source is $X$. Subsequently, the \emph{estimation error} $f(X, \hat{X}_1^i)$ after round $i$ will be the source in round $i+1$.}
\label{fig:mdfb}

\end{figure}

In this paper, we consider a multi-round variant of the MD problem, where the encoder sends $K$ packets in parallel during each round, after which (perfect) feedback as to which packets arrived is made available to the encoder. Next, based on the feedback, a refinement step is performed, followed by another round of transmission of $K$ new packets, and so forth, up to a total number of $M$ rounds; see Fig.~\ref{fig:mdfb}. The main insight revealed is that in this problem, at least for the important case of a Gaussian source with mean-square error distortion, as well as for the case of a one-sided exponential source with one-sided  error distortion, by taking the rate $R$ used for compression by each encoder to be sufficiently small, while there is an inherent gap to the ideal RDF, this gap can be made as small as desired, for \emph{any} erasure pattern. 
Specifically, one can approach ideal performance for any finite fixed total sum-rate $MKR$, as $R \rightarrow 0$ in each round (where accordingly $M \rightarrow \infty$).

To that end, 
 we introduce the property of \emph{unconditional incremental refinements} (UIRs), where each refinement is encoded at a very low rate and independently of the other refinements. 
Each refinement is individually rate distortion (R-D) optimal, and any subset of the refinements is jointly nearly R-D optimal. In the language of MD, for any $K$ descriptions, the system has no excess marginal rate \cite{zhang:1995} and at the same time has almost no excess sum rate \cite{ahlswede:1985}.
 
\subsection{Incremental information in channels and sources}
To obtain some intuition for this low-rate optimality, let us look on a dual phenomenon in channel coding. 
In the realm of channel coding, it is well known that for Gaussian channels, repetition loses very little in terms of mutual information, when operating at a very low signal-to-noise ratio (SNR) (the wideband regime) \cite{verdu:1990}. Namely 
\begin{align*}
    \frac{1}{2} \log(1+K\cdot \SNR) \approx K \cdot \frac{1}{2}  \log_2(1 + \SNR),
\end{align*}
as long as $K \cdot \SNR \ll 1$.
This linear behavior is in fact much more general as observed  by Shulman~\cite{shulman:2003} where the behavior of  $K$-repetition codes used over discrete memoryless channels was studied. Shulman considered the very noisy regime where $Y_i$ was the output of the $i$th ``parallel channel" when applying a repetition code. He showed that assuming that $Y_1,\dotsc, Y_K$ are independent outcomes of the channel, then $I(X;Y_1,\dotsc, Y_K)  \approx \sum_i I(X;Y_i)$, see \cite[Chapter 6]{shulman:2003} for details. 

As a consequence of this behavior, it was concluded in \cite{shulman:2003} that repetition coding is an effective means to achieve rateless coding over general DMCs in the regime where the (maximal) rate of transmission is low. In \cite{erez2012rateless}, it was demonstrated that for the case of Gaussian channels, this intuition can be leveraged to construct rateless codes that are not limited to the low-rate regime, by incorporating layered encoding and successive decoding. 

In the present work, we explore how this behavior of mutual information can be leveraged in the realm of lossy source coding. 
Let $X$ be the source and $Y$ the output of the source coder, and assume that we are using the source coder twice on the ``same" (see discussion below) source to obtain the outputs $Y_1$ and $Y_2$, where $Y_1-X-Y_2$ satisfy a Markov chain. If the source coder is \emph{unconditionally incrementally refineable}, it means that $I(X;Y_1,Y_2)  \approx  I(X;Y_1) + I(X;Y_2)$ at low rates. Furthermore, the quality of the reconstruction $\hat{X}(Y_1,Y_2)$ should be roughly equal to the case where $(X, Y_1, Y_2)$ do not have to satisfy a Markov chain constraint. 

From a channel coding perspective we know that if $Y_1$ and $Y_2$ are ``repetitions" of the source $X$ observed over parallel channels, then $Y_1-X-Y_2$ satisfy a Markov chain. 
An important difference between the source coding and channel coding scenarios arises when it comes to the operational significance of the mutual information relations. In channel coding, one can literally use repetition coding since the channel noise is independent in the repetition branches. In contrast, in lossy source coding, the noise, i.e., the distortion, is artificial and thus applying the same quantization operation multiple times is of no use. A means to circumvent this problem and still use the ``same code" is to employ some form of dithered (randomized) codebook. Indeed, dithered quantization has been widely employed in the MD problem \cite{frank2002dithered,chen2006multiple,ostergaard:2009}.

Another difference, which again stems from the fact that we are the ones creating the ``noise",  is that there may be many possible test channels that one can use, which satisfy asymptotic optimality (in the limit of low rate).
As we will observe, different test channels correspond to drastically different compression schemes.

\subsection{Notation and definitions}
Throughout the paper, we use capital letters such as $X$ and $Y$ for random variables and small letters for their realizations, e.g., $x$ and $y$.  Markov chains are denoted by $Y-X-Z$, { and $X \indp Z$ means that $X$ and $Z$ are statistically independent. $\mathbb{E}_Z$ denotes statistical expectation with respect to the random variable $Z$. The sequence $X_1,\dotsc, X_i$ will frequently be denoted $X_1^i$.
The conditional variance of $X$ given $Y$ is written as:
\begin{equation}
\mathrm{var}(X|Y) \triangleq \mathbb{E}_X[ (X - \mathbb{E}_X[ X| Y] )^2 |Y],
\end{equation}
and $\mathbb{E}_Y[\mathrm{var}(X|Y)]$ is then the minimum MSE (MMSE) due to estimating $X$ using $Y$.

The distortion-rate function (DRF) is denoted $D_X(R)$, and the rate-distortion function (RDF) is denoted $R_X(D)$. 

\subsection{Encoding and decoding policies}
We consider a random vector $X\in \mathbb{R}^n$ with $n$  i.i.d.\ elements. The marginal distribution of $X$ is allowed to be arbitrary. The vector $X$ is encoded in $M$ rounds into $K$ descriptions in each round.
The encoder in round $i=1,\dotsc, M,$ is defined as:
\begin{equation}
    \mathcal{F}_i : \mathcal{X}^i \times \mathcal{S}^i \to  \mathcal{B}_i^{K},
\end{equation}
where $\mathcal{X}, \mathcal{S}$, and $\mathcal{B}$ denote the alphabets of the source, side information, and codewords, respectively. The side information, known at both transmission ends, will in our work mainly denote an external random signal such as a \emph{dither} signal, so that the encoder and  decoder are allowed to be stochastic. 
Let $B^{(j)}(i)$ be the $j$th output of the encoder in the $i$th round. 
The coding rate $R^{(j)}(i) =\frac{1}{n}\mathbb{E}[|B^{(j)}(i)|]$ is defined as the expected length (in bits per sample) of $B^{(j)}(i)$. The sum-rate over the $K$ descriptions in round $i$ is $\sum_{j=1}^{K} R^{(j)}(i)$, and the total accumulated sum-rate over all $M$ rounds is:
\begin{equation}
    \bar{R}_M = \sum_{i=1}^M \sum_{j=1}^{K} R^{(j)}(i).
\end{equation}
We will often assume that $R^{(j)}(i) = R$, i.e., the coding rate is the same for all descriptions over all rounds. In this case, the total accumulated sum-rate is simply given by $\bar{R}_M=MKR$.

Let $\mathcal{I}(i) \subseteq \{1,\dotsc, K\}$ denote the set of indices of the received descriptions in round $i$. For example, if $K=4$ and only $B^{(2)}(i)$ and $B^{(3)}(i)$ are received, then $\mathcal{I}(i)=\{2,3\}$. Let $\mathcal{I}^i =\{ \mathcal{I}(1), \dotsc, \mathcal{I}(i)\}$. 
The decoder to be used in round $i$ depends upon the sequence of received description, that is:
\begin{equation}
\mathcal{G}_i^{\mathcal{I}^i} :  \mathcal{B}^{\mathcal{I}^i} \times  \mathcal{S}^{\mathcal{I}^i} \to \hat{\mathcal{X}}, \quad \forall \, \mathcal{I}^i,
\end{equation}
where $\hat{\mathcal{X}}$ is the reproduction alphabet, which could be equal to the source alphabet. 

The expected distortion $\bar{D}_i$ after round $i$ is:
\begin{equation}
    \bar{D}_i=\mathbb{E}[d(X, \mathcal{G}_i^{\mathcal{I}^i}(B^{\mathcal{I}^i},S^{\mathcal{I}^i}))],
\end{equation}
where $d(\cdot,\cdot)$ denotes a single-letter distortion measure, and where the expectation is with respect to $X$ and $\mathcal{I}^i$.

The following new joint source-channel coding problem with block acknowledgement is defined in this paper. Let $R_X(\cdot)$ be the RDF of the source $X$ under some single-letter distortion measure $d(\cdot,\cdot)$. Assume that the source is successively encoded over $M$ rounds and with $K$ descriptions in each round. Assume that the encoder is informed about which of the $K$ descriptions that are received by the decoder in each round. Then, the problem is to establish the existence of a family of encoders and decoders, where for a given source and distortion measure, and asymptotically in the number of rounds, the intermediate achievable sum-rate $\bar{R}_i$ and distortion $\bar{D}_i$ satisfy:
%
%
\begin{align}
\lim_{M\to \infty} \bigg[ \bar{R}_{i} - R_X(\bar{D}_{i}) \bigg] = 0, \forall i.
\end{align}

We will mainly base our analysis on mutual information measures and we will refer to the general test channels shown in Fig.~\ref{fig:setup} rather than the explicit coding policies defined above. In our analysis, we will assume the existence of stochastic encoding and decoding policies that can produce random signals with certain desired distributions. Towards that end, it is common in the information theoretic literature to simply define $X$ to be a vector of i.i.d. variables, which are to be jointly encoded but where one is only interested in the marginal distributions. In the present work, we will follow this common strategy and simply assume that one would employ high-dimensional vector quantizers on a long sequence of i.i.d.\ samples.

\subsection{Paper Organization}
The paper is organized as follows:
Section II focuses on the case of a single round with $K$ independent encodings. Here we prove asymptotic optimality of the rate ratio, a quantity we shall refer to as \emph{efficiency} in the sequel, in the limit of vanishing sum-rates for the quadratic Gaussian case, the one-sided exponential source under a one-sided error distortion,  generalized Gaussian sources under $p$-th power distortions, general sources and Gaussian coding noise, and general sources in the regime of very noisy test channels.

Section III focuses on the case of multiple rounds, see Fig.~\ref{fig:setup}. 
Here, we confine our treatment to the two special cases mentioned above, the quadratic-Gaussian scenario or that of an exponential  with a one-sided distortion measure (defined in detail below). The reason for this narrowing of scope is that for these two special cases, we have the beneficial property that the estimation error has precisely the same distribution (up to scaling) as the source itself, yielding in turn the same efficiency to hold for all rounds. This property is key in allowing one to derive performance guarantees that hold for a finite, non-vanishing, total rate. In particular, it allows us to show that for these two sources (with the associated distortion measures), for a fixed total rate $MKR$, the efficiency approaches one as $R \rightarrow 0$ (and accordingly $M\rightarrow \infty$.)


For these two sources, we also provide a non-asymptotic analysis as a function of $R, K$ and $M$. In all these analyses, we implicitly assume infinite block lengths.
We furthermore analyze general sources with Gaussian coding noise (i.e., Gaussian test channels)
and under MSE distortion.
In particular, we show that for general (``smooth") sources under MSE distortion, the efficiency of the multi-round incremental-refinement scheme approaches one in the high resolution limit where $R \rightarrow 0$; specifically, $R \rightarrow 0$ and $MR \rightarrow \infty.$

Section IV builds on the results of Section III, describing the implication (and application) of the latter to the problem motivating this work, namely to a scenario of incremental multiple descriptions with feedback as illustrated in Fig.~\ref{fig:mdfb}. The feedback indicates which $k$ out of $K$ descriptions were received. 
As the erasure pattern can be arbitrary and thus the total ``received rate" can vary anywhere between $0$ and $MKR$ (invalidating high resolution assumptions), in this section we confine attention solely to the two special cases mentioned above.
In round $i$, the error signal due to estimating the source using the $k_i$ received descriptions, forms the new source. The new source in each round is again encoded into $K$ independent encodings (multiple descriptions). We compare the performance to the RDF. We end this section by proposing a simple and practical threshold vector quantizer, which produces $K$ encodings. We prove asymptotic optimality for the case of a Gaussian source and MSE distortion in the limit of low rate quantization per round. 

The conclusions are given in Section V, and longer proofs are in the appendix. 

%
%
%

\section{Single Round of Incremental Refinements}
\label{sec:single-round}

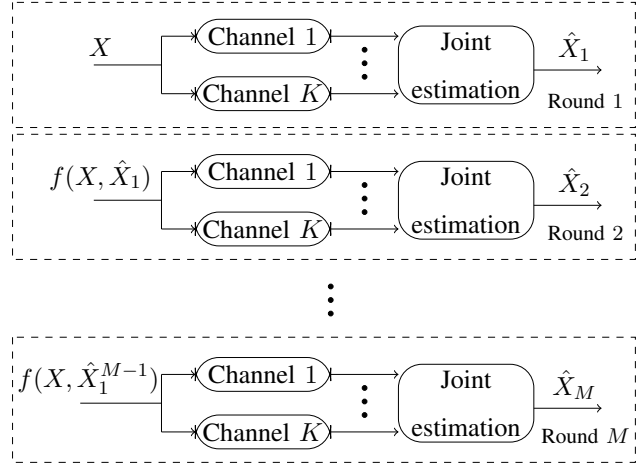
\begin{figure}
\centering
\scalebox{0.9}{
\tikzstyle{every node} = [align=center]
\begin{tikzpicture}

\draw [dashed]  
   (-0.7,3) rectangle (8.5,1.15)
	node[pos=0.92,above]{\footnotesize{Round $1$}};

\draw [->] (1.5,2.5) -- (2,2.5);
\draw [->] (1.5,1.65) -- (2,1.65);
\draw [-] (1.5,2.5) -- (1.5, 1.65);

\draw [-] (0.5,2.5-0.85/2) -- (1.5, 2.5-0.85/2)
	node[pos=0.1, above]{$X$};

\draw [rounded corners=2ex]  
   (2,2.75) rectangle (4,2.25)
	node[pos=0.5]{Channel $1$}; 

\draw [rounded corners=2ex]  
   (2,1.9) rectangle (4,1.4)
	node[pos=0.5]{Channel $K$};

\draw [->] (4,2.5) -- (5,2.5);

\filldraw (4.5, 2.3) circle (1pt);
\filldraw (4.5, 2.1) circle (1pt);
\filldraw (4.5, 1.9) circle (1pt);

\draw [->] (4,1.65) -- (5,1.65);

\draw [rounded corners=2ex]  
   (5,2.65) rectangle (7,1.5)
	node[pos=0.5]{Joint\\[0mm] estimation};

\draw [->] (7,2) -- (8,2)
	node[pos=0.6,above]{$\hat{X}_1$}; 

\draw [dashed]  
   (-0.7,3-1.95) rectangle (8.5,1.15-1.95)
	node[pos=0.92,above]{\footnotesize{Round $2$}};

\draw [->] (1.5,2.5-2) -- (2,2.5-2);
\draw [->] (1.5,1.65-2) -- (2,1.65-2);
\draw [-] (1.5,2.5-2) -- (1.5, 1.65-2);
\draw [-] (0.5,2.5-0.85/2-2) -- (1.5, 2.5-0.85/2-2)
	node[pos=0.1, above]{$f(X,\hat{X}_1)$};

\draw [rounded corners=2ex]  
   (2,2.75-2) rectangle (4,2.25-2)
	node[pos=0.5]{Channel $1$}; 

\draw [rounded corners=2ex]  
   (2,1.9-2) rectangle (4,1.4-2)
	node[pos=0.5]{Channel $K$};

\draw [->] (4,2.5-2) -- (5,2.5-2);

\filldraw (4.5, 2.3-2) circle (1pt);
\filldraw (4.5, 2.1-2) circle (1pt);
\filldraw (4.5, 1.9-2) circle (1pt);

\draw [->] (4,1.65-2) -- (5,1.65-2);

\draw [rounded corners=2ex]  
   (5,2.65-2) rectangle (7,1.5-2)
	node[pos=0.5]{Joint\\[0mm] estimation};

\draw [->] (7,2-2) -- (8,2-2)
	node[pos=0.6,above]{$\hat{X}_2$};


\filldraw (4, 2.3-2-1.5) circle (1pt);
\filldraw (4, 2.1-2-1.5) circle (1pt);
\filldraw (4, 1.9-2-1.5) circle (1pt);

\draw [dashed]  
   (-0.7,3-1.95-3) rectangle (8.5,1.15-1.95-3)
	node[pos=0.92,above]{\footnotesize{Round $M$}};

\draw [->] (1.5,2.5-2-3) -- (2,2.5-2-3);
\draw [->] (1.5,1.65-2-3) -- (2,1.65-2-3);
\draw [-] (1.5,2.5-2-3) -- (1.5, 1.65-2-3);
\draw [-] (0.3,2.5-0.85/2-2-3) -- (1.5, 2.5-0.85/2-2-3)
	node[pos=0.1, above]{$f(X,\hat{X}_1^{M-1})$};

%

\draw [rounded corners=2ex]  
   (2,2.75-2-3) rectangle (4,2.25-2-3)
	node[pos=0.5]{Channel $1$}; 

\draw [rounded corners=2ex]  
   (2,1.9-2-3) rectangle (4,1.4-2-3)
	node[pos=0.5]{Channel $K$};

\draw [->] (4,2.5-2-3) -- (5,2.5-2-3);

\filldraw (4.5, 2.3-2-3) circle (1pt);
\filldraw (4.5, 2.1-2-3) circle (1pt);
\filldraw (4.5, 1.9-2-3) circle (1pt);

\draw [->] (4,1.65-2-3) -- (5,1.65-2-3);

\draw [rounded corners=2ex]  
   (5,2.65-2-3) rectangle (7,1.5-2-3)
	node[pos=0.5]{Joint\\[0mm] estimation};

\draw [->] (7,2-2-3) -- (8,2-2-3)
	node[pos=0.6,above]{$\hat{X}_M$}; 

\end{tikzpicture}
}

\caption{$M$ rounds of $K$ parallel test channels. The source $X$ is input to $K$ parallel test channels and jointly estimated as $\hat{X}_1$ in round 1. In round 2, $f(X,\hat{X}_1)$, which in the Gaussian case is simply the residual $X-\hat{X}_1$, is input to $K$ parallel test channels and jointly estimated as $\hat{X}_2$.  After $M$ rounds, the final estimation of $X$ is $\hat{X}_M$. }
\label{fig:setup}
\end{figure}

In this section, we will consider a source which is \emph{separately} encoded into $K$ independent encodings and \emph{jointly} decoded. This situation can be modelled as a source $X$ which is transmitted over set of $K$ parallel test channels and then jointly estimated as $\hat{X}_1$ using all the channel outputs, see the top part of Fig.~\ref{fig:setup}. The figure also illustrates the case of multiple rounds of $K$ parallel test-channels, where the source is jointly estimated as $\hat{X}_M$ after $M$ rounds. The situation with multiple rounds is considered in the next section.

\begin{definition}
We define $Y_1, \dotsc, Y_K$, to be independent encodings of the common source $X$ and with respect to some distortion measure $d(\cdot,\cdot)$ used at the decoder, if and only if:
\begin{enumerate}
    \item $Y_1^K$ are jointly conditionally mutually independent given $X$. Form all the pairs of complementary sets $\mathcal{J}=\{(J,J^C)\}$, where $J\subseteq \{1,\dotsc, K\}$ and $J^C = \{1,\dotsc, K\}\backslash J$,  and let $Y_{J}$ and $Y_{J^C}$ be the sets of $Y_i$'s that are indexed by $J$ and $J^C$, respectively. Then the following Markov chain applies: $Y_J - X - Y_{J^C}$, for all pairs $(J, J^C) \in \mathcal{J}$.
    \item Let $X\to Y$ denote the optimum test-channel for the given source under the distortion measure $d(\cdot,\cdot)$. Then, for $i=1,\dotsc, K$, $Y_i$ has the same distribution as $Y$.
\end{enumerate}
\end{definition}

There are several ways to show that the coding schemes are asymptotically rate-distortion optimal. For example, if the slope of the operational RDF due to using $K$ descriptions, coincides with the slope of the true RDF in the limit of low rate, then the scheme is asymptotically optimal as the  coding rate vanishes.  For non-zero coding rates, 
this  means  that  the true RDF  and  the  operational  RDF are equivalent up till a first-order approximation. 

One can also, for example, form a so-called worst case efficiency, where the operational sum-rate is divided by the true RDF. If this ratio tends to 1, then the scheme is asymptotically optimal. Finally, 
we will also model coding schemes via so-called test channels, and in this case we assess optimality via the rate-loss in terms of excess mutual information across the test-channels but without addressing the distortion. We will make use of these approaches to assess asymptotic optimality of the different test channels and  sources under different distortion measures in the sequel.

\subsection{Gaussian Source and MSE}\label{sec:qg}
The RDF of an i.i.d. Gaussian source $X \sim \mathcal{N}(0,\sigma_X^2)$
under a quadratic distortion measure is given by
\begin{align}
    R_X(D)=\frac{1}{2} \log(\sigma_X^2/D).
    \label{eq:gaussian_rdf}
\end{align}
To draw the analogy to the channel coding problem counterpart, 
it is convenient to define the ``unbiased" signal-to-distortion ratio
(uSDR) as 
\begin{align*}
    \uSDR=\frac{\sigma^2}{D}-1.
\end{align*}
This amounts to the equivalent SNR in a test channel followed by Wiener estimation. 

With this notation, \eqref{eq:gaussian_rdf} becomes
\begin{align}
    R_X(\uSDR_{\rm opt})=\frac{1}{2} \log(1+\uSDR_{\rm opt}).
    \label{eq:gaussian_rdf1}
\end{align}
Suppose now that we pass the source through parallel (optimal) dithered quantizers to obtain reproductions $Y_i$, where
the differences $X-Y_i$  will all be of distortion $D$ and further will be mutually independent.
Denoting now the unbiased SDR attained by averaging the individual reconstructions by $\uSDR$, we will obtain 
a combined (central) reconstruction with an unbiased SDR of $K \cdot \uSDR$. 
Consequently, just as in the case of channel coding, as long as 
$K\cdot \uSDR \ll 1$, reconstruction based on any subset of outputs of a set of dithered quantizers is close to individual RDF optimality, where the maximal loss in terms of bit rate can be captured by defining 
the \emph{worst-case} efficiency:
\begin{align}\label{eq:worst_eta}
    \eta_K(R) & \triangleq \frac{\log(1+K\cdot \uSDR_{\rm opt})}{K \log(1+\uSDR_{\rm opt})} 
    = \frac{\log\left(1+K\left(2^{2R}-1\right)\right)}{2KR}.
\end{align}
It can be noticed that $\eta_K(R) \in [0,1]$, and $1/ \eta_K(R)$ is the excess rate (as a fraction of the optimal rate) beyond the minimal rate required to attain the combined distortion using conditional source coding strategies such as successive refinement or MD coding. If $\eta_K(R) \approx 1$ then it is nearly rate-distortion optimal to separately encode the source into $K$ descriptions. On the other hand, if $\eta_K(R) \ll 1$ then the resulting sum-rate of the $K$ descriptions is significantly greater than the rate-distortion function.

Let us assume the availability of a source coding scheme that produces Gaussian quantization noise $N$, which is independent of the source $X$. For example, this is  asymptotically the case for good subtractively dithered lattice vector quantizers in the limit where the dimension tends to infinity~\cite{zamir:2014}. Thus, we assume that we can model the output of the source coder by $Y=X+N$, where both $X$ and $N$ are Gaussian distributed. We refer to such a source encoder as a \emph{Gaussian} source coder.

The following theorem shows that near zero rate, the mutual information over the additive white Gaussian test channel is linear in the number of descriptions. 
\begin{theorem}\label{theo:gaussian}
Let the source $X$ be Gaussian distributed with variance $\sigma_X^2$, and let us use a Gaussian source coder $K$ times on $X$ to obtain $Y_i = X + N_i, i=1,\dotsc, K,$ where $N_i\indp X, i=1,\dotsc, K,$
are zero-mean and mutually independent Gaussian each with variance $\sigma_N^2$. 
Then, for any finite $\sigma_X^2>0$:
\begin{align*}
\lim_{\sigma_N^2\to \infty}
\frac{I(X;X+N_1, \dotsc, X+N_{K})}
{\sum_{j=1}^K I(X;X+N_j)}  = 1.
\end{align*}
\end{theorem}

\begin{IEEEproof}
It is easy to show that\footnote{As stated in Lemma~\ref{theo:oversampling}  \eqref{eq:mi-g} is  also true  for arbitrarily distributed sources $X$, and Gaussian noise $N_i$. Lemma~\ref{theo:oversampling} is known as the Theorem of Irrelevance in digital communications, cf.~\cite{gallager:2006}}
\begin{equation}\label{eq:mi-g}
I(X;X+N_1, \dotsc, X+N_{K}) = I(X;X+ \frac{1}{\sqrt{K}} N_1).
\end{equation} 
To prove the theorem, we observe that:
\begin{align}
\lim_{\sigma_N^2\to \infty}\!\!
\frac{I(X;X+ \frac{1}{\sqrt{K}} N_1)}
{K I(X;X+N_1)} = 
\!\lim_{\sigma_N^2\to \infty}\!\!
\frac{\frac{1}{2}\log_2(K\sigma_X^2/\sigma_N^2 + 1)}
{\frac{K}{2}\log_2(\sigma_X^2/\sigma_N^2 + 1)} = 1,
\end{align}
where the left-hand-side is identical to $\lim_{R\to 0} \eta_K(R)$, and 
where the last equality follows since $\lim_{c\to 0} c^{-1}\log_2(1+c) = 1$. 
\end{IEEEproof}

In terms of the worst-case efficiency in \eqref{eq:worst_eta}, then Theorem \ref{theo:gaussian} shows that it asymptotically tends to one as the sum-rate tends to zero, i.e., $1/\eta_K(R) \to 1$ as $R\to 0$, where $R = I(X; X + N_1)$. }


\subsection{One Sided Exponential Source Under One-Sided Error Criterion}
\label{sec:expsource}

The one-sided exponential source is defined as \cite{Ver96}:
\begin{equation}\label{eq:source}
f_X(x) = \lambda e^{-\lambda x}, \quad x\geq 0,
\end{equation}
where $\mathbb{E}{[X]} = 1/\lambda$, and $\lambda>0$. 

Let $y$ denote the reproduction of the coder, and let the one-sided  error criterion be given by:
\begin{equation}\label{eq:distortion}
  d(x,y) = \begin{cases}
  x-y, & \text{if}\  x\geq y, \\
  \infty, &  \text{if}\ x<y. 
  \end{cases}
\end{equation}

The rate-distortion function for the exponential source with one-sided error criterion is given by \cite{Ver96}:
\begin{equation*}
R_X(D) = \begin{cases}
-\log(\lambda D), & 0\leq D\leq \frac{1}{\lambda}, \\
0, & D> \frac{1}{\lambda}.
\end{cases}
\end{equation*}

Let $X=Z+Y$ denote the backward test channel whose optimal conditional output distribution is given by \cite{Ver96, si:2014}:
\begin{equation}\label{eq:cond}
f_{X|Y}(x|y) = \frac{1}{D} e^{-\frac{(x-y)}{D}}, \quad x\geq y\geq 0.
\end{equation}
Given $K$ independent encodings, $Y_1^K$. The estimator that selects the maximum $\hat{Y} = \max \{ Y_1,\dotsc, Y_K\}$ will be denoted the select-max estimator of $X$. 
\begin{lemma}[Optimality of the select-max estimator: Lemma 2 in \cite{erez:2020}]\label{lem:selectmax}
Let  $Y_1,\dotsc, Y_K,$ be  $K>0$ independent encodings of the one-sided expontial source $X$. The select-max estimator $\hat{Y} = \max \{ Y_1,\dotsc, Y_K\}$ is an optimal estimator under the one-sided  distortion measure.
\end{lemma}

\begin{lemma}[Distortion of the select-max estimator: Lemma 3 in \cite{erez:2020}]\label{lem:D}
Let $Y_1,\dotsc, Y_K$, be $K>0$ independent encodings of the one-sided expontial source $X$ with parameter $\lambda>0$. The total expected distortion $\bar{D}$ due to using the select-max estimator on the  outputs $Y_1,\dotsc, Y_K$ is given by:
\begin{equation}\label{eq:odrfk1}
\bar{D} = \frac{D}{K-(K-1)\lambda D}.
\end{equation}
\end{lemma}

The following three lemmas show that the estimation error $\tilde{Z}= X - \hat{Y}$ has the same statistical properties as the optimum noise in the backward test channel, i.e., $\tilde{Z}$ is exponentially distributed and independent of $\hat{Y}$. Moreover, $\hat{Y}$ is a sufficient statistic for $X$ from $(Y_1, \dotsc, Y_K)$. 

\begin{lemma}[Distribution of estimation error: Lemma 1 in \cite{erez:2021}]\label{lem:exp_error}
Let $Y_1,\dotsc, Y_K$, be $K>0$ independent encodings of the one-sided expontial source $X$ with parameter $\lambda>0$. Moreover, 
let $\delta = \frac{1}{D} - \lambda, 0\leq D\leq 1/\lambda$. 
Then, the estimation error $\tilde{Z}=X - \hat{Y}$, due to estimating the source $X$ by the select-max estimator $\hat{Y}$, is one-sided exponentially distributed with parameter $\lambda' = \lambda+ K\delta$, i.e.:
\begin{equation*}
f(\tilde{z}) = (\lambda+K\delta) e^{-(\lambda+K\delta)\tilde{z}},\quad  \tilde{z}\geq 0.
\end{equation*}
\end{lemma}

\begin{lemma}[Sufficient statistic: Lemma 2 in \cite{erez:2021}]
\emph{Let $X$ be a one-sided exponential source, and let $X \to Y_1, \dotsc, X \to Y_K$ be $K$ parallel (RDF achieving) test channels. 
Moreover, let  $\hat{Y} = \max \{Y_1,\dotsc , Y_K\}$. 
Then,
\begin{align}
X -  \hat{Y} - Y_1^K
\end{align}
form a Markov chain, i.e., $\hat{Y}$ is a sufficient statistic for $X$ from $Y_1^K$.}
\end{lemma}

\begin{lemma}[Orthogonality principle: Lemma 3 in \cite{erez:2021}]
\emph{Let $X$ be a one-sided exponential source, and let $X \to Y_1, \dotsc, X \to Y_K$ be $K$ parallel (RDF achieving) test channels. Finally, let  $\hat{Y} = \max \{Y_1,\dotsc , Y_K\}$. 
Then, the backward channel $X = \hat{Y} + \tilde{Z}$ is additive, i.e., the estimation error $\tilde{Z}$ is independent of the estimator $\hat{Y}$. Moreover, $\tilde{Z}$ is independent of the joint output vector $(Y_1,\dotsc, Y_K)$. }
\end{lemma}

The following theorem assesses the asymptotic distortion-rate performance due to using $K$ independent encodings as the sum-rate tends to zero. 

\begin{theorem}\label{theo:expo}
Let the source be one-sided exponential with parameter $\lambda>0$, and let $\bar{D}$ in \eqref{eq:odrfk1} be the distortion due to using the select-max estimator on $K$ independent encodings each of rate $R$. Moreover, let $D_X(R)$ denote the DRF of the source under the one-side error distortion. Then, 
\begin{equation}
\lim_{R \to 0} \frac{ \frac{\partial}{\partial R} D_X(KR)}
{\frac{\partial}{\partial R}\bar{D}}
= 1,
\end{equation}
\end{theorem}

\begin{IEEEproof}
Let the distortion be given by $D=\frac{1-\epsilon}{\lambda}$, for any $1>\epsilon>0$. Then the rate for a single description is given by the RDF:
\begin{equation*}
R_X(D) = -\log_2(D\lambda) = -\log_2( 1-\epsilon ). 
\end{equation*}
The DRF is given by:
\begin{align*}
D_X(R) = \frac{1}{\lambda} 2^{-R}.
\end{align*}
Inserting the rate of $K$ times the single-description rate into the DRF leads to a distortion of:
\begin{align}\label{eq:drf}
D_X(KR) &= \frac{1}{\lambda} 2^{ K\log_2(1-\epsilon)}= \frac{1}{\lambda} (1-\epsilon)^{K}.
\end{align}
On the other hand, the operational DRF \eqref{eq:odrfk1} using $K$ independent encodings  and the select-max estimator yields:
\begin{align}\label{eq:odrfk}
\bar{D} = \frac{1-\epsilon}{\lambda( 1 + \epsilon(K- 1))}.
\end{align}
The ratio of the derivatives of \eqref{eq:drf} and \eqref{eq:odrfk} wrt.\ $\epsilon$ tends to $1$ as $\epsilon \to 0$:
\begin{align*}
\lim_{\epsilon \to 0} \frac{ \frac{\partial}{\partial \epsilon} \frac{1}{\lambda} (1-\epsilon)^{K} }
{
\frac{\partial}{\partial \epsilon} \frac{1-\epsilon}{\lambda( 1 + \epsilon(K- 1))}
}
= 1,
\end{align*}
which proves the theorem. 
\end{IEEEproof}

Theorem \ref{theo:expo}
 shows that in the limit of vanishing rates (large distortions), the slope of the operational DRF using $K$ independent encodings each of rate $R$, coincides with the slope of the true DRF using a single channel at rate $KR$. 
Thus, at small rates and under the one-sided error criterion, it is nearly rate-distortion optimal, to encode the one-sided exponential source into $K$ independent encodings and then reconstruct using the simple select-max estimator. 

\begin{figure}[tb]
\begin{center}
\includegraphics[width=7cm]{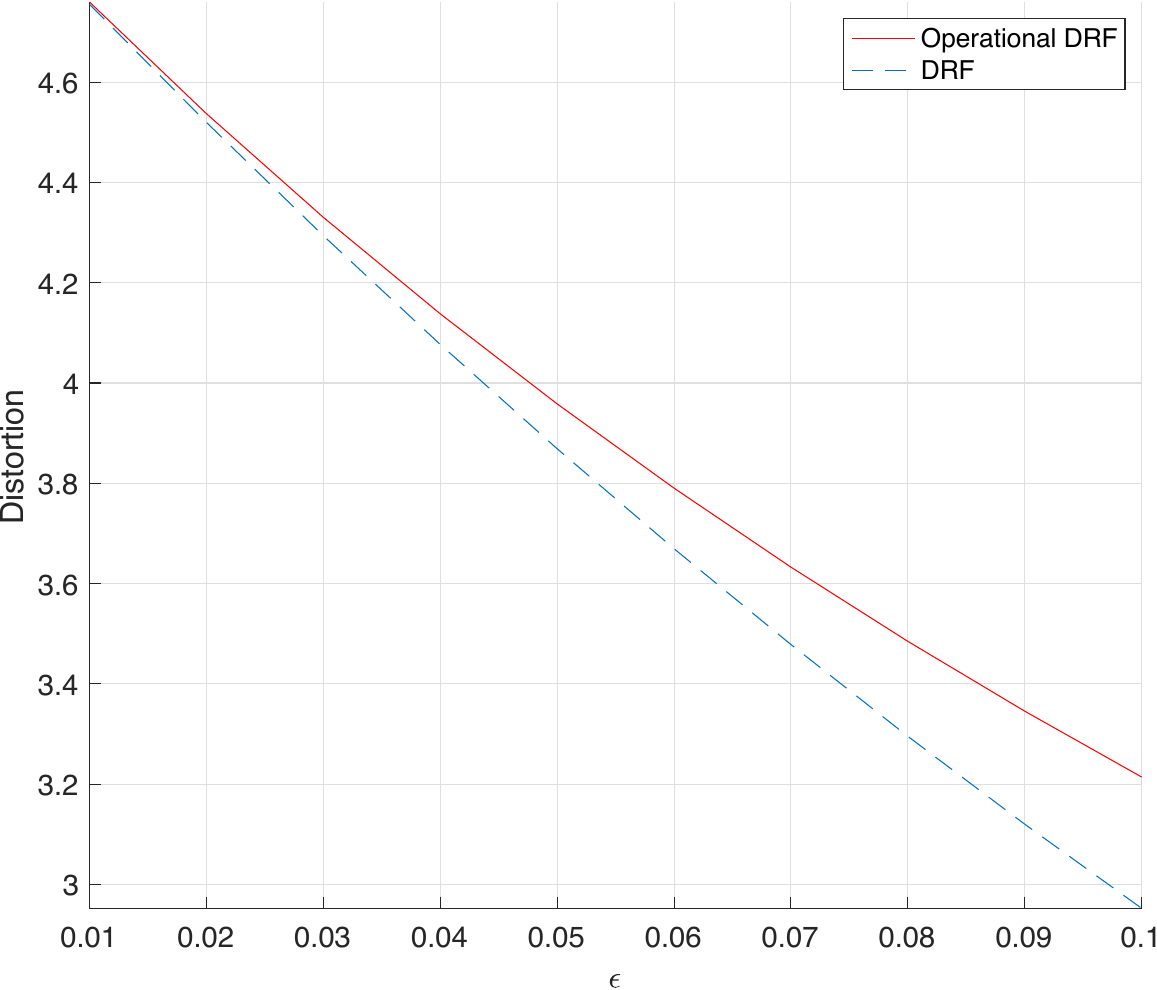} 
\caption{Operational DRF and DRF for the source with $\lambda=0.2$ and $K=5$ descriptions. The rates are equal for the operational DRF and the DRF and given by $-K\log_2(1-\epsilon)$. Recall that $\mathbb{E}[X] = \lambda^{-1} = 5$. }
\label{fig:odrf}
\end{center}
\end{figure}

To illustrate this behaviour near zero rate, we have shown the DRF \eqref{eq:drf} and the operational DRF \eqref{eq:odrfk} using $K=5$ independent encodings in Fig.~\ref{fig:odrf}. In the figure, we have used a one-sided exponential source with parameter $\lambda=0.2$.
The rates are equal for the DRF and the operational DRF and are given by $-K\log_2(1-\epsilon)$. As  $\epsilon \to 0$, the distortion tends to its maximum, i.e., $D\to \mathbb{E}[X] = \lambda^{-1} = 5$. From the figure, one may observe that the slope of the operational DRF approaches that of the DRF as $\epsilon$ becomes small. This demonstrates that the DRF  and the  operational  DRF  are equivalent near $R=0$ up till a first-order approximation.

\subsection{Generalized Gaussian Source under $p$-th Power Distortion Criterion}
The distribution $\mathcal{N}_q(0,\alpha^q)$ of a zero-mean generalized Gaussian random variable with parameters $\alpha,q$ is given by \cite{dytso:2017}:
\begin{align}
p_X(x)  &= \frac{c_q}{\alpha} e^{-\frac{|x|^q}{2\alpha^q}}, \\  \label{eq:cq}
c_q&= \frac{q}{2^{\frac{q+1}{q}}\Gamma\bigg(\frac{1}{q}\bigg)},
\end{align}
for all $x\in\mathbb{R}$. 

We consider the $p$-th power distortion measure, i.e.:
\begin{align}
d(x,\hat{x}) = |x-\hat{x}|^p, \quad p\geq 1.
\end{align}

The generalized Gaussian distribution maximizes the Shannon entropy and the Renyi entropy under a $p$-th absolute moment constraint \cite{dytso:2017}. Specifically, 
$h(X) \leq \frac{1}{q}\log(\frac{\alpha^q e}{c_q^q})$ with equality if and only if $X\sim \mathcal{N}_q(0,\alpha^q)$. 

The Shannon lower bound for $X\sim \mathcal{N}_q(0,\alpha^q)$ and $(p,q)\in \mathbb{R}^{2}_+$ under the $p$-th power distortion measure is given by \cite{dytso:2017}:
\begin{align}
R_{q,p}(\alpha, D) &= \inf  \ I(X;Y) \\
&\geq \left[ \log\big(\frac{\alpha}{D}\big) + \log\bigg( \frac{c_p}{c_q} e^{\frac{1}{q}-\frac{1}{p}} \bigg)\right]^+,
\end{align}
where the infimum is over all conditional distributions $p_{Y|X}$ satisfying the $p$-th power distortion constraint given by:
\begin{equation}
\mathbb{E}[ |X-Y|^p] \leq \frac{2}{p}D^p.
\end{equation}

For any $\alpha\geq 1$, $1\leq q\leq 2$, and $1\leq p < 2$, Shannon's lower bound is tight and equality is obtained in the backward test channel using $Z\sim \mathcal{N}_p(0,D^p)$ independent of $Y$, where $Y$ has a mixture distribution given by \cite{dytso:2017}:
\begin{equation}\label{eq:pY}
p_{Y}(y) = \bigg(\frac{D}{\alpha}\bigg)^{p+1} \delta(y) + \bigg(1 - \bigg(\frac{D}{\alpha}\bigg)^{p+1}\bigg) g(y),
\end{equation}
where $\delta(y)$ is the Dirac delta function and $g(y)$ is the pdf of a random variable with characteristic function:
\begin{equation}
\phi(t) = \frac{\phi_p(\alpha t) - \big(\frac{D}{\alpha}\big)^{p+1} \phi_p(D t)}{\big(1 - \big(\frac{D}{\alpha}\big)^{p+1}\phi_p(Dt)\big)},
\end{equation}
and where $\phi_p$ is the characteristic function of a random variable with distribution $\mathcal{N}_p(0,1)$, which is given by:
\begin{equation}
\phi_p(t) = 2c_p \int_{0}^{\infty} \cos(t  x) e^{-\frac{x^p}{2}} \, dx.
\end{equation}

%

In the following we assume that $p=q > 0$, and we will be interested in finding the forward test channels. 

The conditional distribution $p_{X|Y}(x|y)=p_Z(z)$ of the noise $Z$ is given by:
\begin{align} \label{eq:x_given_y1}
p_{X|Y}(x|y) &= \frac{c_p}{D} e^{-\frac{|x-y|^p}{2D^p}} \\ \label{eq:x_given_y2}
&= \frac{c_p}{D} e^{-\frac{|z|^p}{2 D^p}}.
\end{align}

\begin{lemma}[Conditional distortion for a generalized Gaussian source]\label{lem:cond_dist}
Let the noise $Z$ in the backward test channel $X=Z+Y$, be independent of $Y$ and distributed as $Z\sim \mathcal{N}_p(0,D^p)$. Then, for any finite $p>0$, the conditional distortion $\int p_{X|Y}(x|y)|x-y|^p \, dx$ is independent of $y$ and given by:
\begin{equation}\label{eq:cond_dist}
\int_\mathbb{R} p_{X|Y}(x|y)|x-y|^p \, dx = \frac{2}{p}D^p.
\end{equation}
\end{lemma}

\begin{IEEEproof}
From Proposition 2 in \cite{dytso:2017} it can be deduced that:
\begin{equation}\label{eq:abs_moment}
\int_\mathbb{R} e^{-c |z|^p}|z|^p dz  = \frac{2}{p^2} \Gamma\bigg(\frac{1}{p}\bigg) c^{-\frac{p+1}{p}}.
\end{equation}
By a change of variables, $z=x-y$ in \eqref{eq:cond_dist}, where $dz = dx$, one can readily verify that \eqref{eq:cond_dist} is obtained by inserting $c=\frac{1}{2 D^p}$ into \eqref{eq:abs_moment} and normalizing by $c_p/D$. 
\end{IEEEproof}

The conditional distribution $P_{Y|X}(y|x)$ of the forward test channel is given by:
\begin{align}
p_{Y|X}(y|x) &= \frac{p_{X|Y}(x|y)p_Y(y)}{p_X(x)} \\ \notag
&= \frac{\alpha}{D}  e^{-\frac{|x-y|^p}{2 D^p}}  e^{\frac{|x|^p}{2\alpha^p}}\bigg(
\bigg(\frac{D}{\alpha}\bigg)^{p+1} \delta(y) \\
&\quad + \bigg(1 - \bigg(\frac{D}{\alpha}\bigg)^{p+1}\bigg) g(y)\bigg),
\end{align}
which is a mixture distribution.

We now assume the presence of $K$ parallel forward test channels having outputs $y_1,\dotsc, y_K$, which satisfy the properties of being independent encodings of $x$.
 Clearly the joint distribution of $(Y_1^K,X)$ satisfies:
\begin{align}\label{eq:markov_channels}
p(y_1,\dotsc, y_K,x) &= p_{Y|X}(y_1|x)\cdots p_{Y|X}(y_K|x)p(x) \\
&= p_{Y^K|X}(y^K|x) p(x).
\end{align}

\begin{lemma}[Probability of an all-zeros output]\label{lem:prob_zero}
Let $X\sim \mathcal{N}_p(0,\alpha^p)$, and let $Y_i, i=1,\dotsc, K$ be the outputs of $K$ parallel channels that jointly with $X$ satisfy \eqref{eq:markov_channels}. Moreover, let the marginal distribution of $Y_i, \forall i$, be a mixture distribution given by:
\begin{equation}\label{eq:mix}
p_{Y_i}(y_i) = \bigg( \frac{D}{\alpha}\bigg)^{p+1} \delta(y_i) + \bigg(1 - \bigg(\frac{D}{\alpha}\bigg)^{p+1}\bigg)\tilde{g}(y_i),
\end{equation}
where $\tilde{g}(y_i)$ is a well-defined pdf, and where $\exists \epsilon > 0$ so that $\tilde{g}(y_i) < \infty$ for all $y_i<\epsilon$.  
Then, the probability that all $K$ outputs are zero is given by:
\begin{equation}\label{eq:p0}
P( \sum_{i=1}^K |y_i|= 0) = D \bigg(  \frac{D}{\alpha}\bigg)^{pK}  (D^p + K \alpha^p - K D^p)^{-\frac{1}{p}}.
\end{equation}
\end{lemma}

\begin{remark}
We note that several mixture distributions on the form given by \eqref{eq:mix} were provided in \cite{dytso:2017}. One example is that of \eqref{eq:pY}.
\end{remark}

\begin{IEEEproof}[Proof of Lemma~\ref{lem:prob_zero}]
Due to the delta distribution in \eqref{eq:mix} and the fact that $\tilde{g}$ is bounded in a neighbourhood around $y_i=0$, we only need to evaluate the joint distribution of $Y^K$ at $y_i=0, \forall i$. In order to do this, we marginalize \eqref{eq:markov_channels} over $x\in \mathbb{R}$ and insert $y_i=0, \forall i$.  
{\allowdisplaybreaks
\begin{align}
&P( \sum_{i=1}^K |y_i|= 0) = 
\int_\mathbb{R} (p_{Y|X}(0|x))^K p(x) \, dx \\
&=\frac{c_p}{ \alpha}  \bigg(\frac{D}{\alpha}\bigg)^{pK} 
\int_\mathbb{R} 
\bigg(
e^{-\frac{|x|^p}{2 D^p}}  e^{\frac{|x|^p}{2\alpha^p}}\bigg)^K
 e^{-\frac{|x|^p}{2\alpha^p}} \, dx \\
&=2\frac{c_p}{ \alpha}  \bigg(\frac{D}{\alpha}\bigg)^{pK} 
\int_0^{\infty}
e^{-\frac{|x|^p}{2 \alpha^p D^p}(K\alpha^p -K D^p + D^p)} \, dx \\
&\overset{(a)}{=} 2\frac{c_p}{\alpha} \bigg(\frac{D}{\alpha}\bigg)^{pK} \frac{1}{p}2^{\frac{1}{p}}D \alpha (D^p + K \alpha^p - K D^p)^{-\frac{1}{p}}\Gamma\bigg(\frac{1}{p}\bigg) \\
&\overset{(b)}{=} D \bigg(  \frac{D}{\alpha}\bigg)^{pK}  (D^p + K \alpha^p - K D^p)^{-\frac{1}{p}},
\end{align}
where $(a)$ follows since $\int_0^\infty \exp( -c x^p) \, dx = c^{-\frac{1}{p}}p^{-1}\Gamma(p^{-1})$, and $(b)$ is obtained by using \eqref{eq:cq}.
}
\end{IEEEproof}

Let $Y_i,i=1,\dotsc, K,$ be the outputs from $K$ channels that jointly with $X$ satisfy \eqref{eq:markov_channels}. Then $Y_1,\dotsc, Y_k$ are independent encodings of $X$. Let $\mathcal{I} \subseteq \{1,\dotsc, K\}$
be a set containing the indices for the outputs, which are non zero.  Moreover, if $\mathcal{I}$ is non-empty, we let $i^* = \min i\in \mathcal{I}$ denote the minimum of the indices in $\mathcal{I}$. We define the \emph{select-non-zero} estimator as $\hat{Y} = Y_{i^*}$ if $|\mathcal{I}|>0$ and otherwise $\hat{Y}=0$. 
At general coding rates, the select-non-zero estimator is a sub-optimal estimator. However, in the limit of vanishing small rates, it becomes asymptotically optimal.

\begin{lemma}[Distortion for a generalized Gaussian source]\label{lem:dist_gg}
Let $X\sim \mathcal{N}_p(0,\alpha^p)$, and let $Y_1, \dotsc, Y_K$, be $K$ independent encodings of $X$.  Moreover, let the marginal distributions of $Y_i$ be given by \eqref{eq:mix} for all $i=1,\dotsc, K$. Then, the joint distorton $\bar{D}$ under the $p$-th power distortion measure and using the select-non-zero estimator is given by:
\begin{align}\label{eq:ggauss_dist}
\bar{D} &= 
 P( \sum_{i=1}^K |y_i|= 0)
\frac{2}{p}\alpha^pD^{p} \big( D^p + K\alpha^p - KD^p\big)^{-1} \\  \label{eq:Dk}
&\quad + (1-P( \sum_{i=1}^K |y_i|= 0))\frac{2}{p}D^p.
\end{align}
\end{lemma}

\begin{IEEEproof}[Proof of Lemma~\ref{lem:dist_gg}] 
Let $\hat{y}=\hat{y}(y_1,\dotsc,y_K)$ be the select-max estimator of $x$ based on all $K$ channels outputs $y_i, i=1,\dotsc, K$. With this, the joint distortion $\bar{D}$ can be written as:
\begin{align}
\bar{D} &= \int_{\mathbb{R}^K} \int_\mathbb{R} p_X(x) p_{Y^K|X}(y^K|x) |x-\hat{y}|^p \, dx \,dy^K\\ \label{eq:lim}
&= \underbrace{\int_\mathbb{R}  p_X(x) p_{Y^K|X}(y^K=0|x) |x|^p \, dx}_{\triangleq D_0} \\
&\quad +\underbrace{\int_{\mathbb{R}^K\backslash\{y^K=0\}} \int_\mathbb{R} p_X(x) p_{Y^K|X}(y^K |x) |x-\hat{y}|^p \, dx \,dy^K}_{\triangleq D_1}.
\end{align}
The left term, $D_0$, in \eqref{eq:lim} is the distortion when all channel outputs are zero, in which case the optimal estimator is $\hat{y} = 0$. When at least one of the channel outputs is non-zero, the estimator $\hat{y}$ is not necessarily optimal. The right term, $D_1$, is the distortion when at least one of the channel outputs is non zero. If we use the select-non-zero estimator, i.e., $\hat{y}=y_{i^*}$, we can express $D_1$ in closed-form: 
{\allowdisplaybreaks
\begin{align}
D_1 &=  (1- P( \sum_{i=1}^K |y_i| = 0)) \\
&\quad\times \int_{\mathbb{R}_+} \int_\mathbb{R} p_X(x) p_{Y|X}(y_{i^*}|x) |x-y_{i^*}|^p \, dx \,dy_{i^*} \\ \notag
&= (1- P( \sum_{i=1}^K |y_i| = 0)) \frac{2}{p}D^p,
\end{align}
}
where we used \eqref{eq:cond_dist} to get to the last equality.

We will now show that $D_0$ in \eqref{eq:lim} can be expressed in closed form:
{\allowdisplaybreaks
 \begin{align}
D_0 &= \int_\mathbb{R} \bigg(\frac{\alpha}{D}  e^{-\frac{|x|^p}{2D^p}}  e^{\frac{|x|^p}{2\alpha^q}}
\bigg(\frac{D}{\alpha}\bigg)^{p+1} \bigg)^K
\frac{c_p}{\alpha} e^{-\frac{|x|^p}{2\alpha^p}} |x|^p \, dx\\
&=\frac{c_p}{\alpha}\bigg(\frac{D}{\alpha}\bigg)^{pK} \int_\mathbb{R}\bigg( e^{-\frac{|x|^p}{2D^p}}  e^{\frac{|x|^p}{2\alpha^p}}\bigg)^K
 e^{-\frac{|x|^p}{2\alpha^p}} |x|^p\, dx\\
&= \frac{2}{p}\alpha^pD^{p+1} \bigg(\frac{D}{\alpha}\bigg)^{pK}
\big( D^p + K\alpha^p - KD^p\big)^{-\frac{p+1}{p}}  \\
&=P( \sum_{i=1}^K |y_i|= 0)
\frac{2}{p}\alpha^pD^{p} \big( D^p + K\alpha^p - KD^p\big)^{-1}.
\end{align}
}

Since $\bar{D}= D_0  + D_1$, the lemma is proved.
\end{IEEEproof}

The two lemmas below, show that it is  asymptotically optimal to encode the generalized Gaussian source into $K$ independent encodings and using the select-non-zero estimator, as the rate tends to zero.

For a fixed rate $R>0$ and a fixed number of descriptions $K$, let the distortion redundancy factor $\rho_D(R,K) \triangleq \frac{\bar{D}}{D_\mathrm{opt}(KR)}$ describe the ratio of the joint distortion $\bar{D}$ due to using $K$ descriptions each at rate $R$ over the optimum distortion $D_\mathrm{opt}(KR)$ due to using $KR$ bits in a single description.

\begin{lemma}[Distortion redundancy of a generalized Gaussian source]\label{lem:lim_dist_gg} 
Let $X\sim \mathcal{N}_p(0,\alpha^D)$  and let $Y_1,\dotsc, Y_K$, be independent encodings of $X$.  Moreover, let the marginal distributions of $Y_i$ be given by \eqref{eq:mix} for all $i=1,\dotsc, K$. 
Then, under the $p$-th power distortion measure and using the select-non-zero estimator:
\begin{equation}
\lim_{D\to \alpha} \rho_D(R,K) = 1.
\end{equation}
\end{lemma}

\begin{IEEEproof}[Proof of Lemma~\ref{lem:lim_dist_gg}] 
The Shannon lower bound for the case of $q=p$ and distortion $D'=\frac{2}{p}D^p$ can be written as \cite{dytso:2017}
\begin{equation}
R_{\mathrm{slb}}(D) = \log\big(\frac{\alpha}{D}\big),
\end{equation}
which implies that 
\begin{equation}
D_X(R)=\alpha 2^{-R}.
\end{equation}
Since the description rate is $R=-\log_2(\alpha/D)$, it follows that using a rate which is $K$ times $R$  gives the conditional (joint) distortion 
\begin{align}\label{eq:opt_D11}
D_X(KR) = \frac{2}{p}(D^K\alpha^{1-K})^p.
\end{align}
The distortion redundancy factor $\rho_D(K)$ can then be expressed as $\bar{D}$ divided by \eqref{eq:opt_D11}, that is:
\begin{align}
\rho_D(K) &= \frac{\bar{D}}{\frac{2}{p}(D^K\alpha^{1-K})^p} \\ \notag
&=  \frac{P( \sum_{i=1}^K |y_i|= 0)
\alpha^pD^{p} \big( D^p + K\alpha^p - KD^p\big)^{-1}}
{(D^K\alpha^{1-K})^p} \\
&\quad + \frac{(1-P( \sum_{i=1}^K |y_i|= 0))D^p}{(D^K\alpha^{1-K})^p}
\end{align}
which for any finite $K$ tends to one as $D\to \alpha$.
\end{IEEEproof}

\begin{theorem}
[Asymptotic slope optimality of $K$ independent encodings of the generalized Gaussian source and select-non-zero estimation]\label{theo:slope_gg}
The slope of the operational RDF of the generalized Gaussian source under the $p$-th power distortion criterion and when using the select-non-zero estimator on $K$ independent encodings, coincides with the  slope of Shannon's RDF at zero rates, and is given by:
\begin{equation}\label{eq:partial_rd}
\frac{\partial}{\partial \bar{D}} R(\bar{D}) = - \frac{1}{2\alpha^{p-2}}.
\end{equation}
\end{theorem}

\begin{IEEEproof}
Let $\tilde{D}_1^p \triangleq \frac{p}{2} \bar{D}$, where $\bar{D}$ is the distortion of the $K$-description scheme and given by \eqref{eq:ggauss_dist}, i.e.:
\begin{align}
\tilde{D}_1^p \triangleq  p_0
\alpha^p D^{p} \big( D^p + K\alpha^p - KD^p\big)^{-1} + (1-p_0)D^p,
\end{align}
where $p_0 \triangleq  P( \sum_{i=1}^K |y_i|= 0)$ is a function of $D$ and is given  by \eqref{eq:p0}. 
The derivative of $\tilde{D}_1$ wrt.\ $D$ at the point $D=\alpha$ is given by:
\begin{align}\label{eq:d_der}
\frac{\partial}{\partial D} {\tilde{D}_1^p}\vert_{D=\alpha} &= p K \alpha^{p-1},
\end{align}
which implies that  the slope of $\bar{D}$ is $2K\alpha^{p-1}$. 

The derivative of the rate  function $K\log(\alpha/D)$ of the $K$-description scheme is given by:
\begin{align}\label{eq:r_der}
\frac{\partial}{\partial D} K\log(\alpha/D) &= -K D^{-1}.
\end{align}
The ratio of \eqref{eq:r_der} over \eqref{eq:d_der} (and then normalizing by $2/p$) is the derivative of the operational rate-distortion function of the $K$-description single round scheme, and is given by \eqref{eq:partial_rd}. 

From Shannon's single-description RDF we have that $R_X(D) = \log(\alpha/D)$, where the distortion under the $p$-th power distortion criteria is $\bar{D}=\frac{2}{p}D^p$. If we write the DRF as 
$D_X(R) = \alpha 2^{-R}$, and replace $R$ by $KR$, we obtain the distortion achieved when the rate is increased  $K$ times. Thus, we get $D_X(KR) = \alpha 2^{-K(\log(\alpha/D))} = D^K\alpha^{-K+1}$, which implies that the slope of the resulting $p$-th power distortion $\bar{D}$ is
\begin{equation}\label{eq:d_der1}
\frac{\partial}{\partial D}  \bar{D}(2R) |_{D=\alpha} = \frac{\partial}{\partial D} \frac{2}{p} (D^K\alpha^{-K+1})^p|_{D=\alpha} = 2K\alpha^{p-1}.
\end{equation}
The corresponding RDF is given by $R_X(\bar{D}) = \log(\alpha/(D^K\alpha^{1-K}))$, which has slope:
\begin{equation}\label{eq:r_der1}
\frac{\partial}{\partial D} R = -K\alpha^{-1}.
\end{equation}
Dividing \eqref{eq:r_der1} by \eqref{eq:d_der1} yields \eqref{eq:partial_rd}, which proves the theorem.
\end{IEEEproof}

%
%
%
%
%

\subsection{General Sources and Additive White Gaussian Noise Channels}
In Section~\ref{sec:qg}, we considered the case of a Gaussian source and a source coding scheme that produced Gaussian noise. In this section, we extend it to the case of arbitrarily distributed sources and source coding schemes that produce Gaussian noise. 
However, we do not claim that the efficiency tends to one asymptotically in the limit of low resolution as we did in the previous section. 
Instead, we first introduce an additive RDF, and prove asymptotic efficiency with regards to the additive RDF. Then we consider two different type of test-channels and assess their asymptotic behavior as the mutual information over the channels tends to zero. In this latter case we do not assess the distortion.

\subsubsection{I-MMSE}
Guo et al.~\cite{guoshamai:2005} was able to establish an explicit connection between information theory and estimation theory by using an \emph{incremental} Gaussian channel. For
future reference, we include one of their results below:
\begin{theorem}[\cite{guoshamai:2005}]\label{theo:gsv}
Let $N$ be zero-mean Gaussian of unit variance, independent of $X$, and let $X$ have an arbitrary distribution that satisfies $\mathbb{E}[X^2] < \infty$. Then
\begin{equation}\label{eq:immse}
\frac{\mathrm{d}}{\mathrm{d}\gamma} I(X;\sqrt{\gamma} X + N) = \frac{\log_2(e)}{2} \mmse(\gamma),
\end{equation}
where
\begin{equation*}
\mmse(\gamma) = \mathbb{E}[ (X - \mathbb{E}[ X| \sqrt{\gamma}X + N]
)^2] = \mathbb{E}[\mathrm{var}(X|\sqrt{\gamma}X+N)].
\end{equation*}
\end{theorem}

We now model the output of the source coder by $Y_i = \sqrt{\gamma}X + N_i$, where $\gamma$ is equal to the SNR if $N_i$ is unit normal. We will show that we can embed the unconditional incremental refinement property within the I-MMSE relation. Before doing this, we first need the following lemma,
which is inspired by oversampling for multiple descriptions~\cite{ostergaard:2009}. Specifically, if one oversamples a source $K$ times, and adds mutually independent noises to the oversampled source sequence, then the resulting noisy source sequence satisfies the property of linearity in the mutual information:
\begin{lemma}[Mutual information in noisy oversampling: Lemma 2 in \cite{ostergaard:2011}]\label{theo:oversampling}
Let $X$ be arbitrarily distributed with
variance $\sigma_X^2$, and let $N_i\indp X, i=0,\dotsc, K-1,$ be a sequence
of zero-mean mutually independent Gaussian sources each with variance $\sigma_N^2$. Then
\begin{equation*}
I(X;X+N_0, \dotsc, X+N_{K-1}) = I(X;X+ \frac{1}{\sqrt{K}} N_0).
\end{equation*}
\end{lemma}

\begin{lemma}[I-MMSE for an oversampled source: Lemma 3 in \cite{ostergaard:2011}]\label{theo:uncond}
Let $Y_i = \sqrt{\gamma}X + N_i, i =0,\dotsc, K-1$, where $X$ is arbitrarily
distributed with variance $\sigma_X^2$, and $N_0,\dotsc, N_{K-1},$
are zero-mean unit-variance i.i.d.\ Gaussian distributed independent of $X$.  Then
\begin{align}\label{eq:jointd21}
\lim_{\gamma \to 0}\frac{1}{\gamma} I(X;Y_0^{K-1}) &= K
\lim_{\gamma \to 0}\frac{1}{\gamma} I(X;\sqrt{\gamma}X + N_0) \\ \notag 
&= \frac{K\log_2(e)}{2}\sigma_X^2
\end{align}
and
\begin{equation}\label{eq:jointd}
\lim_{\gamma \to 0}\frac{1}{\gamma}\bigg[\frac{1}{\mathbb{E}[\mathrm{var}(X|Y_0^{K-1})]} -\frac{1}{\sigma_X^2}\bigg] = K.
\end{equation}
\end{lemma}

If $X$ is Gaussian, then \eqref{eq:jointd21} follows from the linearity of the mutual information in the SNR at low SNR, while the conditional variance in \eqref{eq:jointd} is equal to $(\sigma_X^{-2} + K\gamma)^{-1}$, for all $\gamma$. Thus, Lemma~\ref{theo:uncond} demonstrates that the same expressions hold even if the source is \emph{not} Gaussian in the limit of low SNR.

Lemma~\ref{theo:uncond} is concerned with the limit where $\gamma\to 0$. For the case of a small but non-zero $\gamma$, we can turn Lemma~\ref{theo:uncond} into an approximation by using the asymptotic expansion of the mutual information given in \cite[Eq. (92)]{guoshamai:2005}:
\begin{corollary}
Let the setup be the same as in Lemma~\ref{theo:uncond}. For $\gamma\approx 0$, we have:
\begin{equation}\label{eq:approx}
 I(X;Y_0^{K-1}) =  \frac{K\log_2(e)}{2}\sigma_X^2 \gamma- \frac{K^2 \log_2(e)}4 \sigma_X^4 \gamma^2 + o(\gamma^3).   
\end{equation}
\end{corollary}

\subsubsection{The Additive RDF}
The ARDF describes the best R-D performance achievable
by an additive test channel followed by optimum estimation, including
the possibility of time sharing (convexification)~\cite{zamir:1996}. 
We will here restrict attention to Gaussian noise, MMSE
estimation (MSE distortion), and no time-sharing, so we take the ``freedom'' to use
the notation \emph{additive RDF}, $R_X^{\text{add}}(D)$, for this special case (i.e.\ no minimization over free parameters).
Let the additive noise $N$ be zero-mean Gaussian
distributed with variance $0<\theta<\infty$. Then,  we define the simplified version of the ARDF in the following way:
\begin{equation*}
R_X^{\text{add}}(D) \triangleq I(X;Y),
\end{equation*}
where the noise variance $\theta$ is chosen such that $D=\mathbb{E}[\mathrm{var}(X|Y)]$, and $Y=\sqrt{\gamma}X + N$, for some $\gamma \geq 0$.

\subsubsection{Asymptotic optimality of the additive RDF}
We will now show that the slope of $R_X^{\text{add}}(D)$ at $D=D_{\text{max}}$ for
a source $X$ with variance $\sigma_X^2$ is independent of the
distribution of $X$. In fact, the slope is identical to the slope of the RDF of a Gaussian source $X'$ with
variance $\sigma_{X'}^2=\sigma_X^2$. 

\begin{lemma}[Slope of  additive RDF: Lemma 1 in \cite{ostergaard:2011}]\label{theo:slope_fx}
Let $Y=\sqrt{\gamma}X + N,$ where $N \indp X$, $X$ is arbitrarily
distributed with variance $\sigma_X^2$ and $N$ is Gaussian
distributed according to $\mathcal{N}(0,1)$. Moreover, let $R_X^{\text{add}}(D)$
be the additive RDF. Then 
\begin{equation}\label{eq:slope}
\lim_{D\to D_{\mathrm{max}}}\frac{\mathrm{d}}{\mathrm{d}D} R_X^{\text{add}}(D)= -\frac{\log_2(e)}{2\sigma_X^2},
\end{equation}
irrespective of the distribution on $X$.
\end{lemma}

\begin{IEEEproof}[Proof of Lemma~\ref{theo:slope_fx}]
This proof is also given in \cite{ostergaard:2011}. We repeat it here in order to be able to make an explicit reference to \eqref{eq:mmse_g}, which is needed in the proof of Lemma~\ref{lem:rateloss}.
Note that $D$ is a function of $\gamma$, i.e., $D(\gamma) = \mathbb{E}[\mathrm{var}(X|\sqrt{\gamma}X + N)]$. Thus, the additive RDF is defined parametrically as $R_X^{\text{add}}(D(\gamma))$.
%
From the derivative of a composite function, it follows that 
\begin{equation*}
\frac{\mathrm{d}}{\mathrm{d}D} R_X^{\text{add}} =
\frac{\frac{\mathrm{d}}{\mathrm{d}\gamma} R_X^{\text{add}}}{
\frac{\mathrm{d}}{\mathrm{d}\gamma} D
}.
\end{equation*}
From~\cite{guoshamai:2005}, we identify that $R_X^{\text{add}}(D(\gamma)) = I(\gamma)$ and $D(\gamma) = \mathrm{mmse}(\gamma)$, and that $I(\gamma)$ and $\mathrm{mmse}(\gamma)$ have the following convergent Taylor series expansions:
\begin{align}\label{eq:ig}
&I(\gamma) = \log_2(e)\bigg[\frac{1}{2}\gamma\sigma_X^2 -\frac{1}{4}\gamma^2\sigma_X^4 +
\frac{1}{6}\gamma^3\sigma_X^6 \\ \notag
& - \frac{1}{48}\bigg[ (\mathbb{E}X^4)^2 -
  6\mathbb{E}X^4 - 2(\mathbb{E}X^3)^2 + 15 \bigg]\gamma^4\sigma_X^8 + o(\gamma^5)\bigg], \\  \label{eq:mmse_g}
&\mmse(\gamma) = \sigma_X^2-\gamma\sigma_X^4 +
\gamma^2\sigma_X^4 +\frac{1}{6}\gamma^3\sigma_X^6 + o(\gamma^4).
\end{align}
Since the series expansions are convergent, we can do term-wise differentiation from which it follows that the higher order terms vanish as $\gamma\to 0$.
Thus, from~(\ref{eq:ig}) we obtain:
\begin{equation*}
\lim_{\gamma\to 0}
\frac{\mathrm{d}^2}{\mathrm{d}\gamma^2}  I(\gamma) = -\frac{\log_2(e)}{2}\sigma_X^4,
\end{equation*}
and  $\lim_{\gamma\to 0}\frac{\mathrm{d}}{\mathrm{d}\gamma}
D = 2\frac{\mathrm{d}^2}{\mathrm{d}\gamma^2}  I(\gamma) =
-\sigma_X^4$. 
Moreover, since $\lim_{\gamma\to 0}\frac{\mathrm{d}}{\mathrm{d}\gamma}
R_X^{\text{add}} = \frac{\log_2(e)}{2}\mmse(\gamma)=\log_2(e)\sigma_X^2/2$ and since $\gamma\to 0$ implies $D\to D_{\mathrm{max}}$, we have that the slope of $R_X^{\text{add}}(D)$ with respect to
$D$ at $D=D_\mathrm{max}$ is 
\begin{equation*}
\lim_{D\to D_{\mathrm{max}}}\frac{\mathrm{d}}{\mathrm{d}D}R_X^{\text{add}} = -\frac{\log_2(e)}{2\sigma_X^2}. 
\end{equation*}
\end{IEEEproof}


Comparing Lemma~\ref{theo:slope_fx} to Lemma~\ref{theo:uncond} we deduce that the incrementally refineable source coder proposed above is asymptotically ARDF optimal. This, however, does not mean that it is  asymptotically rate-distortion optimal since the ARDF may suffer from a rate loss with respect to the true RDF. Since we are operating at very small rates, it is meaningful to consider the \emph{multiplicative} rate loss of the ARDF instead of the \emph{additive} rate loss. In Lemma~\ref{lem:rateloss}, we show that there exists sources, where the multiplicative rate loss is unbounded. 
\begin{lemma}[Asymptotic multiplicative rate loss of the additive RDF]\label{lem:rateloss}
The multiplicate rate loss between the ARDF and the RDF may be unbounded, i.e., there exists finite variance sources, where 
\begin{equation*}
\lim_{D\to \sigma_X^2} \frac{R_X^{\text{add}}(D)}{R_X(D)}\to\infty.
\end{equation*}
\end{lemma}

\begin{IEEEproof}
We provide the complete proof here, since only a partial proof was given in \cite{ostergaard:2011}.
Let $X$ be a Gaussian mixture source with a density $P_X(x)$ given by
$P_X(x) = P_0 \mathcal{N}(0,\sigma_0^2) + P_1 \mathcal{N}(0,\sigma_1^2)$,
where $P_0+P_1=1$.
The variance $\sigma_X^2$ of $X$ is $\sigma_X^2 = P_0 \sigma_0^2 + P_1\sigma_1^2$.
The components contribution can be parametrized by $\lambda\in [0;1]$ as
follows: $P_0\sigma_0^2 = \lambda \sigma_X^2, P_1\sigma_1^2 = (1-\lambda)\sigma_X^2$. It will be convenient to let $\sigma_X^2 = 1$ and
$\lambda=\frac{1}{2}$. Moreover, we shall assume that $\sigma_1^2 > 1
> \sigma_0^2 \geq  \frac{1}{2}$.
Notice that as $\sigma_1^2\to\infty$ we have that $P_1\to 0, P_0\to
1$, and $\sigma_0^2\to \frac{1}{2}$.

At this point, let $S = 0$ with probability $P_0$ and $S=1$ with
probability $P_1$, and let $S$ be an indicator of the two components,
i.e., $X \sim \mathcal{N}(0,\sigma_0^2)$, if $S=0$, and
$X\sim \mathcal{N}(0,\sigma_1^2)$,  if $S=1$.
The RDF, conditional on the indicator $S$, is:
\begin{equation*}
R_{X|S}(D) = 
\begin{cases}
\displaystyle\frac{1}{2}\sum_{i\in \{0,1\}} P_i \log_2(\sigma_i^2 / D), &
\text{if $0<D\leq \sigma_0^2$}, \\[5mm]
\displaystyle\frac{P_1}{2}\log_2\bigg( \frac{P_1\sigma_1^2}{D - P_0\sigma_0^2} \bigg), &  \text{if $\sigma_0^2 < D < 1$}.
\end{cases}
\end{equation*}
Thus, the slope of $R_{X|S}(D)$ w.r.t.\ $D$ is given by
\begin{align*}
\lim_{D\to \sigma_X^2}\frac{\mathrm{d}}{\mathrm{d}D} R_{X|S}(D) &=
 -\frac{P_1}{4\ln(2)\sigma_X^2},
\end{align*}
which tends to zero as $\sigma_1^2\to \infty$ and $P_1\to 0$. It
follows that the ratio of the slope of the conditional RDF and the
slope of the ARDF grows unboundedly as $\sigma_1^2\to
\infty$. 

Informally, as $\sigma_1^2\to \infty$, $\sigma_0^2\to
\frac{1}{2}$, it becomes increasingly easier for the uninformed encoder and decoder to guess the correct mixture component of the source. Thus, the conditional RDF converges towards the true RDF $R_X(D)$, from which it follows that the
ratio $\lim_{\sigma_1^2/\sigma_0^2\to \infty}\lim_{D\to \sigma_X^2} R_X^{\text{add}}(D)/R_X(D)\to\infty$.

We will now formally prove that the encoder can guess (with probability one) the mixture component distribution, which the source sample belongs to.
To do so, 
we assume that the encoder has to guess each sample independently of each other. Thus, at the encoder, we form a hypothesis test deciding whether sample $X$ belongs to a normal distribution of variance $\sigma_0^2$ or of variance $\sigma_1^2$. 

Let $M_i$ denote the event that $X\sim \mathcal{N}(0,\sigma_i^2)$ and let $p_i$ denote the prior probability on $M_i$. Moreover, let $p(M_i|X)$ denote the probability that a given $X=x$ is drawn from $\mathcal{N}(0,\sigma_i^2)$. Then the ratio test can be written as:
\begin{align}
r &\triangleq \frac{p(M_0| X=x)}{p(M_1|X=x)} =
 \frac{ p(X|M_0)p(M_0)}{p(X|M_1)p(M_1)}
\\ \label{eq:ratiotest}
&= \sqrt{\frac{\sigma_1^2}{\sigma_0^2}} \frac{p_0}{p_1} \exp\big(x^2(\frac{1}{2\sigma_1^2}-\frac{1}{2\sigma_0^2})\big).
\end{align}
To find the thresholds $\pm \xi$ upon which to decide in favor of $M_0$ if $|x|\leq \xi$ or in favor of $M_1$ if $|x|>\xi$, we let $r=1$ and solve for $\xi=x$ in~(\ref{eq:ratiotest}), i.e.
\begin{align*} 
\xi &= \pm 2\sqrt{ \frac{\sigma_0^2\sigma_1^2}{\sigma_0^2-\sigma_1^2}\ln\big( \sqrt{\frac{\sigma_0^2}{\sigma_1^2}}\frac{p_1}{p_0}\big)} 
=\pm 2\sqrt{ \frac{3\sigma_0^2\sigma_1^2}{\sigma_0^2-\sigma_1^2}\ln\big( \sqrt{\frac{\sigma_0^2}{\sigma_1^2}}\big)},
\end{align*}
where the last equality follows since $\sigma_0^2p_0 = \sigma_1^2p_1$ and $p_0+p_1=1$, which implies that $p_i = \sigma_j^2/(\sigma_0^2+\sigma_1^2), i\neq j$.
%
For large $\sigma_1^2$ and fixed $\sigma_0^2$, it follows that 
\begin{equation*}
\xi \approx 2\sqrt{3\sigma_0^2 \ln(\sigma_1)}
\end{equation*}
and, thus, $\lim_{\sigma_1^2\to\infty}  \xi \to \infty$.

Let $f_i$ denote the Normal distribution of the $i$th component. Then, for a fixed and finite $\sigma_0^2$, it follows that
\begin{align*}
\lim_{\sigma_1^2\to \infty} \frac{2p_1\int_0^{\xi}  f_1(x) dx}{2p_0\int_{0}^{\xi}  f_0(x) dx} \to 0,
\end{align*}
since $p_1\to 0$ and the remaining terms are positive and bounded. Thus, the probability of observing any symbol $X$ where $|X|<\xi$ and where $X$ is drawn from $f_1$ tends to zero as $\sigma_1^2\to \infty$. 
On the other hand, we show next that
\begin{align}\label{eq:limit}
\lim_{\sigma_1^2\to \infty} \frac{2p_0\int_\xi^{\infty}  f_0(x) dx}{2p_1\int_{\xi}^{\infty}  f_1(x) dx} \to 0.
\end{align}
Thus, the probability of observing a symbol $X$ where $|X|>\xi$ and where $X$ is drawn from $f_0$ also tends to zero. 
To prove~(\ref{eq:limit}), we first note the interesting fact that while  $\lim_{\sigma_1^2\to \infty} \int_\xi^{\infty}  f_0(x) dx = 0$, the integral in the denominator of~(\ref{eq:limit}) is non-zero. Without loss of generality, let $\xi'=\sqrt{\sigma_1^2}$, where for large $\sigma_1^2, \xi'>\xi$, then 
\begin{equation*}
\int_{\xi'}^{\infty}  f_1(x) dx = \frac{1}{2}-\frac{1}{2}\mathrm{erf}\bigg(\frac{1}{2}\sqrt{2}\bigg) 
\approx 0.1587.
\end{equation*}
Thus, we just need to show that $\lim_{\sigma_1^2\to \infty} p_1^{-1}\int_\xi^{\infty}  f_0(x) dx = 0$. Clearly, $p_1 = \frac{\sigma_0^2}{\sigma_0^2+\sigma_1^2} \approx \sigma_1^{-2}$. On the other hand, it is known that~\cite[(7.1.13)]{abramowitz:1972}
\begin{equation}\label{eq:bound_err}
\int_{\xi}^{\infty} e^{-t^2}dt  < \frac{1}{2\xi e^{\xi^2}}.
\end{equation}
Letting $\sigma_0^2=\frac{1}{2}$, and using~(\ref{eq:bound_err}), it follows that 
\begin{align*}
&\int_{\xi=c\sqrt{\ln(\sigma_1^2)}}^{\infty}  f_0(x) dx 
< c' \frac{1}{2c\sqrt{\ln(\sigma_1^2)} e^{(c\sqrt{\ln(\sigma_1^2)})^2}} \\
&= c' \frac{1}{2c\sqrt{\ln(\sigma_1^2)} (\sigma_1^2)^{c^2}} 
< c' \frac{1}{2c\sqrt{\ln(\sigma_1^2)} \sigma_1^2},
\end{align*}
where $c,c'$ are constants greater than one. We have now proved that $ p_1^{-1}\int_\xi^{\infty}  f_1(x) dx = o(\frac{1}{\sqrt{\ln(\sigma_1^2)}})$, which tends to zero as $\sigma_1^2\to \infty$.

The encoder knows (with high probability), which mixture components is active. This information should also be conveyed to the decoder. 
However, when encoding a large block of samples, the entropy $H(S)$ of the component indicator is vanishing small in the limit where $\sigma_1^2\to\infty$ and $p_1\to 0$. Specifically, 
\begin{align} \label{eq:rate}
I(X;Y) &= I(X,S;Y) - I(S;X|Y) \\ \notag
&= I(X;Y|S) + I(S;Y) - I(S;Y|X) \\ \notag
&\leq I(X;Y|S) + I(S;Y) \\ \notag
&= I(X;Y|S) + H(S) - H(S|Y)  \\ \label{eq:cond_rate}
&\leq I(X;Y|S) + H(S),
\end{align}
where $H(S) = -p_0\log_2(p_0) - p_1\log_2(p_1)$ so that 
\begin{align} \label{eq:H(s)}
\lim_{\sigma_1^2\to\infty} H(S) &= \lim_{\sigma_1^2\to\infty} - p_1\log_2(p_1) = 0.
\end{align}
Using~(\ref{eq:H(s)}) in~(\ref{eq:cond_rate}), shows that 
\begin{equation*}
\lim_{\sigma_1^2\to\infty} I(X;Y) \leq \lim_{\sigma_1^2\to\infty} I(X;Y|S)
\end{equation*}
and the encoder may transmit using a coding rate equal to the conditional RDF $R_{X|S}(D)$.\footnote{The coding rate overhead due to informing the decoder about the locations of the improbable components is also upper bounded by $\lceil(\log_2 \binom{L}{ p_1 L}\rceil$, where $L$ denotes the block length. When $p_1\to 0$, this overhead clearly vanishes.}
\end{IEEEproof}

\subsection{General Sources and Very Noisy Channels}
As is standard in rate distortion theory, we consider test channels as a means to obtain the encoding functions. Towards that end, we introduce  so-called \emph{noisy} channels that can model data compression at low rates. 
At vanishing small data rates, we show that the channels exhibit similar asymptotic behaviour from a mutual information perspective.
In particular, if the  source $X$ is independently encoded into $Y$ and $Z$, so that $X\to Y$ and $X\to Z$ both satisfy certain properties of the noisy test channels, 
then their  mutual informations satisfy $I(X;Y,Z) \approx I(X;Y) +  I(X;Z)$. 
If the Markov chain $Y - X - Z$ holds, then  $I(X;Y,Z) = I(X;Y) + I(X;Z) - I(Y;Z)$, and to establish the aforementioned approximation, we will then only need to show that at low coding rates, i.e., when both $I(X;Y)$ and $I(X;Z)$ are small, then $I(Y;Z)$ is even smaller. 

\subsubsection{Discrete-Discrete $\epsilon/1$-channel}
Recall the very noisy channel (VNC), which was used by Gallager in~\cite{gallager:1965} in order to derive a universal exponent-rate curve in the limit of low SNR. A similar notion of VNCs was proposed by Lapidoth et al.\ in~\cite{lapidoth:2003}. Here the motivation was
to study noisy wideband broadcast channels. This VNC has both discrete input and output alphabets. 
We show below, that the mutual information in this test channel is  linear in the number of descriptions at low rates.

Let $P_X$ be the input distribution and $P_Y$ the marginal distribution of the channel output. 
Then, a VNC is defined by~\cite{reiffen:1963,gallager:1965,lapidoth:2003}:
\begin{equation}\label{eq:DVNC}
P_{Y|X}(y|x) \triangleq P_Y(y) \big( 1 + \epsilon \psi(y | x) \big), \quad
\forall\, x \in\mathcal{X}, y\in\mathcal{Y},
\end{equation} 
where $\psi(y|x)$ is any function satisfying $\sum_y  P_Y(y) \psi(y|x) =0, \forall x,$ in order to guarantee  that (\ref{eq:DVNC}) is a well-defined conditional distribution. In some sense $\psi(y|x)$ describes the shape or direction of a local pertubation of the joint distribution on $\mathcal{X}$ and $\mathcal{Y}$, and $\epsilon$ decribes its size  \cite{abbe:2008}. 
It was shown in~\cite{lapidoth:2003,gallager:1965} that if $X\rightarrow Y$ defines a VNC, then the mutual information
$I(X;Y)$ is proportional to $\epsilon^2$:
\begin{equation}\label{eq:epsMI}
\lim_{\epsilon\to 0} \frac{1}{\epsilon^2} I(X;Y) = \frac{1}{2} \sum_y
P_Y(y) \sum_x P_X(x) \eta(y|x)^2,
\end{equation}
where
\begin{equation*}
\eta(y|x) = \psi(y|x) - \sum_{x'} P_X(x')\psi(y|x').
\end{equation*}
Notice that the r.h.s.\ of~(\ref{eq:epsMI}) is independent of
$\epsilon$ and non-zero if $\psi(y|x)$ depends upon $x$.

The following lemma shows that if $Y$ and $Z$ are conditionally independent given $X$, and if $X\to Y$ and $X\to Z$ are VNCs, then the cascading (product) channel $Y\to Z\ (P_{Y|Z}=P_{Y|X}P_{X|Z})$ is much more noisy than the individual channels $X\to Y$ and $X\to Z$. This implies that $I(Y;Z)$ tends to zero faster than either of $I(X;Y)$ and $I(X;Z)$ as $\epsilon \to 0$. 

\begin{lemma}[Mutual information over very noisy discrete channels]\label{theo:noisy_mutual1}
Let $X,Y,Z$ be discrete random variables that satisfy the Markov chain: $Y - X - Z$. Moreover, let $X$ be arbitrarily distributed  with $P(x) > 0, \forall x$, and let $X\to Y$ and $X\to Z$ be VNCs so that $P_{Y|X}(y|x) =
P_Y(y)(1+\epsilon \psi_Y(y|x))$ and $P_{Z|X}(z|x) =
P_Z(z)(1+\epsilon \psi_Z(z|x))$. Then,
\begin{equation*}
\lim_{\epsilon\to 0} \frac{1}{\epsilon^2} I(X;Y,Z) = \lim_{\epsilon\to 0} \frac{1}{\epsilon^2} \left[ I(X;Y)+I(X;Z) \right].
\end{equation*}
\end{lemma}
\begin{IEEEproof}
See the appendix.
\end{IEEEproof}


%
\subsubsection{Continuous-Discrete $\epsilon/1$-channel}\label{sec:eps-1}
We now introduce a test-channel with a continuous input alphabet and a discrete output alphabet. We refer to this channel as a continuous-discrete $\epsilon/1$-channel if  one particular output symbol, say $y'\in\mathcal{Y}^d$, has almost all the probability.
%
Thus a very peaky distribution is induced on the channel output symbols. 
Let $0<\epsilon\ll 1$, then:
\begin{equation}\label{eq:VNC2}
P_Y(y') = \int_{x\in \mathcal{X}} P_X(x) P_{Y|X}(y'|x) \, dx=  1 - \epsilon.
\end{equation}

This model is motivated by the fact that for many sources and distortion criteria, the optimal reconstruction alphabet contains very few elements near $D_\mathrm{max}$~\cite{rose:1994}. 
In particular, for common distortion measures (e.g., Hamming, MSE), $D_\mathrm{max}$ is achieved by a unique channel, that maps all inputs into a single output $y* = \min_y E[d(X,y)]$ \cite{rose:1994,verdu:1990}. Moreover, the optimal test channel changes continuously with $D$, and near $D=D_\mathrm{max}$ the optimal channel becomes an $\epsilon/1$-channel \cite{rose:1994}.
We will consider the asymptotic case where $\epsilon_Y,\epsilon_Z\to 0$  and their  ratio $\epsilon_Y/\epsilon_Z$ is fixed.

\begin{lemma}[Mutual information over very noisy continuous-discrete channels]\label{theo:epsilon}
Let $X$ be a  random variable with pdf $P_X$ and let $Y,Z$ be discrete random variables that satisfy the Markov chain: $Y - X - Z$. Moreover, let $X\to Y$ and $X\to Z$ be two continuous-discrete $\epsilon_Y/1$ and $\epsilon_Z/1$ channels given by $P_{Y|X}$ and $P_{Z|X}$, respectively. Then, the following asymptotic expression holds:
\begin{equation}\label{eq:mi_ratio}
\lim_{\epsilon_Y,\epsilon_Z\to 0} \frac{I(X;Y,Z)}{I(X;Y)+I(X;Z)} = 1.
\end{equation}
\end{lemma}
\begin{IEEEproof}
See the appendix.
\end{IEEEproof}
This shows that near zero rate, the mutual information over the continuous-discrete $\epsilon/1$ test channel is also linear in the number of descriptions, as was the case for the AWGN channel as well as for the discrete-discrete $\epsilon/1$ test channel.

\section{Multiple Rounds of Incremental Refinements}
In the previous section, we considered coding a source into $K$ descriptions in a single round. Optimality was then assessed in the limit, where the coding rate tended to zero. This will necessarily lead to a large distortion. To decrease this distortion, we turn our attention to multi round encoding in this section. In each round, we can use a small coding rate, and then incrementally improve the distortion, see e.g., Fig.~\ref{fig:setup}. 
To establish optimality, we will consider the asymptotical limit, where the number of rounds $M$ tends to infinity. Moreover, the coding rate in each round tends to zero, yet the total sum-rate over all rounds is finite and positive. For a given distortion, we then assess asymptotic optimality via the rate-loss of the operational RDF compared to the true RDF. 

A necessary condition for a small rate loss in the multiple round case is that the source be successively refinable, or even better -- from a practical viewpoint -- additively successively refinable \cite{tuncel:2003}. But this is not sufficient:  in order to attain an overall efficiency $MKR/R_X(D)$ that is close to one, the efficiency in most rounds must be close to one. Furthermore, to achieve a small (additive) excess rate $MKR-R_X(D)$, we need to keep the accumulated rate loss over all $(M\to \infty)$ rounds small.


\subsection{Gaussian Sources and MSE Distortion}
Let us consider the situation of a zero-mean unit-variance memoryless
Gaussian source $X$, which is to be encoded successively in $M$
rounds with $K$ descriptions per each round. 


We begin the first round by encoding $X$ unconditionally into $K$ descriptions $Y_{1,j}, j=1,\dotsc, K$. Let each description yield distortion $d_1 \triangleq \mathbb{E}[\mathrm{var}(X|Y_{1,j})]  = \sigma_X^2/(1+\gamma\sigma_X^2)$, which implies that $\gamma = (\sigma_X^2 - d_1)/(d_1\sigma_X^2)$. It follows that each description uses rate $r_1 = \frac{1}{2}\log_2(\sigma_X^2/d_1)$, and 
the joint distortion $\bar{D}_1$ when using all $K$ descriptions is:
\begin{align}\label{eq:D1L}
\bar{D}_1 &=  \mathbb{E}[\mathrm{var}(X|Y_{1,1},\dotsc, Y_{1,K})] = 1/(\sigma_X^{-2} + K\gamma) \\ \label{eq:D1L1}
&= \frac{d_1}{K - (K-1)d_1/\sigma_X^2},
\end{align}
where the r.h.s.\ of \eqref{eq:D1L} follows by the arguments presented below \eqref{eq:jointd}, and \eqref{eq:D1L1} follows by inserting the expression for $\gamma$.
In round 2, we form the residual source $\mathbb{E}[X|Y_{1,1},\dotsc, Y_{1,K}]$ obtained by estimating $X$ from the $K$ descriptions obtained in the first round. Notice that the variance of this residual source is $\bar{D}_1$. We then encode the residual source into $K$ descriptions $Y_{2,1},\dotsc, Y_{2,K}$, where each description yields the distortion $d_2 = \bar{D}_1/(1+\gamma \bar{D}_1)$. Similarly, 
in round $i$, $K$ descriptions $Y_{i,j}, j=1,\dotsc, K$, are
constructed unconditionally of each other. Thus, the joint distortion $\bar{D}_M$ of the $M$th round when coding the residual source of variance $\bar{D}_{M-1}$ is
given by
\begin{equation}\label{eq:Dsuc}
\bar{D}_M  = \frac{d_M}{K - (K-1)d_M / \bar{D}_{M-1}}
\end{equation}
and the sum-rate at round $M$ is given by
\begin{equation}\label{eq:Rsuc}
R_M = K\sum_{j=1}^M  r_M = K  \sum_{j=1}^M  \frac{1}{2} \log_2( \bar{D}_{j-1} / d_j ).
\end{equation}

\begin{figure}[th]
\begin{center}
\includegraphics[width=8cm]{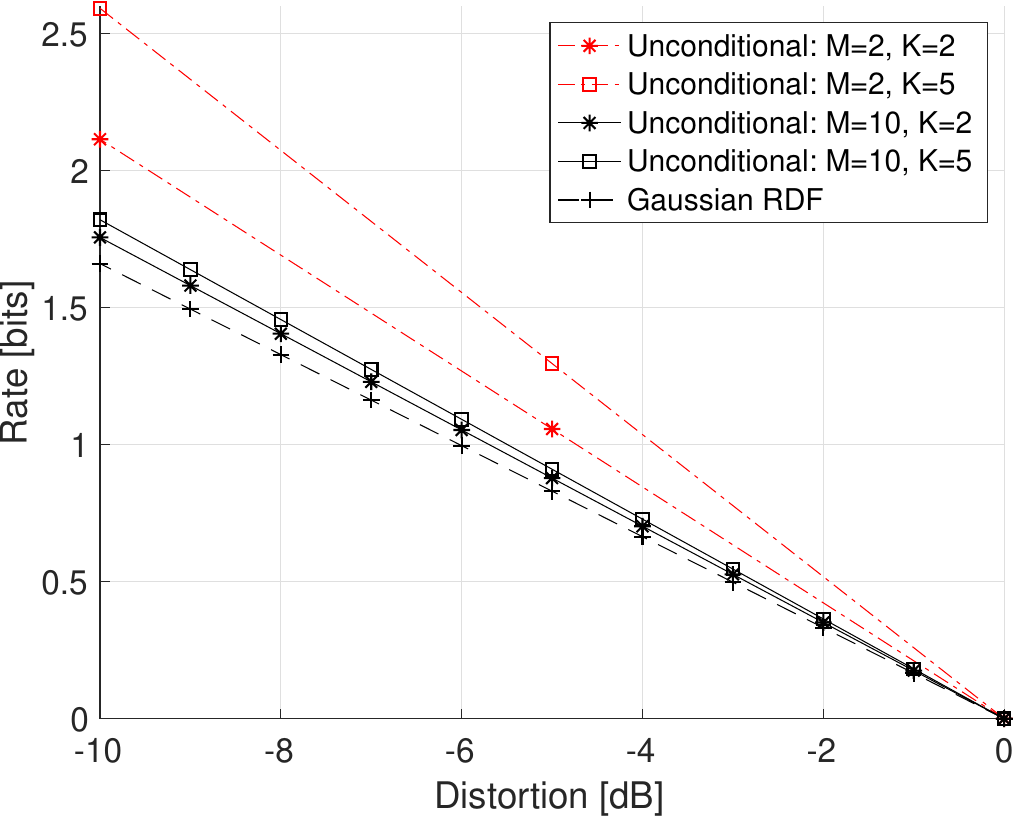}
\vspace{-2mm}
\caption{Unconditional and Gaussian (conditional refinement coding) using $M$ rounds and $K$ descriptions in each round. }
\label{fig:uncond1}
\end{center}
\vspace{-4mm}
\end{figure}

Since the Gaussian
source is successively refinable, constructing $K$ descriptions in each round, that are conditioned upon each other,  will
achieve the true RDF with a sum-rate at round $M$ of $R_M^* = \frac{1}{2}\log_2( \bar{D}_{M-1}/\bar{D}_{M} )$,
where $\bar{D}_M$ is given by~(\ref{eq:Dsuc}).\footnote{Using $K$ \emph{conditional} descriptions at rate $r$ leads to the same joint distortion as that of a single description at rate $Kr$.} On the other hand, the rate
required when unconditional coding is used in each round is given
by~(\ref{eq:Rsuc}). For comparison, we have illustrated the
performance of unconditional coding and conditional coding when the source is
encoded into $K\in \{2,5\}$ descriptions per round, 
for the case of 
of $M\in \{2,10\}$ rounds, see
Fig.~\ref{fig:uncond1}. In this example,
$\sigma_X^2=1$ and $\bar{D}_M = 0.1$. 
The red dashed curve illustrates the case when we use $M=2$ rounds with $\bar{D}_1 = 0.3162, \bar{D}_2=0.1$ (marked by stars) and $K=2,5$. The black solid curves illustrate the cases when using $M=10$ rounds (and $K=2,5$) to get to the same resulting distortion $\bar{D}_{10}=0.1$. The  distortions $\bar{D}_i, i=1,\dotsc, 10$ are uniformly distributed on the dB scale. 
Notice that when using smaller
increments, i.e., when $M=10$  as compared to when $M=2$ (red curve), the resulting
rate loss due to using unconditional coding is significantly reduced. On the other hand, using less descriptions per round (smaller $K$) also reduces the rate loss.

The efficiency $\eta_K(R) = R_X(D)/(KR)$, for the example in Fig.~\ref{fig:uncond1} on Gaussian multi-round coding is illustrated in Table~\ref{tab:eff}.
It can be noticed that the efficiency  depends upon the number of descriptions in each round as well as the number of rounds. 
\begin{table}[th]
\begin{center}
\begin{tabular}{|c|c|c|c|}\hline
$M=2,K=2$ & $M=2,K=5$ & $M=10,K=2$ & $M=10, K=5$ \\ \hline
0.7854 & 0.6407 & 0.9457 & 0.9121 \\ \hline
\end{tabular}
\caption{Efficiency: $\eta_K(R) = R_X(D)/(KR)$. For unconditional Gaussian multi-round coding with $M$ rounds and $K$ descriptions in each round. }
\label{tab:eff}
\end{center}
\end{table}

\begin{theorem}[Asymptotic optimal sum-rate for a Gaussian source]\label{theo:sumrate-g}
Consider a Gaussian source, MSE distortion, and $K$ independent encodings in each round. Moreover, assume that the rate of each description in each round is equal. Then, 
for any fixed distortion $D>0$, the total accumulated sum-rate $R_M$ is asymptotically optimal as the number of rounds tends to infinity, and the rate in each round tends to zero. Specifically, 
\begin{equation}
\lim_{M\to \infty} R_X(D) - R_M =  0.
\end{equation}
\end{theorem}

\begin{IEEEproof}
To obtain an equal sum-rate in each round (and thereby an equal description rate), it follows from \eqref{eq:Rsuc} that we must choose $d_i$ such that $\bar{D}_{i-1}/d_{i} = \bar{D}_{i}/d_{i+1}, \forall i$. Thus, $d_{i+1} = d_i \bar{D}_{i}/\bar{D}_{i-1}$. 
Inserting this relationship into \eqref{eq:Dsuc} leads to a resulting joint distortion after $M$ rounds given by:
\begin{align}
\bar{D}_{M} = \frac{d_1^M \bar{D}_{0}}{(\bar{D}_{0}K - (K-1)d_1)^M} = \frac{d_1^M \sigma_X^2}{(\sigma_X^2 K - (K-1)d_1)^M}.
\end{align}
On the other hand, the resulting total sum-rate after $M$ rounds is given by:
\begin{align}
R_M &=  \frac{K M}{2} \log_2 (\bar{D}_{0}/d_1) = \frac{K M}{2} \log_2 (\sigma_X^2/d_1). 
\end{align}
In order to obtain a finite rate and a non-zero distortion, when $M\to \infty$, we let the  distortion $d_1$ of round 1 depend upon $M$ through the distortion parameter $d$, i.e.:
\begin{align}
d_1 = \bigg(\frac{\sigma_X^2}{d}\bigg)^{-1/M}\sigma_X^2, 
\end{align}
where $\sigma_X^2\geq d>0$. 
The resulting sum-rate for any $M$ rounds is therefore fixed at:
\begin{align}\label{eq:result_sumrate}
R_M = \frac{K }{2} \log_2 \big(\frac{\sigma_X^2}{d} \big),
\end{align}
and the resulting distortion is:
\begin{align}
\bar{D}_M =  \frac{ d (\sigma_X^2)^M }{(\sigma_X^2 K - (K-1) \bigg(\frac{\sigma_X^2}{d}\bigg)^{-1/M}\sigma_X^2 )^M},
\end{align}
which has the following limit:
\begin{align}\label{eq:limit_M}
\lim_{M\to \infty} \bar{D}_M = \sigma_X^2 \bigg( \frac{d}{\sigma_X^2}\bigg)^K. 
\end{align}

The optimal distortion at rate $R_M$ is:
\begin{align}
D^* &=  D_X(R_M) = \sigma_X^2 2^{-2 R_M} \\
&=  \sigma_X^2 2^{-2 (\frac{K }{2} \log_2 \big(\frac{\sigma_X^2}{d} \big))} \\ \label{eq:smalld}
&= \sigma_X^2 \bigg( \frac{d}{\sigma_X^2}\bigg)^K, 
\end{align}
which is equal to \eqref{eq:limit_M}. 

For any fixed distortion $D$, choosing $d = \sigma_X^2( D/\sigma_X^2)^{1/K}$ leads to the sum-rate $R_M$ \eqref{eq:result_sumrate}, which is asymptotically optimal as the number of rounds tends to infinity and the resulting
operational distortion $\bar{D}_{M} \to D$.
\end{IEEEproof}

\subsection{One Sided Exponential Source Under One Sided Error Criterion}
In Section \ref{sec:expsource}, the one-sided exponential source was encoded into $K$ independent encodings and decoded using the select-max estimator. It was shown that such an approach was asymptotically optimal under the one-sided error criterion in the limit of zero rate. In this section, we extend this result to multiple rounds, where the  estimation error in round $M$ becomes the source in round $M+1$. 
In each round, we assume the source is encoded into $K$ descriptions each of rate $R$. Thus, the sum-rate of each round is $KR$. The total sum-rate over $M$ rounds is then given by $MKR$. 

\begin{lemma}[Successive refinability of an exponential source]\label{lem:exp_sr}
Let $X$ be one-sided exponentially distributed. Then, for $K=1$ encodings, $X$ is successive refinable under the one-sided error distortion for any coding rate $R$ and any rounds $M$. 
\end{lemma}

\begin{IEEEproof}
Let $\lambda_M$ denote the source distribution in round $M$, then the distribution of the error $z_M$ in round $M+1$ is 
$\lambda_{M+1}=\lambda_M + K\delta_M$, where $\delta_M = D_M^{-1} - \lambda_M$, and $D_M$ is the single-channel (test-channel) distortion achieved by each individual description. 
From Lemma~\ref{lem:exp_error} the source in round $M+1$ is  exponentially distributed with parameter $\lambda_{M+1}$ given as:
\begin{align*}
\lambda_{M+1} &= \lambda_{M}+K(D_M^{-1} - \lambda_{M}).
\end{align*}
The total distortion $\bar{D}_M$ in round $M$ is:
\begin{align}\label{eq:D1}
\bar{D}_{M} = \frac{D_{M}}{K-(K-1)D_{M}\lambda_{M}}.
\end{align}

We now fix the rate of each test channel in each round to be $R$. 
Using that $R_X(D_M) = - \log_2(\lambda_{M} D_M)$, we obtain  $\lambda_{M} D_M = 2^{-R}$, which after using \eqref{eq:D1} yields:
\begin{align*}
\bar{D}_M &=  \frac{D_M}{K-(K-1)D_M\lambda_M} \\
&=   \frac{\lambda_M^{-1}2^{-R}}{K-(K-1)D_M\lambda_M}  \\
&= \lambda_{M+1}^{-1}. 
\end{align*}


Let $R'$ be the total rate after $M$ rounds each having $K$ descriptions so that the rate per description is $R=R'/(KM)$. Thus, after $M$ rounds we have:
\begin{align*}
\lambda_{M+1} &= \lambda_M + K(D_M^{-1}-\lambda_M) \\
&=  \lambda_M(1 + K(2^{R'/(K M)}-1)) \\
&= \lambda_{M-1} (1 + K(2^{R'/(K M)}-1))^2 \\
&= \lambda_{1} (1 + K(2^{R'/(K M)}-1))^M.
\end{align*}

For $K=1$, we obtain 
\begin{align*}
\lambda_{M+1} &= \lambda_{1} 2^{MR},
\end{align*}
which shows that for $K=1$, the source is successive refinable for any $R$ and any number of rounds $M$.  
\end{IEEEproof}

Lemma~\ref{lem:exp_sr} proves successive refinability for $K=1$ and any coding rate. For $K>1$, the source is generally not successive refinable. However, for any finite positive sum-rate, the source is asymptotically successively refinable in the limit, where the number of rounds tends to infinity, and the rate in each round tends to zero. This can be deduced from the following theorem.

\begin{theorem}[Asymptotic optimality of independent encodings of exponential sources]\label{lem:exp_opt}
For any finite total sum-rate and under the one-sided error distortion measure, it is asymptotically rate-distortion optimal to encode a one-sided exponential source into $K\geq 1$ independent encodings and using the select-max decoder in each round, as the number of rounds $M$ tends to infinity.
\end{theorem}

\begin{IEEEproof}
The proof for the case of $K=1$ follows from Lemma~\ref{lem:exp_sr}. 
For $K>1$, we consider the case of an asymptotically large number of rounds $M$. Specifically, we multiply $\lambda_{M+1}$ by $2^{-R'}$ and show that this product tends to $\lambda_1$ as $M\to \infty$, which implies that $\lambda_{M+1} \to \lambda_1 2^{R'}$ as $M\to \infty$, that is:
\begin{align*}
&\lim_{M\to \infty} 2^{-R'} \lambda_{M+1} = \lim_{M\to \infty}2^{-R'}   \lambda_{1} (1 + K(2^{R'/(K M)}-1))^M \\
&= \lim_{M\to \infty}  \lambda_{1} (2^{-R'/M}  + K(2^{-R'/(K M)(K-1)}- 2^{-R'/M}))^M \\
&= \lambda_1.
\end{align*}
Since $\bar{D}_M = \lambda_{M+1}^{-1}$, it follows that for any finite $K\geq 1$, and any total rate  $R'>0$, the total distortion $\bar{D}_{M} \to \lambda_{1}^{-1} 2^{-R'}$ asymptotically in $M$, which is equivalent to the rate-distortion behaviour of ideal successive refinement. 

Since the total sum-rate is finite, it follows that the sum-rate in each round must tend to zero as the number of rounds tends to infinity. 
In each round, using $K$ independent encodings followed by the select-max decoder is asymptotically optimal as the rate tends to zero. The proof now follows immediately since the source becomes asymptotically successive refinable as the number of rounds tends to infinity. Thus, the rate loss within each round due to using independent encodings vanishes, and due to the source being asymptotically (in the number of rounds) successive refinable, the accumulated rate loss over many rounds also vanishes.
\end{IEEEproof}

\subsection{General Sources, Gaussian Noise Channels, and MSE Distortion}

\subsubsection{Multi Round Non Gaussian Source Coding}\label{sec:incremental_side}

In the single round case, $K$ independent encodings (descriptions) are constructed. In the multi round case, the new source is the residual from the previous round. This residual could be dependent upon descriptions from previous rounds. To address this issue,  we extend Lemma~\ref{theo:uncond} to include the possibility of having "side information" available at the encoder and decoder, which is dependent upon the source. The case of  side information available at the encoder and the decoder was not considered in the I-MMSE relation~\cite{guoshamai:2005}. 
With Lemma \ref{theo:condmi} and Corollary~\ref{cor:uncond} below, we generalize Theorem~\ref{theo:gsv} and Lemma~\ref{theo:uncond} to include side information.
\begin{lemma}[Conditional I-MMSE relation]\label{theo:condmi}
Let $Y=\sqrt{\gamma}X+N$ where $N\sim \mathcal{N}(0,1)$ and 
$X$ is
arbitrarily distributed, independent of $N$ and of variance
$\sigma_X^2$. 
Let $Z$ be arbitrarily
distributed and correlated with $X$ but independent of $N$. Then
\begin{align*}
I(X;Y|Z) &= \frac{\log_2(e)}{2}\big(\gamma \mathbb{E}[ \mathrm{var}{(X|Z)}]  - \frac{1}{2} \gamma^2 \mathbb{E}[ \mathrm{var}{(X|Z)}^2]\big) \\
&\quad+ o(\gamma^2).
\end{align*}
and in the limit $\gamma\to 0$ we have:
\begin{equation*}
\lim_{\gamma\to 0} \frac{1}{\gamma}I(X;Y|Z) = \frac{\log_2(e)}{2}\mathbb{E}[ \mathrm{var}(X|Z)].
\end{equation*}
\end{lemma}
\begin{IEEEproof}
See the appendix.
\end{IEEEproof}

\begin{corollary}\label{cor:uncond}
Let $Y_i = \sqrt{\gamma}X + N_i, i =0,\dotsc, K-1$, where 
$X$ is arbitrarily
distributed with variance $\sigma_X^2$ and  $N_0,\dotsc, N_{K-1},$ are 
zero-mean unit-variance i.i.d.\ Gaussian distributed. 
Let $Z$ be arbitrarily jointly 
distributed with $X$, where $(X,Z)$ are independent of $N_i, \forall i$. Then
\begin{equation*}
\lim_{\gamma \to 0}\frac{1}{\gamma} I(X;Y_0,\dotsc, Y_{K-1}|Z) = \frac{K\log_2(e)}{2}\mathbb{E}[\mathrm{var}(X|Z)].
\end{equation*}
\end{corollary}

\subsubsection{Conditions for Asymptotic Optimality of Linear Estimation}
It was  shown by Akyol et al.~\cite{akyol:2010}, that for an
arbitrarily distributed source $X$, contaminated by Gaussian noise $N$, the
MMSE estimator of $X$ given $Y=\sqrt{\gamma}X+N$, converges in
probability to a linear estimator, in the limit where $\gamma\to 0$. 
We show below that in general the conditional MMSE
estimator $\mathbb{E}[X|Y,Z]$ with side information $Z$, 
where $Z$ is independent of $N$ but is arbitrarily correlated with
$X$ becomes asymptotically linear in $X$ but not in  $Z$ (unless $Z$ is Gaussian).

\begin{lemma}[Asymptotic linearity of the conditional MMSE estimator]\label{theo:nonlinear}
Let $Y=\sqrt{\gamma}X + N,$ where $N \indp X$, $X$ is arbitrarily
distributed with variance $\sigma_X^2$ and $N$ is Gaussian
distributed according to $\mathcal{N}(0,1)$. Moreover, let $Z$ be
arbitrarily distributed, independent of $N$ but arbitrarily correlated
with $X$.  Then the conditional MMSE estimator $\mathbb{E}[X|Y,Z]$ is asymptotically 
linear as $\gamma\to 0$ if and only if 
\begin{equation}\label{eq:mmse_equiv}
\mathbb{E}_Z[\mathrm{var}(X|Z)^2] = (\mathbb{E}_Z[\mathrm{var}(X|Z)])^2.
\end{equation}
\end{lemma}

\begin{IEEEproof}
We provide the complete proof, which fixes a typo in the proof provided in \cite{ostergaard:2011}. 
We first consider the unconditional case, where $Z=\emptyset$. Let us
assume that $\mathbb{E}X = \mu_X \neq 0$. Recall that $Y =
\sqrt{\gamma}X + N$, where $\mathbb{E}N=0$ and $\sigma_N^2=1$. 
For small $\gamma$, the optimal estimator is linear, and we have that
\begin{equation}\label{eq:linest}
\mathbb{E}[X|Y] = \mu_X + \alpha(Y - \mathbb{E}Y) + o(\sqrt{\gamma}),
\end{equation}
where $\alpha$ is the optimal Wiener coefficient for $Y-\mathbb{E}Y=Y-\sqrt{\gamma}\mu_x$ given by 
$\alpha  \sqrt{\gamma}\sigma_X^2 + o(\sqrt{\gamma})$.
From~(\ref{eq:mmse_g}), we know that the MMSE behaves as:
\begin{equation}\label{eq:var_uncond}
\mathbb{E}_Y[\mathrm{var}(X|Y)] = \sigma_X^2 - \gamma \sigma_X^4 + o(\gamma).
\end{equation}
On the other hand, in the conditional case with side information $Y$,
for each $Z=z$ the source has mean $\mathbb{E}[X|Z=z]$ and variance
$\mathrm{var}(X|Z=z)$. Using this in~(\ref{eq:linest}), and fixing $Z=z$, leads to
\begin{equation*}
\mathbb{E}[X|Y,Z=z] = \mathbb{E}[X|Z=z] + \alpha_z (Y - \mathbb{E}[Y|Z=z])
+ o(\sqrt{\gamma}),
\end{equation*}
where the Wiener coefficient depends on  $z$, i.e.,
$\alpha_z = \sqrt{\gamma}\mathrm{var}(X|Z=z) + o(\sqrt{\gamma})$. 
Using~(\ref{eq:mmse_g}) for a fixed $Z=z$ yields
\begin{align*}
\mathbb{E}_Y[\mathrm{var}(X|Y,Z=z)] &= \mathrm{var}(X|Z=z) - \gamma\, \mathrm{var}(X|Z=z)^2 \\
&\quad + o(\gamma).
\end{align*}
Taking the average over $Z$ results in
\begin{align}\notag
&\mathbb{E}_{Y,Z}[ \mathrm{var}(X|Y,Z)] \\ \label{eq:var_cond}
&= \mathbb{E}_Z[ \mathrm{var}(X|Z)] 
- \gamma\,\mathbb{E}_Z[\mathrm{var}(X|Z)^2] + o(\gamma).
\end{align}
By Jensen's
inequality, it follows that
\begin{equation*}
\mathbb{E}_Z[\mathrm{var}(X|Z)^2] \geq (\mathbb{E}_Z[\mathrm{var}(X|Z)])^2,
\end{equation*}
with equality if and only if the conditional variance $\mathrm{var}(X|Z=z)$ is
independent of the realization of $z$ as would be the case for a linear estimator.
Thus, comparing~(\ref{eq:var_cond})
to~(\ref{eq:var_uncond}) shows that the linear estimator is generally
not optimal.
\end{IEEEproof}

\subsection{General Sources under MSE Distortion and at High Resolutions}

Most i.i.d.\ sources are successive refinable at high resolutions under the MSE distortion~\cite{lastras:2001}. Moreover, 
for a fixed variance, it is well known that the Gaussian source is the hardest to code under the MSE distortion~\cite{cover:2006}.
We exploit these properties in Theorem \ref{theo:high_rate} below in order to show that the Gaussian source is asymptotically in rate the hardest to UIR code under the MSE.  

\begin{theorem}[A Gaussian source is asymptotically the worst to code using independent encodings]\label{theo:high_rate}
Let $X$ be arbitrarily distributed, let $\tilde{X}$ be Gaussian distributed, and let both $X$ and $\tilde{X}$ have finite differential entropy and finite variance. Moreover, let $\tilde{X}$ have the same mean and variance as $X$.
Then under the MSE distortion measure:
\begin{align}\notag
    \lim_{D_1,D_2\to 0, \frac{D_1}{D_2}=c}&R(\tilde{X},D_1,D_2,K,M) - R(X,D_1,D_2,K,M)  \\
    &\geq 0.
\end{align}
where $R(\phi,D_1,D_2,K,M)$ denotes the total sum-rate for UIR encoding a source $\phi$ going from an initial distortion $D_1$ to a final distortion $D_2<D_1$ using $M$ rounds and $K$ descriptions in each round. 
\end{theorem}

\begin{IEEEproof}
Let the source $X$ be encoded and reconstructed by the MMSE estimator $\hat{X}$. Moreover, let the coding error be given by the difference $E=X-\hat{X}$. In successively refinable coding, the error $E$ is independent of $\hat{X}$ \cite{equitz:1991}. Due to this independence, in UIR source coding at high resolution (for Gaussian or non-Gaussian sources) one can ignore the conditioning on the previous round, and
directly encode the error signal. 

For a fixed variance, it is well known that the Gaussian source is the hardest to code under the MSE distortion~\cite{cover:2006}. It follows that for the same distortion, 
the rate for coding an arbitrarily distributed error $E$ of variance $\sigma_E^2$ is smaller than or equal to that obtained due to coding a Gaussian error of the same variance $\sigma_E^2$. At general resolution, the error $E$ for a non-Gaussian source is not Gaussian. However, $E$ becomes asymptotically Gaussian in a Divergence sense, i.e., $D(E||G)\to 0$, where $G$ is Gaussian, in the limit of high resolution \cite{linder:1994}. 


Let us first prove that for a single-description scheme, the  coding rate going from an initial distortion $D_1$ to a final distortion $D_2<D_1$ is always smaller than or equal to the  rate required if the source was Gaussian. 
%
%
Note that this also holds even if the source $X$ is not successive refinable. 
 
The RDF of a general source $X$ with finite differential entropy $h(X)$ and under a difference distortion is lower bounded by Shannon's lower bound (SLB) \cite{berger:1971}:
\begin{align}\label{eq:SLB}
R(D) &\geq R_\text{SLB}(D) = h(X) - \frac{1}{2} \log_2(2 \pi e D),
\end{align}
where equality is achieved in the limit where $D\to 0$ \cite{linder:1994}. 
On the other hand, the RDF is clearly upper bounded by the rate obtained due to using a Gaussian coding scheme (instead of an optimal coding scheme), where the source is encoded as $X+N$ and $N \sim \mathcal{N}(0,D)$:
\begin{align}
R(D) &\leq I(X;X+N) \\ \label{eq:upper}
&= h(X+N) - \frac{1}{2} \log(2 \pi e D).
\end{align}
If we now upper bound $R(D_2)$ using \eqref{eq:upper} and lower bound $R(D_1)$ using $R_\text{SLB}$ \eqref{eq:SLB}, where $D_2<D_1$, we can establish the following inequality, which upper bounds the maximal rate required going from $D_1$ to $D_2$:
\begin{align} \notag
R(D_2) - R(D_1) &\leq h(X+N_2) - \frac{1}{2} \log(2 \pi e D_2) \\
&\quad - \big(h(X) - \frac{1}{2} \log_2(2 \pi e D_1) \big) \\ \label{eq:diff}
&= \frac{1}{2}\log_2\frac{D_1}{D_2}  + h(X+N_2) - h(X).
\end{align}
At this point, we let $D_1,D_2\to 0$, keeping the ratio $D_1/D_2=c>1$ constant and finite, which when used in \eqref{eq:diff} yields:
\begin{align}
\lim_{D_1,D_2\to 0, \frac{D_1}{D_2}=c}&\bigg[ \frac{1}{2}\log_2\frac{D_1}{D_2}  + h(X+N_2) - h(X) \bigg] \\
&=\frac{1}{2}\log_2\frac{D_1}{D_2},
\end{align}
which follows from the fact that in the limit of $D\to 0$, $h(X+N)\to h(X)$ for sources with a finite differential entropy and finite variance~\cite{linder:1994}.
It follows that asymptotically $R(D_2) - R(D_1) \leq \frac{1}{2}\log_2(D_1/D_2)$, where the right-hand-side is the rate required if the source was Gaussian.

To prove the theorem, we  repeat the above arguments for the case of $K$ descriptions. Assume we have $K$ independent encodings of $X$, which individually leads to distortion $D'_1$ and when combined yields distortion $D_1\leq \tilde{D}_1$, where $\tilde{D}_1$ would be joint distortion for $K$ independent encodings of the Gaussian source $\tilde{X}$, which individually leads to the same distortion $\tilde{D}'_1=D'_1$. 
Then we can lower bound the sum-rate of the $K$ encodings by the sum of the individual lower bounds, i.e.,  
\begin{equation}\label{eq:lowa}
\sum_{i=1}^K  R_\mathrm{SLB}(D'_1) = K h(X) - \frac{K}{2}\log_2(2\pi e D'_1 ).
\end{equation}
Similarly, we can use a $K$ descriptions Gaussian coding scheme, $X+\tilde{N}_i, i=1,\dotsc, K,$ to get the upper bound, where each description leads to a distortion $D'_2 = \tilde{D}'_2$ and when combined the joint distortion is $D_2\leq \tilde{D}_2$. The following upper bound on the sum-rate can then be established:
\begin{equation}\label{eq:upa}
\sum_{i=1}^K R(D'_2) \leq \sum_{i=1}^K h(X+\tilde{N}_i) - \frac{1}{2}\log_2(2\pi e D'_2). 
\end{equation}

Subtracting \eqref{eq:lowa} from \eqref{eq:upa} yields the following upper bound:
\begin{align} \notag
\sum_{i=1}^K R(D'_2) - \sum_{i=1}^K R(D'_1) &\leq  \sum_{i=1}^K \big(h(X+\tilde{N}_i) - \frac{1}{2}\log_2(2\pi e D'_2)\big) \\ \notag
&\quad -  K h(X) + \frac{K}{2}\log_2(2\pi e D'_1 ) \\ \notag
&\overset{\lim }{=}\frac{K}{2}\log_2(D'_1 / D'_2 ),
\end{align}
where the last equality holds in the limit where $\tilde{D}_1,\tilde{D}_2 \to 0$ and $\tilde{D}_1/\tilde{D}_2 = c>1$. Since this last equality is non-negative and achieved for a Gaussian coding scheme, the theorem is proved.
\end{IEEEproof}

\section{Incremental Multiple Descriptions with Feedback}
In the previous sections, we considered coding a source into $K$ independent encodings. Then, all $K$ independent encodings were combined to form an estimate of the source. In the next round, this estimate was further refined by being coded into $K$ new independent encodings. It is interesting to consider the case, where instead of using all $K$ encodings in each round, one may consider the case of only using a subset of the $K$ encodings. Indeed, this can then be considered as a special case of multiple descriptions with feedback. For example, an encoder forms $K$ independent encodings (descriptions), which are transmitted over an erroneous communications network. 
At the decoder, only a subset of the descriptions are received, and the decoder then informs the encoder which descriptions were received. The encoder can then successively refine the source using only the descriptions that are received by the decoder. The new source is then further encoded into $K$ descriptions, and so on, see Fig.~\ref{fig:mdfb}.
The following theorem shows that the accumulated sum-rate of the received descriptions after any number of rounds in the system of Fig.~\ref{fig:mdfb}, is asymptotically optimal as the number of rounds tends to infinity, and the rate in each round tends to zero. 

\begin{theorem}
Let $R_X$ be the RDF for the source $X$ under the distortion measure $d(\cdot,\cdot)$, and consider $M$ rounds, each having $K$ independent encodings, and either of the following two cases:
\begin{enumerate}
    \item $X$ is one-sided exponential, $d(\cdot,\cdot)$ is the one-sided distortion measure, and the decoder uses the select-max estimator,
    \item $X$ is Gaussian, $d(\cdot,\cdot)$ is the MSE distortion measure, and the decoder averages the received descriptions.
\end{enumerate} 
Let $R$ be the rate of each description, let $0\leq k_i \leq K$ denote the number of received descriptions in round $i$, and let $\bar{D}_i>0$ be the resulting distortion in round $i$, for $i=1,\dotsc, M$. 
Finally, let $m_\ell$ denote the number of times $\ell$ descriptions were received in $M$ rounds, where $m_\ell/M > 0$ for $M\to\infty, \ell = 1,\dotsc, K$. 
Then

\begin{equation}
  \lim_{\mathclap{\substack{M \to \infty\\ 
                            m_\ell/M > 0, \forall \ell \\
                              R \to 0\\
                              }}}
\bigg[                              R_X(\bar{D}_i) 
                              - R \sum_{i'=1}^i k_{i'}\bigg] = 0, \quad \forall i.
\end{equation}

\end{theorem}

\begin{IEEEproof}
We only need to focus on the cases where a non-zero number of descriptions is received, since the case of $k_i = 0$ for some $i$, will not affect the total sum-rate of the received descriptions, and nor will it affect the resulting distortion. 

\subsection{Case 1: Exponential source}

Let us first consider a single round, where the source, say $X_1$, is to be encoded into $K$ independent encodings, $Y_1,\dotsc, Y_K$, which each are rate-distortion optimal resulting in an individual distortion of $D_1$.
Assume that we receive $k_1$ out of $K$ encodings and that we use the select-max estimator $Z_1 = \max_i \{Y_i\}$.
Then, from Lemma \ref{lem:exp_error} it follows that the estimation error $\tilde{X}_1=X_1-Z_1$ is exponentially distributed with parameter $\lambda_2$, where
\begin{equation}
    \lambda_2 =  (k_1 - (k_1-1)c_1) D_1^{-1},
\end{equation}
where $c_1 = \lambda_1 D_1$. Moreover, the resulting expected distortion $\bar{D}_1$ using $Z_1$ as the estimator is obtained using Lemma \ref{lem:D} as:
\begin{align}
\bar{D}_1 &= \frac{D_1}{k_1-(k_1-1)D_1\lambda_1} 
= \frac{c_1 \lambda_1^{-1}}{k_1-(k_1-1)c_1}. 
\end{align}

Since the rate of the individual descriptions are given by $R_1 = -\log(D_1 \lambda_1)$, it follows that in order to keep the rate equal in all rounds, we need to have $D_1\lambda_1 = D_i\lambda_i = c_i = c, \forall i$. Moreover, to keep the accumulated sum-rate over $M$ rounds finite, when $M$ tends to infinity, we let the rate per description be 
$-\frac{1}{M}\log(D_1\lambda_1)$ for all rounds, which implies that:
\begin{equation}
c=(\lambda_1 D_1)^{1/M}.
\end{equation}
Let $k_1^M$ be the sequence of received descriptions. Let $\bar{R}_M(k_1^M)$ denote the total rate of this sequence, which is given by:
\begin{equation}
\bar{R}_M(k_1^M) = -\frac{1}{M}\log(D_1\lambda_1)\sum_{i=1}^M k_i.
\end{equation}
Since the DRF is given by $D_X(R) = \lambda^{-1}2^{-R}$, we find the optimum distortion given $\bar{R}_M(k_1^M)$ to be:
\begin{align}
    D_X(\bar{R}_M(k_1^M)) &= \lambda_1^{-1} 2^{\frac{1}{M}\log(\lambda_1D_1) \sum k_i}\\ \label{eq:opt_D}
    &= \lambda_1^{-1}(\lambda_1D_1)^{\frac{1}{M}\sum_{i=1}^M k_i}.
\end{align}

Let us now assume that $k_1$ and $k_2$ independent encodings are received in rounds 1 and 2, respectively. The source in round 2 is $X_2 = \tilde{X}_1$, the estimation error of round 1. This source is encoded into $K$ independent encodings each leading to an individual optimum distortion of $D_2$. The resulting expected distortion after round 2, when using the select-max estimator on the $k_2$ successively transmitted encodings is then given by:
\begin{align}
\bar{D}_2 &= \frac{D_2}{k_2-(k_2-1)D_2\lambda_2} \\
&= \frac{c}{k_2-(k_2-1)c}\lambda_2^{-1} \\
&= \frac{c^2 \lambda_1^{-1}}{(k_2-(k_2-1)c)(k_1 - (k_1-1)c)}.
\end{align}
After $M$ rounds we therefore obtain an expected distortion $\bar{D}_M$ given by:
\begin{align} \label{eq:MroundsD}
    \bar{D}_M &= 
    \frac{c^M\lambda_1^{-1}}{\prod_{i=1}^{M}(k_i-(k_i-1)c)} \\ 
    &=  \frac{c^M\lambda_1^{-1}}{\prod_{\ell=1}^{K}(\ell-(\ell-1)c)^{m_\ell}},
\end{align}
where $m_\ell$ is the number of times $\ell$ encodings are successively received over a total of $M$ rounds. We will assume that the ratio $M/m_\ell = c_\ell$ is finite, as $M\to \infty$. 

Focusing on the limit of the $\ell$th term in the product, we obtain:
\begin{align}
    \lim_{m_\ell\to \infty} (\ell-(\ell-1)c)^{m_\ell} 
    &=\lim_{m_\ell\to \infty} (\ell-(\ell-1) (\lambda_1 D_1)^{\frac{1}{m_\ell c_\ell}})^{m_\ell} \\
    &= \big(\lambda_1 D_1\big)^{-\frac{\ell -1}{c_\ell}}.
\end{align}

Taking the limit of $\bar{D}_M$ leads to:
\begin{align}
    \lim_{M\to \infty}
    \frac{c^M\lambda_1^{-1}}{\prod_{\ell=1}^{K}(\ell-(\ell-1)c)^{m_\ell c_\ell}} &=
    (\lambda_1 D_1)\lambda_1^{-1}
    \big(\lambda_1 D_1\big)^{\sum_{\ell=1}^K\frac{\ell -1}{c_\ell}} \\ \label{eq:operation_D}
    &= \lambda_1^{-1}
    (\lambda_1 D_1)^{\sum_{\ell=1}^K\frac{\ell}{c_\ell}}, 
\end{align}
where we used the fact that $\frac{1}{M}\sum_{i=1}^M k_i = \frac{1}{M}\sum_{\ell=1}^K \ell m_\ell = \sum_{\ell=1}^K \ell c_\ell^{-1}$, and $\sum_{\ell=1}^K  c_\ell^{-1} = 1$.
Since \eqref{eq:operation_D} is equal to \eqref{eq:opt_D}, we have proved that the total accumulated sum-rate is asymptotical optimal as $M\to \infty$. 

Consider now the accumulated sum-rate $\bar{R}_{M'}$ and distortion $\bar{D}_{M'}$ obtained after the first $M'$ rounds out of the $M= \infty$ rounds. From \eqref{eq:MroundsD} we can write the operational distortion $\bar{D}_{M'}$ achieved in round $M'$ after $M$ rounds:
\begin{align}
 \lim_{M\to \infty} \bar{D}_{M'}  &=  \lim_{M\to \infty} \frac{c^{M'}\lambda_1^{-1}}{\prod_{i=1}^{M'}(k_i-(k_i-1)c)} \\
  &= \lim_{M\to \infty} \frac{(\lambda_1D_1)^{M'/M}\lambda_1^{-1}}{\prod_{i=1}^{M'}(k_i-(k_i-1) (\lambda_1 D_1)^{1/M})} \\
  &= \lim_{M\to \infty}\frac{(\lambda_1D_1)^{M'/M}\lambda_1^{-1}}{\prod_{i=1}^{M'} (\lambda_1 D_1)^{-k_i} \lambda_1 D_1} \\ \label{eq:final_D}
  &= \lim_{M\to \infty}(\lambda_1D_1)^{1/M} \lambda_1^{-1} (\lambda_1D_1)^{\sum_{i=1}^{M'} k_i}. 
\end{align}
The above distortion is obtained by using a sum-rate of $R\sum_{i=1}^{M'} k_i$. Replacing the sum $\sum_{i=1}^{M} k_i$ in \eqref{eq:opt_D} by $\sum_{i=1}^{M'} k_i$ shows that the 
operational distortion obtained in \eqref{eq:final_D} becomes identical to the DRF. 
This proves that asymptotically as the number of rounds tends to infinity, the resulting distortion achieved after any $M'$ rounds tends to the DRF.

\subsection{Case 2: Gaussian source}
Let the rate $R$ per description in each round to be related to the number of rounds $M$ in the following way:
\begin{equation}
R = \frac{1}{2M}\log_2\bigg(\frac{\sigma_X^2}{d}\bigg),
\end{equation}
where $0< d \leq \sigma_X^2$. The total sum-rate $R_M$ for the received descriptions after $M$ rounds in then given by:
\begin{equation}
R_M = R\sum_{i=1}^M k_i = \frac{1}{2M}\log_2\bigg(\frac{\sigma_X^2}{d}\bigg)\sum_{i=1}^M k_i.
\end{equation}
The optimum distortion $D^*_M$ at rate $R_M$ is obtained by inserting into the DRF:
\begin{align}
D^*_M &= \sigma_X^2 2^{-2R_M} \\ \label{eq:opt_DM}
&= \sigma_X^2 \bigg(\frac{d}{\sigma_X^2}\bigg)^{\frac{1}{M}\sum k_i}.
\end{align}
On the other hand, the operational distortion after round $i$ depends directly upon $k_i$ in the following way:
\begin{align}
\bar{D}_i &= \frac{d_i}{k_i - (k_i-1) d_i/\bar{D}_{i-1}},
\end{align}
where $d_i$ was introduced in Section III-A as the conditional variance of the source in the $i$th round  given one of the $K$ descriptions. 

Consider now a fixed sequence $k_1,\dotsc, k_M$, where $k_i \in \{1,\dotsc, K\}$. Then, the resulting distortion $\bar{D}_1(k_1)$ after round 1 due to receiving $k_1$ descriptions is given by:
\begin{align}
\bar{D}_1(k_1) &= \frac{d_1  \bar{D}_0}{\bar{D}_0k_1 - (k_1-1)d_1}.
\end{align}
Similarly, the resulting distortion $\bar{D}_2(k_1^2)$ after round 2 due to receiving $k_2$ packets in round 2 and $k_1$ packets in round 1 is:
\begin{align}
\bar{D}_{2}(k_1^2) &= \frac{d_2(k_1)  \bar{D}_1(k_1)}{\bar{D}_1(k_1)k_2 - (k_2-1)d_2(k_1)}\\
&= \frac{d_1}
{\bar{D}_0 k_2 - (k_2-1)d_1}\bar{D}_1(k_1) \\
&=\frac{d_1^2 \bar{D}_0}
{(\bar{D}_0 k_2 - (k_2-1)d_1)(\bar{D}_0 k_1 - (k_1-1)d_1)},
\end{align}
where we used the following relation, which guarantees that the sum-rate is the same for all rounds:
\begin{align}
d_{i+1}(k_1^{i}) = d_i(k_1^{i-1}) \frac{\bar{D}_i(k_1^i)}{\bar{D}_{i-1}(k_1^{i-1})}. 
\end{align}
After $M$ rounds we obtain:
\begin{equation}
\bar{D}_{M}(k_1^M) = \frac{d_1^M \bar{D}_{0}}{ \prod_{\ell=1}^K (\bar{D}_{0}\ell - (\ell-1) d_1)^{m_\ell}
},
\end{equation}
where $\sum_{l=1}^K m_\ell = M$, and where $m_\ell$ denotes the number of times that $\ell$ descriptions were received in the $M$ rounds.  

In order to obtain a finite rate and a non-zero distortion, when $M\to \infty$, we let the  distortion $d_1$ of round 1 depend upon $M$ through the distortion parameter $d$, i.e.:
\begin{align}
d_1 = \bigg(\frac{\sigma_X^2}{d}\bigg)^{-1/M}\sigma_X^2, 
\end{align}
where $\sigma_X^2\geq d>0$. 
With this the distortion in round $M$ can be written as:
\begin{align}
\bar{D}_{M}(k_1^M) = \frac{d(\sigma_X^2)^M}{\prod_{\ell=1}^K \big(\sigma_X^2\ell - (\ell-1)  \big(\frac{d}{\sigma_X^2}\big)^{1/M} \sigma_X^2 \big)^{m_\ell}},
\end{align}
where we inserted $\bar{D}_0= \sigma_X^2$.

Let $M/m_\ell = c_\ell$ be constant so that $m_\ell$ tends to infinity at the same rate as $M$. Then, the limit of the ratio of $(\sigma_X^2)^{m_\ell}$ and the $\ell$th term in the product can be written as:
\begin{align}
&\lim_{m_\ell \to \infty} (\sigma_X^2)^{m_\ell}\big(\sigma_X^2\ell - (\ell-1) \bigg(\frac{d}{\sigma_X^2}\bigg)^{1/(c_\ell m_\ell)} \sigma_X^2 \big)^{-m_\ell} \\
&= \bigg(\frac{d}{\sigma_X^2}\bigg)^{(\ell-1)/c_\ell} 
\end{align}
which implies that:
\begin{align}
\lim_{M \to \infty} \bar{D}_{M}(k_1^M) &= d \prod_{\ell=1}^{K}\bigg(\frac{d}{\sigma_X^2}\bigg)^{(\ell-1)/c_\ell} \\
&= d \bigg(\frac{d}{\sigma_X^2}\bigg)^{\sum_{\ell=1}^{K}(\ell-1)/c_\ell} \\ \label{eq:op_D_M}
&= \sigma_X^2\bigg(\frac{d}{\sigma_X^2}\bigg)^{\sum_{\ell=1}^{K}\ell/c_\ell},
\end{align}
where we used that $\sum c_\ell^{-1} = 1$. Since $M^{-1}\sum_{i=1}^M k_i =M^{-1}\sum_{l=1}^K \ell m_\ell = \sum_{\ell=1}^{K}\ell/c_\ell$, the optimum final distortion \eqref{eq:opt_DM} is asymptotically equal to the operational distortion \eqref{eq:op_D_M} after $M$ rounds in the limit where $M\to \infty$.

We can use a similar technique as for Case 1, in order to prove that the operational distortion obtained after any finite number of rounds asymptotically tends towards the DRF as the total number of rounds tends to infinity. This proves the theorem.
\end{IEEEproof}

\subsection{Incremental Gaussian Multiple Descriptions}
In the quadratic Gaussian MD problem, a Gaussian random variable $X$ is encoded into $K$ descriptions $Y_1,\dotsc, Y_K$.
If we consider the symmetric case, then the rates of the descriptions are equal, i.e., $R_i=R, \forall i$, and the distortion observed at the decoder, does not depend upon which descriptions that are received but only upon the number of received descriptions; $D_\mathcal{I} = D_{\mathcal{I}'}, \forall \mathcal{I},\mathcal{I}'\subseteq \{1,\dotsc, K\}$ such that $|\mathcal{I}| = |\mathcal{I}'|$. 

\begin{figure}[th]
\begin{center}
\includegraphics[width=7cm]{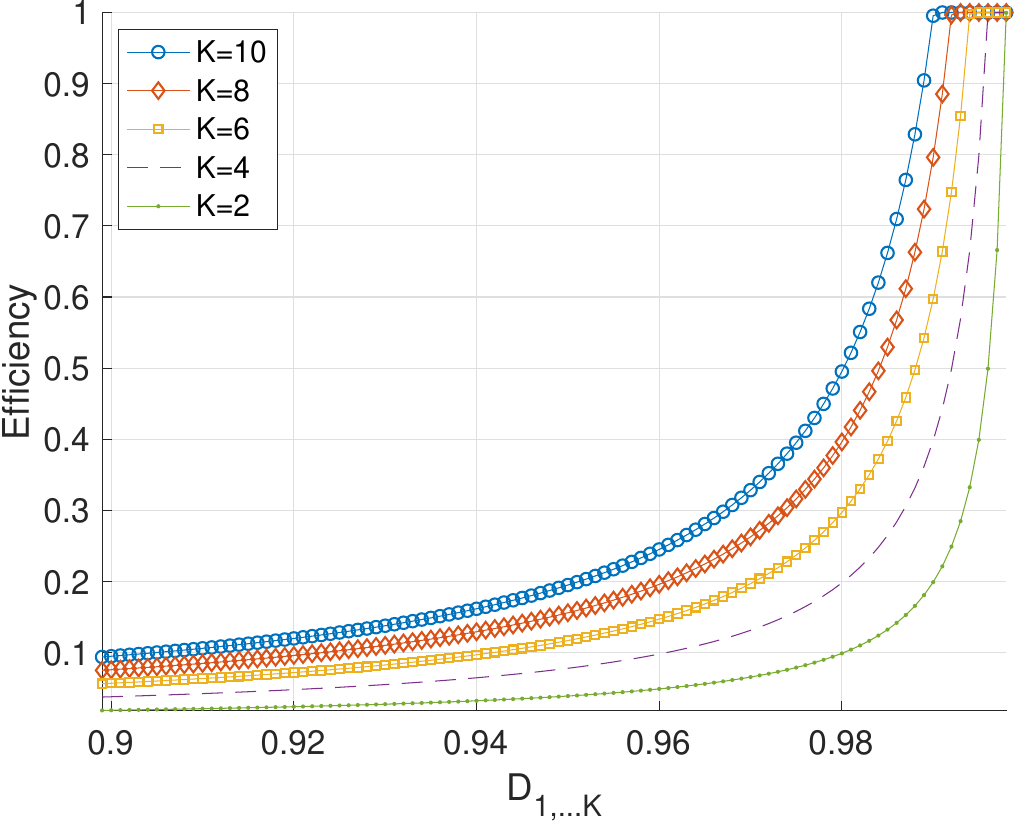}
\caption{Efficiency \eqref{eq:md_eff}  for $|\mathcal{I}|=K$ multiple-description coding of a zero-mean unit-variance Gaussian source. }
\label{fig:md_eff}
\end{center}
\vspace{-4mm}
\end{figure}

The rate-distortion region for the quadratic Gaussian MD problem is only known for some specific cases. For example, for the two-receiver case, where one is only interested in the distortion when receiving $k<K$ or all $K$ descriptions, the optimal sum-rate region (for Gaussian vector sources) was given as an optimization problem in~\cite{wang:2009}. For the symmetric case (and scalar Gaussian sources), the  optimization problem of~\cite{wang:2009}  was explicitly solved in \cite{mashiach:2010}. In this subsection, we focus on the two-receiver case of $k=1$ or $k=K$ descriptions. Let the distortion due to receiving a single description be $D_i=D, \forall i$, and let $D_{1,\dotsc, K}$  denote the distortion due to receiving all $K$ descriptions. Then, the optimal per description rate $R$ as a function of $(D, D_{1,\dotsc, K})$ is given by \cite{mashiach:2010}:
\begin{align} \notag
R(D, D_{1,\dotsc, K}) &=
\frac{1}{2}\log_2 \bigg(  \frac{(K-1)(\sigma_X^2-D_{1,\dotsc, K})}{K(D - D_{1,\dotsc, K})} \bigg) \\ 
& + 
\frac{1}{2K}\log_2 \bigg(  \frac{ \sigma_X^2(D - D_{1,\dotsc, K}) }{(K-1)D_{1,\dotsc, K}(\sigma_X^2-D)} \bigg).
\end{align}
If we now let $D=D_{\max} - \epsilon = \sigma_X^2-\epsilon$ for some $\epsilon>0$, and vary $D_{1,\dotsc, K}$, it is possible to illustrate the efficiency for the case where we are only interested in $|\mathcal{I}|\in \{1,K\}$:
\begin{align}\label{eq:md_eff}
 \eta_K(\mathcal{I},R) 
=  \begin{cases}
\frac{R(D)}{R(D, D_{1,\dotsc, K}) }, & |\mathcal{I}| = 1, \\
\frac{R(D_{\mathcal{I}})}{KR(D, D_{1,\dotsc, K}) }, & |\mathcal{I}| = K, 
\end{cases}
\end{align}
since $R_i = R(D, D_{1,\dotsc, K})$. This is illustrated in Fig.~\ref{fig:md_eff} for the case of $\epsilon=1/1000, \sigma_X^2=1$, and $K\in\{2,4,\dotsc, 10\}$.

\begin{figure}[th]
\begin{center}
\includegraphics[width=7cm]{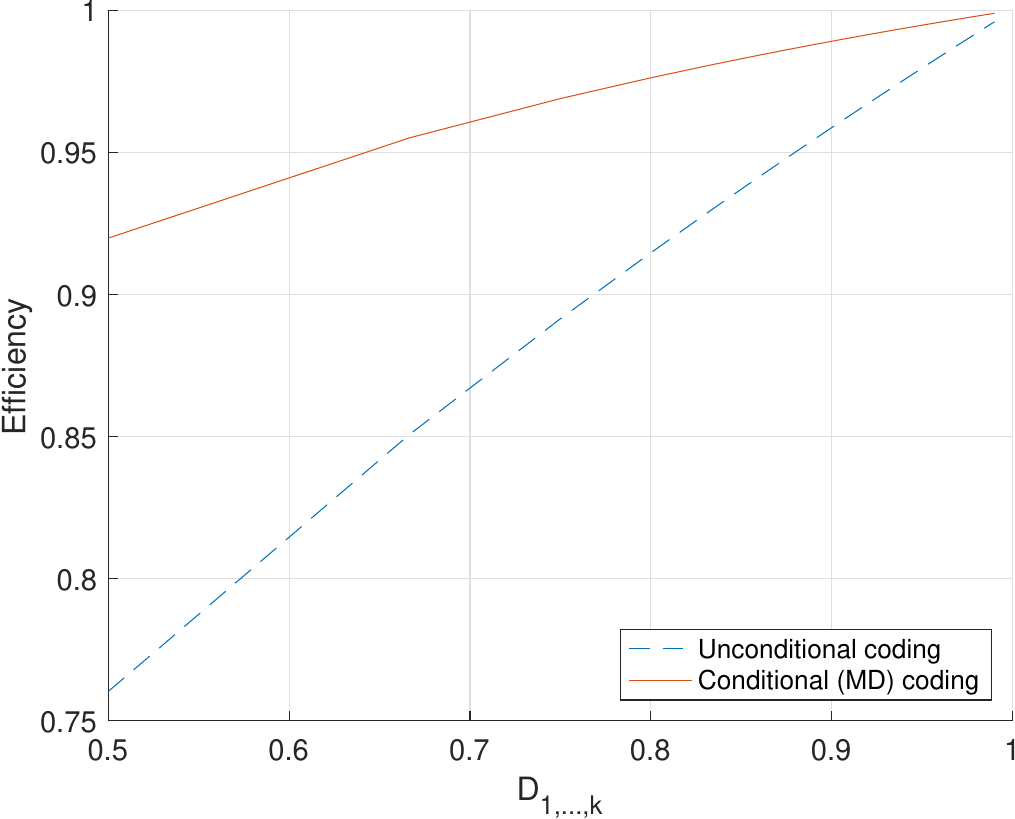}
\caption{Efficiency \eqref{eq:md_eff} for conditional coding (multiple-description coding) versus unconditional coding of a zero-mean unit-variance Gaussian source.}
\label{fig:md_sd_eff}
\end{center}
\end{figure}

One extreme solution to the MD coding problem is when the distortion due to receiving a single description is minimized, i.e., the case of "no excess marginal rate".
Indeed, in the quadratic Gaussian case, generating independent descriptions $Y_i = X + N_i$ with mutually independent noises will then be optimal. However, if we compromise the distortion due to receiving a single description, in order to get an even smaller distortion when receiving more descriptions, then the use of use mutually independent noises is often suboptimal \cite{ozarow:1980}. Thus, in order to better illustrate the loss of unconditional coding (e.g., using mutually independent noises) over that of conditional coding (MD coding), we provide the following example. 
For the case of unconditional coding, we let $Y_i = X+N_i, i=1,\dotsc, K$, where $N_i$ are mutually independent Gaussian random variables of equal variance $\sigma_N^2$, and $X$ is standard Gaussian. Combining $K$ descriptions yields a distortion of:
\begin{equation*}
D_{1,\dotsc, K} = \frac{\sigma_X^2\sigma_N^2}{\sigma_N^2 + K\sigma_X^2}.
\end{equation*}
The sum-rate for the $K$ descriptions $\{Y_i\}_{i=1}^{K}$ is given by $\frac{K}{2}\log_2(\sigma_X^2/D_1)$, where $D_1 = \frac{\sigma_X^2\sigma_N^2}{\sigma_N^2 + \sigma_X^2}$. On the other hand, when using Gaussian MD coding, we can slightly increase the individual distortions $D_i$ due to receiving a single description, in order to achieve the desired $D_{1,\dotsc, K}$ at a lower sum-rate than that of unconditional coding. The efficiency of unconditional versus conditional (MD) coding is shown in Fig.~\ref{fig:md_sd_eff}. Notice that at high distortions of $D_{1,\dotsc, K}$, the efficiency of unconditional coding is not far from that of MD coding.

\subsection{Incremental Multiple Descriptions Vector Quantization}

We now introduce a  multiple descriptions vector quantizer that can be modelled by the $\epsilon/1$ channel. Specifically, we propose a very practical binary (threshold) vector quantizer, which is asymptotically rate-distortion optimal in the limit of zero rate in the quadratic Gaussian case. Our quantizer is inspired by the binary scalar quantizer in \cite{marco:2006}. 

\subsubsection{Threshold Vector Quantizer}
Let $X\in\mathbb{R}^n$ be zero-mean and Gaussian distributed with covariance matrix $\Sigma_X = \sigma^2I$, where $I\in \mathbb{R}^{n\times n}$ is the identity matrix. Moreover, let $x\in\mathbb{R}^n$ be a realization of $X$ and let $x(i)$ denote the $i$th element of the vector $x$. 
Let $\mathcal{Q}_{i,\xi}$ be an $n$-dimensional binary quantizer with its single decision region being the hyperplane described by $x(i)=\xi$ for some \emph{fixed} $i$.\footnote{The index $i$ could alternatively be found as the component of the source that decreases the MSE the most, which would yield a rate $1/n + \log(n)/n$.  The rate $\log(n)$ is needed to describe the desired component (location) within the vector and the 1 bit describes whether it is above or below the threshold.}

Let $c_j\in\mathbb{R}^n, j=0,1,$ denote the reproduction points of the quantizer. The Voronoi cells $V_j, j=0,1$ are given by:
\begin{equation*}
V_j \triangleq \{ x\in\mathbb{R}^n : \|x - c_j\|^2_2 \leq \|x - c_{j'}\|_2^2, j\neq j' \}.
\end{equation*}
At this point, we let $c_0 = \mathbf{0}\in\mathbb{R}^n$ be the all zero vector, and we let $c_1$ be the centroid of $V_1$. We note that $c_1 = [0, \dotsc,0,  \hat{z}, 0,\dotsc, 0]^T$ is the all zero vector except that the $i$th element is given by the centroid $\hat{z}$ of the $i$th interval:
\begin{equation*}
\hat{z} = \frac{1}{\int_{\xi}^{\infty} f(z) dz}  \int_{\xi}^{\infty} f (z) z\, dz = \frac{\sqrt{2}\sigma \,{{\rm e}^{-{\frac {{\xi}^{2}}{{2\sigma }^{2}}}}}}{\sqrt{\pi } \big(1- {{\rm erf}\big({\frac { \sqrt{2}\xi}{2\sigma }}\big)} \big)},
\end{equation*}
where $f$ denotes the marginal distribution of $X$, 
$\mathrm{erf}$ denotes the error function, and $\sigma^2$ is the variance of the individual elements of $X$.
Thus, given a vector $x\in\mathbb{R}^n$, the output $\hat{x}$ of the quantizer is given by:
\begin{equation*}
\hat{x} = \mathcal{Q}_{i,\xi}(x) = \begin{cases}
\mathbf{0}, & \text{if}\ x_i \leq \xi, \\
[0, \dotsc,0,  \hat{z}, 0,\dotsc, 0]^T, & \text{otherwise}.
\end{cases}
\end{equation*}
The probability $p_0$ that $X\in V_0$ is independent of the vector dimension $n$ and given by the scalar integral:
\begin{equation}
p_0 = \int_{-\infty}^\xi f(z)\, dz = \frac{1}{2} + \frac{1}{2}\mathrm{erf}\bigg(\frac{\sqrt{2} \xi}{2\sigma}\bigg).
\end{equation}
Since the quantizer has only two cells, $p_1=1-p_0$. 

The distortion due to using this particular construction of a binary quantizer is clearly equal to the (marginal) variance $\sigma^2$ of the source for the $(n-1)$ dimensions (coordinates) of $c_1$ which are equal to zero. The $i$th coordinate of $c_1$ is non-zero and yields a smaller distortion $D_i< \sigma^2$, which is given by:
\begin{equation}\label{eq:D}
D_i = \int_{-\infty}^\xi f(z) z^2\, dz +  \int_{\xi}^{\infty} f(z) (z-\hat{z})^2\, dz.
\end{equation}
It follows that the total distortion $D$ is equal to 
\begin{align} \label{eq:Dis}
D= D_i + \sigma^2(n-1) 
=\frac{\,{\sigma }^{2} \big( \pi \,n\big( 1 - {{\rm erf}\big({\frac { \sqrt{2}\xi}{2\sigma }}\big)}\big)  - {{\rm e}^{-{\frac {{\xi}^{2}}{{\sigma }^{2}}}}} \big)}{ {\pi } \big( 1- {{\rm erf}\big({\frac { \sqrt{2}\xi}{2\sigma }}\big)} \big) }.
\end{align}

The entropy $R$ of the quantizer is given by:
\begin{equation}\label{eq:R}
R = -p_0\log_2(p_0) - p_1\log_2(p_1).
\end{equation}

From \eqref{eq:R} and \eqref{eq:D} it follows that both $R$ and $D$ are functions of the threshold $\xi$. To find the slope of $R(D)$ w.r.t.\ $D$, we may therefore use the derivative of a composite function:
\begin{equation*}
\lim_{\xi\to \infty} \frac{\partial}{\partial D} R(D) =
\lim_{\xi\to \infty} \frac{ \frac{\partial}{\partial \xi} R}{  \frac{\partial}{\partial \xi} D } =
 - \frac{1}{2\sigma^2 \ln(2)},
\end{equation*}
which coincide with the RDF of a white Gaussian source at zero rate. Thus, the quantizer is asymptotically RDF optimal in the limit of  zero rate.

\subsubsection{Unconditional Refinement}
The only non-zero codeword of the binary vector quantizer $\mathcal{Q}_{i,\xi}$ is parallel to the unit-vector (basis vector) along the $i$th coordinate. 
In an $n$-dimensional vector space, we may form $n$ distinct binary vector quantizers $\mathcal{Q}_{i,\xi}, i=1,\dotsc, n$, where each is non-zero on a different coordinate.\footnote{One may increase the number of quantizers beyond $n$ by, for example, using $-\xi$ in addition to $\xi$.}
Each quantizer is applied independently upon the source vector $X\in \mathbb{R}^n$. This results in $n$ reproduction vectors. Let $\hat{X}_i = \mathcal{Q}_{i,\xi}(X)$ denote the reproduction vector due to using the $i$th quantizer on the source vector $X$. At the decoder, we may receive a subset $\mathcal{I} \subseteq \{1,\dotsc, n\}$ of the $n$ descriptions.
Let $\hat{X}_{\mathcal{I}}$  denote the joint reconstruction using the set of reproductions indexed by $\mathcal{I}$. Similarly, let $D_{\mathcal{I}}$ denote the distortion due to approximating the source $X$ by $X_{\mathcal{I}}$. In particular, we form the following reconstruction rule:
\begin{equation}\label{eq:hatx}
\hat{X}_{\mathcal{I}} = \sum_{i\in\mathcal{I}} \hat{X}_i,
\end{equation}
where no averaging is needed since the non-zero codewords of the different quantizers are orthogonal to each other.
The resulting distortion $D_\mathcal{I}$ is then given by:
\begin{align}
D_\mathcal{I} &= \mathbb{E}\| X - \hat{X}_\mathcal{I}\|^2_2 = |\mathcal{I}| D_i + (n-|\mathcal{I}|)\sigma^2 \\ \label{eq:D_lin}
&=
\frac{\,{\sigma }^{2} \big( \pi \,n\big( 1 - {{\rm erf}\big({\frac { \sqrt{2}\xi}{2\sigma }}\big)}\big)  - |\mathcal{I}|{{\rm e}^{-{\frac {{\xi}^{2}}{{\sigma }^{2}}}}} \big)}{ {\pi } \big( 1- {{\rm erf}\big({\frac { \sqrt{2}\xi}{2\sigma }}\big)} \big) },
\end{align}
where $| \mathcal{I}|$ denotes the cardinately of the set $\mathcal{I}$, and $D_i$ is given by \eqref{eq:Dis}. It is clear from \eqref{eq:D_lin}, that the distortion is linearly decreasing in the number $|\mathcal{I}|$ of combined  descriptions. If the $n$ outputs of the different quantizers are independently entropy coded, the resulting coding rate is simply given by $nR$, where $R$ is the entropy of the individual quantizer outputs, see \eqref{eq:R}. 

The rate-distortion performance for $K=10$ uses of the threshold vector quantizer on a zero-mean vector Gaussian source of dimension $n=10$ and with identity covariance matrix is illustrated in Fig.~\ref{fig:xi2} for the case of $\xi=1$ and $\xi=2$. For $\xi=1$ the entropy per dimension is $H=0.0631$ bits/dim., which gives a sum-rate of $0.63$ bits/dim.\ when using $K=10$ descriptions. The resulting distortion is $-2$ dB/dim., when combining all $K=10$ description. To reduce this distortion, we may increase the threshold $\xi$. For $\xi=2$, the entropy per dimension of each description is $H=0.0157$ bits/dim., which leads a sum-rate of $0.16$ bits/dim., when using $K=10$ descriptions. The resulting distortion is $-0.6$ dB/dim. To get below this distortion level, one can introduce multi-round coding as we do below.

\subsubsection{Multi-Round Threshold Vector Quantization}
\begin{figure}[t]
\begin{center}
\includegraphics[width=7.5cm]{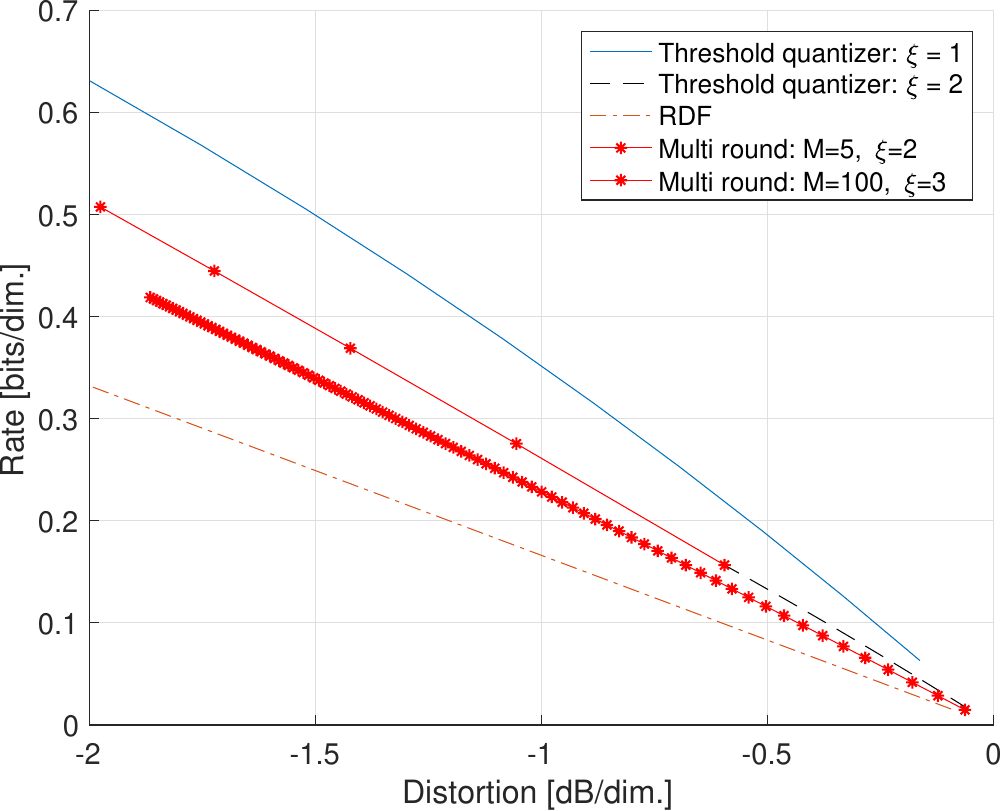}
 \caption{Threshold vector quantization of a Gaussian source under quadratic distortion. Single round coding for $\xi=1$ (solid blue) and $\xi=2$ (dashed) and when combing $K=1,\dotsc, n$ descriptions for $n=10$. For $\xi=2$ the rate per dim.\ is small and the curve stops at $-0.6$ dB for $K=n$ descriptions. To get below this, one can use multi-round coding (stars) and use $K=n$ descriptions in each round. The dash-dotted curve shows the corresponding unit-variance Gaussian RDF.}
  \label{fig:xi2}
  \end{center}
\end{figure}

In the first round, the source is encoded into $K$ descriptions and reconstructed as $\hat{X}$ \eqref{eq:hatx} using a subset of descriptions indexed by $\mathcal{I}$. In the next round, the residual  $X-\hat{X}$ becomes the new source to encode. The variance of this residual is equal to $D_\mathcal{I}$ in \eqref{eq:D_lin}. In Fig.~\ref{fig:xi2}, we have shown the performance of multi-round coding. In this example, we let $\xi=2$ and use $M=5$ rounds. In each round, we use $K=n=10$ descriptions. It can be noticed that multi-round coding makes it possible to further reduce the distortion below that of single-round coding. 

For some $n\in \mathbb{N}$, one may consider to do a single round with $kM\leq n$ descriptions or e.g., $M$ rounds with $K$ descriptions in each round. Assuming the vector is quantized along different dimensions in the $M$ rounds, one can show that in the limit of $\xi\to\infty$, the two approaches lead to the same rate and distortion performance. For a finite $\xi$, using $kM\leq n$ descriptions in a single round is slightly better than using $M$ rounds each of $K$ descriptions.

\subsubsection{Efficiency}
The average efficiency $\eta_K(\mathcal{I},R)$ for $|\mathcal{I}|=K$ evaluated at each of the operating points indicated by * in the multi round coding curves of Fig.~\ref{fig:xi2} is 0.64 and 0.73 for $(M=5,\xi=2)$ and $(M=100,\xi=3)$, respectively.

\begin{figure}[t]
\begin{center}
\includegraphics[width=7cm]{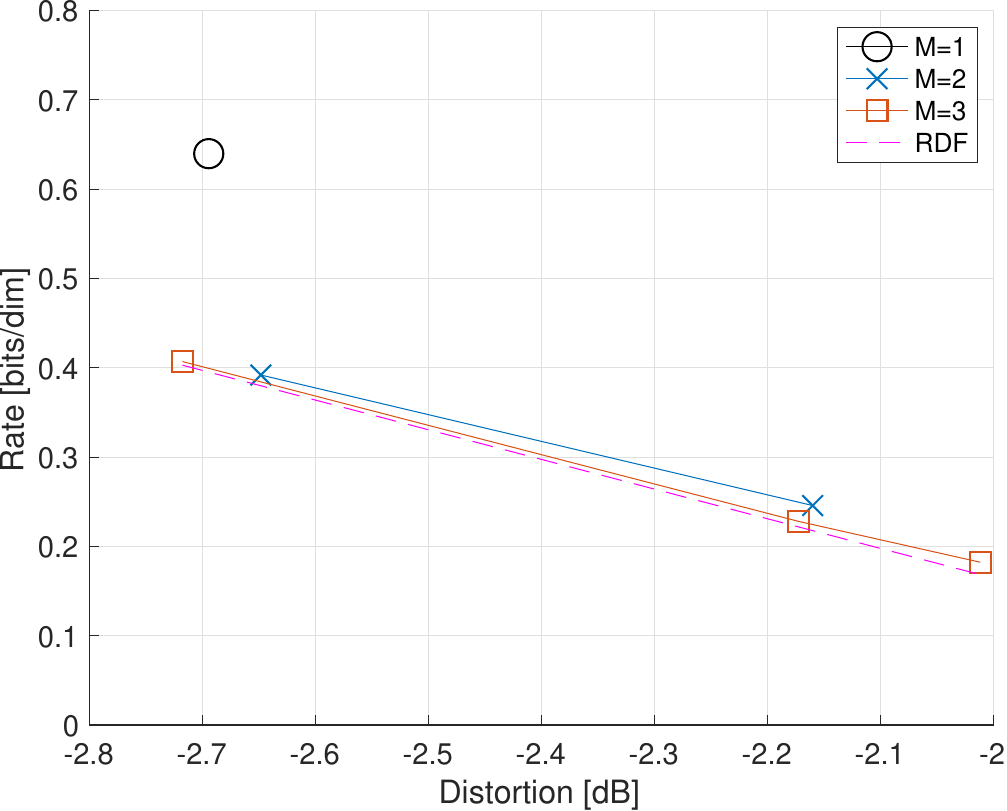}
\caption{Encoding a Laplacian source into $K=20$ descriptions in either $M=1,2$, or $3$ rounds.  }
\label{fig:laplace}
\end{center}
\end{figure}

\subsubsection{Extension to General Sources and Distortion Measures}
Our results indicate that the new idea of encoding a source into independent encodings  and then incrementally improving the reconstruction quality over $M$ rounds is (nearly)  asymptotically optimal for a wide variety of sources and distortion measures. In this section, we will demonstrate how one can apply our proposed threshold vector quantizer in practice to the Laplacian source under the two-sided absolute distortion measure. Specifically, we only have to choose a desired $\xi$ threshold and calculate centroids to form the codebook. Thus, the approach is applicable to quite general sources and distortion measures. We choose the Laplacian source under the two-sided distortion measure, since its RDF is known in closed-form, which makes it possible to compare the operational performance to the optimum.
We compare the performance in terms of the accumulated rate-loss when encoding the source in $M=1,2,$ or $3$ rounds, respectively. In each round, we encode into $K$ descriptions, where we set $K=d$, i.e., the dimension of the source vector. In the simulations, we have used $d=20$ dimensional vectors, and $L=10^6$ realizations. The vectors are i.i.d.\ and the elements within each vector are also i.i.d. The results are presented in Fig.~\ref{fig:laplace}. 
In this simulation we have pre-calculated the centroids for each round, using different realizations of the same source. We also explicitly exploit that the source is symmetric around zero, and that it has zero mean.

For the case of $M=1$, we have used $\xi = 0.5265$. This leads to a distortion of 0.5377, which is equivalent to -2.6946 dB. 
In order to calculate the rate, we count the number of times that the elements of the $\ell$-th source dimension is greater than the threshold $\xi$ and divide by $L$. 
In principle, this corresponds to the probability $p$ of the "rare" symbol in the $\epsilon/1$-channel introduced in Section~\ref{sec:eps-1}. The rate is obtained from the binary entropy: $R= -p\log_2(p) - (1-p)\log_2(1-p)$. Thus, we assume that an efficient entropy encoder will be applied, so that the operational coding rate gets arbitrarily close to the discrete entropy of the quantized output. 

For the case of $M=2$, we have used $\xi = 1.525$ in round 1 and $\xi= -2$ in round 2. In this case where $\xi<0$, we are checking how often the source is less than $\xi$. 
The sum-rate over all $K$ encodings in round $1$ is $K R_1$, and  the sum-rate for round 2 is $R_2$. The total sum-rate (normalized per dimension of the vector) is $\sum_{j=1}^{M=2} R_j$, since $d=K$. 

For the case of $M=3$, we used $\xi = 1.8,  -3.0, 1.7$, for round $j=1,2,3$, respectively. Let $D^{(M)}$ be the resulting distortion after round $M$, then the accumulated rate-loss is given by $\sum_{\ell = 1}^M R_\ell - R(D^{(M)})$, where $R(\cdot)$ denotes the RDF of the source. 

In Fig.~\ref{fig:laplace}, it can be seen that for the same distortion, the rate-loss is significantly decreased by using two rounds instead of a single round. Moreover, by using $M=3$ rounds, the rate-loss is further decreased. In fact, for $M=3$, the operational performance is close to the RDF.

\section{Conclusions}
We introduced a new source coding paradigm, which combines multiple descriptions and successive refinements. Specifically, we introduced the concept of independent encodings, which refer to a set of descriptions, which are conditionally jointly mutually independent given the source. 
The source is first encoded into $K$ independent encodings (descriptions) with zero excess marginal rates. Then, the source is estimated using a subset of the received descriptions. The estimation error is then successively encoded into $K$ new descriptions. In each round, the distortion is further reduced. 
Our main result is that independent encodings and feedback  over multiple rounds lead to nearly ideal multiple-description coding, i.e., simultaneously obtaining zero excess sum-rate and zero excess marginal rates for non-trivial distortions. 

We proved asymptotic optimality for a Gaussian source under the squared error distortion measure, and for a one-sided exponential source under a one-sided distortion measure. We also showed that at high-distortions, and for general sources under the quadratic distortion measure, there is nearly no loss by using $K$ independent encodings each of rate $R$ bits, as compared to forming a single  description using $KR$ bits. 

On the operational side, we proposed a simple threshold \emph{multiple-description} vector quantizer, which outputs $K$ descriptions, each at very low rate. The descriptions are individually as well as jointly  asymptotically rate-distortion optimal for the Gaussian source under the squared error distortion. Our simulations  indicate that using $K$ description in multiple rounds, is nearly optimal also for the case of the Laplacian source under the two-sided absolute error distortion. We believe this holds for a wide range of sources and distortion measures. 
The simple Gaussian \emph{successive refinement} coding scheme that was recently presented in \cite{no:2016}, can be seen as the case of having $K=1$ description in each round. On the other hand, our scheme provides $K$ descriptions in each round, which leads to robustness in case of packet dropouts.

\section{Acknowledgment}
The authors gratefully acknowledge the detailed and constructive comments provided by the associate editor and the anonymous reviewers. 
 
\appendix

\vspace{5mm}





%
%
%
\vspace{0mm}



\begin{IEEEproof}[Proof of Lemma~\ref{theo:noisy_mutual1}]
\label{app:theo:noisy_mutual1}
Mutual information over very noisy discrete channels]
To prove the lemma it is sufficient to show that $I(Y;Z) = o(\epsilon^3)$, since earlier works \cite{gallager:1965,lapidoth:2003} have established that $I(X;Y) = o(\epsilon^2)$ and $I(X;Z) = o(\epsilon^2)$. 
\begin{align}
&I(Y;Z) = 
\sum_{y,z} p(y|z)p(z) \log\left(
\frac{p(y|z)p(z)}{p(y)p(z)} \right) \\ \notag
&\overset{(a)}{=} 
\sum_{x,y,z}p(y|x)p(x|z)p(z) \log\left(
\frac{\sum_{x'} p(y|x')p(x'|z)p(z)}{p(y)p(z)} \right) \\ \label{eq:condpmfs}
&=\sum_{x,y,z} p(y|x) p(z|x)p(x)\log\left(
\frac{\sum_{x'} p(y|x')p(z|x')p(x')}{p(y)p(z)} \right), 
\end{align}
where $(a)$ follows since $p(y|z) = \sum_x p(y|x) p(x|z)$. At this point, we can replace $p(y|x)$ and $p(y)$ by their induced channel conditional $p(y)(1 + \epsilon \psi_Y(y|x))$ and marginal $p(y)(1+\epsilon \psi'_Y(y))$ output distributions, respectively, where
\begin{equation*}
\psi'_Y(y)  = \sum_x p(x) \psi_Y(y|x), \quad \psi'_Z(z)  = \sum_x p(x) \psi_Z(z|x).
\end{equation*}

Thus, the expression within the $\log$ becomes:
\begin{align} \notag
&\frac{\sum_{x'} p(y|x')p(z|x')p(x')}{p(y)p(z)} \\ \notag
&= \frac{\sum_{x'}p(x') p(y)(1+\epsilon \psi_Y(y|x')) p(z)(1+\epsilon\psi_Z(z|x'))}{p(y)(1+\epsilon \psi'_Y(y))p(z)(1 + \epsilon \psi'_Z(z))} \\ \notag
 &= \frac{\sum_{x'}p(x') (1+\epsilon \psi_Y(x'|y)) (1+\epsilon\psi_Z(x'|z))}{(1+\epsilon \psi'_Y(y))(1 + \epsilon \psi'_Z(z))} \\ \label{eq:ratio1}
&=  \frac{1+  \epsilon \big(\psi'_Y(y)+\psi'_Z(z)\big) +\epsilon^2 \sum_{x'}p(x')\psi_Y(y|x')\psi_Z(z|x')  }{1+\epsilon \psi'_Y(y)+ \epsilon \psi'_Z(z) +  \epsilon^2 \psi'_Y(y) \psi'_Z(z)}.
\end{align}
Using the $\log$ expansion $\log(1+c) = c - \frac{1}{2}c^2 + o(c^3)$  on the $\log$ term in \eqref{eq:condpmfs} and inserting \eqref{eq:ratio1} leads to 
{\allowdisplaybreaks\begin{align} \notag
&\log\bigg( \frac{\sum_{x'} p(y|x')p(z|x')p(x')}{p(y)p(z)}\bigg) =  \epsilon \big(\psi'_Y(y)+\psi'_Z(z)\big) \\ \notag 
&\quad +\epsilon^2 \sum_{x'}p(x')\psi_Y(y|x')\psi_Z(z|x')  \\ \notag
&\quad -
\frac{\epsilon^2}{2}\big( \psi'_Y(y)^2  + \psi'_Z(z)^2 \\ \notag
&\quad
+ 2\!\!\sum_{x',x''}\!\! p(x')p(x'')\psi_Y(y|x')\psi_Z(z|x'') \big) \\ \notag
&\quad- \big(\epsilon \psi'_Y(y)+ \epsilon \psi'_Z(z) 
 +  \epsilon^2 \psi'_Y(y) \psi'_Z(z) \big) \\ \notag
&\quad  + \frac{\epsilon^2}{2}\big( \psi'_Y(y)^2+ \psi'_Z(z)^2 + 2 \psi'_Y(y) \psi'_Z(z) \big)  + o(\epsilon^3) \\  \label{eq:logexp}
&= 
 \epsilon^2\sum_{x'}p(x')  \psi_Y(y|x')\psi_Z(z|x')   -  \epsilon^2 \psi'_Y(y) \psi'_Z(z)   + o(\epsilon^3).
\end{align}}
The terms outside the $\log$ in \eqref{eq:condpmfs} can be written in a similar way as the nominator in \eqref{eq:ratio1}, that is
\begin{align}\label{eq:first1}
&\sum_{x,y,z} p(y|x)p(z|x)p(x) =
\sum_{x,y,z}p(x)p(y)p(z) \\  \notag
&\times\big(1+  \epsilon \psi_Y(y|x)+\epsilon\psi_Z(z|x) +\epsilon^2 \psi_Y(y|x)\psi_Z(z|x)  \big).
\end{align}
To complete the expansion of \eqref{eq:condpmfs}, we multiply \eqref{eq:logexp} by \eqref{eq:first1}, which leads to
\begin{align*}
I(Y;Z) &= 
\epsilon^2 \sum_{y,z}p(y)p(z)\big( 
\sum_{x'}p(x') \big( \psi_Y(y|x')\psi_Z(z|x') \big) \\
&\quad -  \psi'_Y(y) \psi'_Z(z) 
\big)  + o(\epsilon^3) = o(\epsilon^3),
\end{align*}
where the last equality follows using the fact that $\psi$ must have zero expected value with respect to the output, i.e., $\sum_{y}p(y)\psi_Y(y|x) = \sum_{z}p(z)\psi_Z(z|x) = 0$, which implies that $\sum_{y}p(y)\psi'_Y(y) = \sum_{z}p(z)\psi'_Z(z)=0$. 
\end{IEEEproof}


\vspace{5mm}

\begin{IEEEproof}[Proof of Lemma~\ref{theo:epsilon}
Mutual information over very noisy continuous-discrete channels] \label{app:theo:epsilon}
Due to the Markov chain, it follows that:
\begin{equation}\label{eq:mi}
I(X;Y,Z) = I(X;Y)+I(X;Z) - I(Y;Z).
\end{equation}
In the following, we will show that $I(X;Y)=o(\epsilon_Y), I(X;Z)=o(\epsilon_Z)$, and $I(Y;Z)=o(\epsilon_Y\epsilon_Z)$, which when combined with~(\ref{eq:mi}) and used in~(\ref{eq:mi_ratio}), proves the lemma.
Let $p_{y_j|x} = P_{Y|X}(y_j|x)$ denote the conditional probability of receiving $y_j$ when $x\in \mathcal{X}$ was transmitted on the channel $P_{Y|X}$. Similarly, let $p_{z_j|x}$ denote the conditional probability of receiving $z_j$ when $x$ was transmitted on the channel $P_{Z|X}$. Let $\mathcal{X}=\mathbb{R}, \mathcal{Y}=\{y_1,\dotsc,y_{|\mathcal{Y}|}\}$, and $\mathcal{Z}=\{z_1,\dotsc,z_{|\mathcal{Z}|}\}$.
Without loss of generality, we let $y_1$ and $z_1$ denote high probability symbols, i.e., $\int_{X} P_{X}(x) p_{y_1|x}dx=1-\epsilon_Y$ and $\int_{\mathcal{X}} P_{X}(x) p_{z_1|x}dx=1-\epsilon_Z$. It follows that $\int_{\mathcal{X}}\sum_{j=2}^{|\mathcal{Y}|}P_{X}(x) p_{y_j|x} dx=\epsilon_Y$ and
$\int_{\mathcal{X}}\sum_{K=2}^{|\mathcal{Z}|}P_{X}(x)\, p_{z_K|x}dx =\epsilon_Z$.

We first define
\begin{equation}\label{eq:epsyj}
\epsilon_{y_j}  \triangleq Pr(Y=y_j) = \int_{\mathcal{X}} P_{X}(x)\,  p_{y_j|x} dx, \quad j=1,\dotsc, |\mathcal{Y}|,
\end{equation}
where $\epsilon_{y_1} = 1-\epsilon_Y$ and $p_{y_j|x}$ denotes the conditional pmf of $y_j$ given $x\in\mathcal{X}$. 
Notice that $\epsilon_Y\to 0$ implies that $p_{y_j|x}\to 0, \forall x,y_j$ where $j\neq 1$, and it furthermore implies that for $j>1$ and any $x\in \mathcal{X}$, 
\begin{equation*}
\lim_{\epsilon_Y\to 0}\frac{p_{y_j|x}}{\epsilon_{y_i}} = a_1,\quad 
\lim_{\epsilon_Y\to 0}\frac{p_{y_j|x}}{\epsilon_Y} = a_2,
\end{equation*}
where $a_1,a_2$ are non-zero and finite constants and therefore independent of $\gamma$.
%
We use the convention that $0\log(0) \triangleq 0$ and $0\log(\frac{0}{0}) \triangleq 0$. Thus, $P_{X}(x)p_{y_j|x} \log\big( \frac{p_{y_j|x}}{\epsilon_{y_j}} \big)$ is well-defined, non-negative, and finite for all $0\leq P_{X}(x)$ and $0\leq p_{y_j|x},\epsilon_{y_j},\leq 1$. With this, we may expand the mutual information in the following way:
{\allowdisplaybreaks
\begin{align*}
&\lim_{\epsilon_Y \to 0 }I(X;Y) =
\lim_{\epsilon_Y \to 0 }
\int_{\mathcal{X}}\!\! P_{X}(x)\big(
\sum_{j=1}^{|\mathcal{Y}|} p_{y_j|x_i} \log\big( \frac{p_{y_j|x_i}}{\epsilon_{y_j}} \big)
\big) dx\\
&=
\lim_{\epsilon_Y \to 0 }
\int_{\mathcal{X}}\!\! P_{X}(x)\big(  p_{y_1|x} \log\big( \frac{p_{y_1|x}}{1-\epsilon_Y} \big) \\
&\quad+
\sum_{j=2}^{|\mathcal{Y}|} p_{y_j|x} \log\big( \frac{p_{y_j|x}}{\epsilon_{y_j}} \big)
\big)dx \\
&\overset{(a)}{=}\lim_{\epsilon_Y \to 0 }
\int_{\mathcal{X}}\!\! P_{X}(x)\big(  p_{y_1|x} \log\big( \frac{p_{y_1|x}}{1-\epsilon_Y} \big)\big) dx +o(\epsilon_Y) \\
&=\lim_{\epsilon_Y \to 0 }
\int_{\mathcal{X}}\!\! P_{X}(x)\big(  p_{y_1|x} \log\big( 1 - \frac{(1-p_{y_1|x}) -\epsilon_Y}{1-\epsilon_Y} \big)\big) dx \\
&\quad+
o(\epsilon_Y) \\
&\overset{(b)}{=} 
\lim_{\epsilon_Y \to 0 }
\int_{\mathcal{X}}\!\! P_{X}(x)\big(  p_{y_1|x} \big(  \frac{\epsilon_Y}{1-\epsilon_Y} -\frac{(1-p_{y_1|x})}{1-\epsilon_Y} + o(\epsilon_Y^2) \big)\big)dx \\
&\quad +
o(\epsilon_Y) \\
&\overset{(c)}{=} o(\epsilon_Y),
\end{align*}
where $(a)$ follows since $|\mathcal{Y}|-1<\infty$ and 
$\lim_{\epsilon_Y\to 0} p_{y_j|x} \log\big( \frac{p_{y_j|x}}{\epsilon_{y_j}} \big) = o(\epsilon_Y)\log(a_1) = o(\epsilon_Y)$, which follows by moving the limit inside the integral, which is possible since $p_{y_j|x}$ is bounded in $x$, and since the linear term dominates the $\log$, which implies that $p_{y_j|x}\log(p_{y_j|x})$ is also bounded. Clearly, $P_X(x)p_{y_j |x} \log p_{y_j|x}$ is therefore also bounded. Thus, we can invoke the dominated convergence theorem.
Step $(b)$ follows by using the asymptotic expansion  of the logarithm, i.e., $\log(1+x) = x + o(x^2)$. Finally, $(c)$ follows since either $p_{y_1|x} = 0$ for any $\epsilon_Y$ or $p_{y_1|x}\to 1$ as $\epsilon_Y\to 0$. The same procedure shows that $I(X;Z)=o(\epsilon_Z)$. 
} 

Let us now consider cascading the two channels $P_{Y|X}$ and $P_{Z|X}$ (with a fixed distribution $f_X$ in the middle) resulting in the product channel $P_{Y|Z}$.
Similar to~(\ref{eq:epsyj}), we define
\begin{equation}\label{eq:epszk}
\epsilon_{z_k}  \triangleq Pr(Z=z_k) = \int_{\mathcal{X}} P_{X}(x)\,  p_{z_k|x}dx, \quad k=1,\dotsc, |\mathcal{Z}|,
\end{equation}
where $\epsilon_{z_1} = 1-\epsilon_Z$. Moreover, we define
\begin{align*}
\epsilon_{y_j|z_k} &\triangleq \frac{Pr(Y=y_j,Z=z_k)}{Pr(Z=z_k)} \\
&\overset{(a)}{=}\int_{\mathcal{X}} \frac{P_{X}(x)\,p_{y_j|x}\,p_{z_k|x}}{\epsilon_{z_k}},
\end{align*}
where $(a)$ is due to the Markovian property, i.e., $p_{y_j|z_k,x} = p_{y_j|x}$. 
As before, we note that $\epsilon_Y,\epsilon_Z\to 0$ implies that $\epsilon_{y_j}\to 0, \epsilon_{y_j|z_k} \to 0$ and that $\frac{\epsilon_{y_j|z_k}} {\epsilon_{y_j}}= a_1$. For $j=k=1$, we have the following result
\begin{align} \notag
\frac{\epsilon_{y_1|z_1}}{\epsilon_{y_1}}  &=
\frac{1-\sum_{j=2}^{|\mathcal{Y}|}\epsilon_{y_j|z_1}}{\epsilon_{y_1}}   \\ \notag
&=\frac{1-\frac{\int_{\mathcal{X}}\sum_{j=2}^{|\mathcal{Y}|}P_{X}(x)p_{y_j|x}p_{z_1|x}dx}{\epsilon_{z_1}}}{\epsilon_{y_j}}   \\ \notag
&=1-\frac{\epsilon_{y_1}-1+\frac{\int_{\mathcal{X}}\sum_{j=2}^{|\mathcal{Y}|}P_{X}(x)p_{y_j|x}p_{z_1|x}dx}{\epsilon_{z_1}}}{\epsilon_{y_1}}   \\ \notag
&=1-\frac{\int_{\mathcal{X}}\sum_{j=2}^{|\mathcal{Y|}}P_{X}(x)p_{y_j|x}(\frac{p_{z_1|x}}{\epsilon_{z_1}}-1)}{\epsilon_{y_1}}   \\ \label{eq:decay}
&=1-\frac{\int_{\mathcal{X}}\sum_{j=2}^{|\mathcal{Y|}}P_{X}(x)p_{y_j|x}(\frac{p_{z_1|x}-\epsilon_{z_1}}{\epsilon_{z_1}})}{\epsilon_{y_1}},
\end{align}
where $\epsilon_{z_1}=1-\epsilon_Z$ and $\epsilon_{y_1}=1-\epsilon_Y$. 
We now first observe that $\lim_{\epsilon_Z\to 0} p_{z_1|x_i} - \epsilon_{z_1} = \lim_{\epsilon_Z\to 0} (p_{z_1|x_i} -1 ) + \epsilon_Z = o(\epsilon_Z)$. Then since $\int 
\sum_{j=2}^{|\mathcal{Y}|}P_X(x)p_{y_j|x} dx= o(\epsilon_Y)$,  we deduce that (\ref{eq:decay}) can be written as $1-o(\epsilon_Y)o(\epsilon_Z)$. 
With this, it follows from~(\ref{eq:decay}) that
\begin{equation*}
\lim_{\epsilon_Y,\epsilon_Z\to 0} \log\big(\frac{\epsilon_{y_1|z_1}}{\epsilon_{y_1}}\big) = o(\epsilon_Y\epsilon_Z),
\end{equation*}
by the $\log$ expansion $\log(1-o(\epsilon_Y)o(\epsilon_Z)) = o(\epsilon_Y)o(\epsilon_Z)$.

\vspace{4mm}

We are now in a position to show the following:
{\allowdisplaybreaks
\begin{align*}
&\lim_{\epsilon_Y,\epsilon_Z \to 0 }
 I(Y;Z) =
\lim_{\epsilon_Y,\epsilon_Z \to 0}
\sum_{k=1}^{|\mathcal{Z}|} \epsilon_{z_k}\big(
\sum_{j=1}^{|\mathcal{Y}|} \epsilon_{y_j|z_k} \log\big( \frac{\epsilon_{y_j|z_k}}{\epsilon_{y_j}} \big)
\big) \\
&= 
\lim_{\epsilon_Y,\epsilon_Z \to 0}\bigg\{
\epsilon_{z_1}  \epsilon_{y_1|z_1} \log\big( \frac{\epsilon_{y_1|z_1}}{\epsilon_{y_1}} \big)  \\
&+
\epsilon_{z_1}
\big(
\sum_{j=2}^{|\mathcal{Y}|} \epsilon_{y_j|z_1} \log\big( \frac{\epsilon_{y_j|z_1}}{\epsilon_{y_j}} \big)
\big)  + \sum_{k=2}^{|\mathcal{Z}|} \epsilon_{z_k}
\big(
\epsilon_{y_1|z_k} \log\big( \frac{\epsilon_{y_1|z_k}}{\epsilon_{y_1}} \big)
\big) \\
&
 + \sum_{k=2}^{|\mathcal{Z}|} \epsilon_{z_k}
\big(
\sum_{j=2}^{|\mathcal{Y}|} \epsilon_{y_j|z_k} \log\big( \frac{\epsilon_{y_j|z_k}}{\epsilon_{y_j}} \big)
\big) \bigg\}\\
&= 
 o(\epsilon_Y\epsilon_Z) \\
&+
\lim_{\epsilon_Y,\epsilon_Z \to 0}\
(1-\epsilon_{Z})
\big(
\sum_{j=2}^{|\mathcal{Y}|} \epsilon_{y_j|z_1} \log\big( 1- \frac{\epsilon_{y_j}-\epsilon_{y_j|z_1}}{\epsilon_{y_j}} \big)
\big) \\
& \quad+
\lim_{\epsilon_Y,\epsilon_Z \to 0} 
 \sum_{k=2}^{|\mathcal{Z}|} \epsilon_{z_k}
\big(
\epsilon_{y_1|z_k} \log\big( 1- \frac{1-\epsilon_Y-\epsilon_{y_1|z_k}}{1-\epsilon_{Y}} \big)
\big) \\
&\quad
 + 
\lim_{\epsilon_Y,\epsilon_Z \to 0} 
\sum_{k=2}^{|\mathcal{Z}|} o(\epsilon_Z)
\big(
\sum_{j=2}^{|\mathcal{Y}|} o(\epsilon_Y) \log\big( a_1 \big)
\big) \\
&=
\lim_{\epsilon_Y,\epsilon_Z \to 0}\
(1-\epsilon_{Z})
\big(
\sum_{j=2}^{|\mathcal{Y}|} \epsilon_{y_j|z_1} \\
&\quad\times \log\big(1  - \frac{ \int_{\mathcal{X}}P_{X}(x)\, p_{y_j|x}dx -\frac{\int_{\mathcal{X}} P_{X}(x)\, p_{y_j|x}\, p_{z_1|x}dx}{\epsilon_{z_1}}}{\epsilon_{y_j}}\big)
\big) \\
& \quad+
\lim_{\epsilon_Y,\epsilon_Z \to 0} 
 \sum_{k=2}^{|\mathcal{Z}|} \epsilon_{z_k}
\big(
\epsilon_{y_1|z_k} \big( \frac{\epsilon_Y}{1-\epsilon_Y} - \frac{1-\epsilon_{y_1|z_k}}{1-\epsilon_{Y}} \\
&\quad + o(\epsilon_Y^2)\big)
\big) 
+ 
o(\epsilon_Y\epsilon_Z) \\
&= \!\!
\lim_{\epsilon_Y,\epsilon_Z \to 0}
(1-\epsilon_{Z})
\big(
\sum_{j=2}^{|\mathcal{Y}|} \epsilon_{y_j|z_1}    \\
&\quad \times\log\big(1- \frac{ \int_{\mathcal{X}}P_{X}(x)\, p_{y_j|x}\big(1 -\frac{p_{z_1|x}}{1-\epsilon_{Z}}\big)dx}{\epsilon_{y_j}}\big)
\big) 
+o(\epsilon_Y\epsilon_Z) \\
&= \!\!
\lim_{\epsilon_Y,\epsilon_Z \to 0}
(1-\epsilon_{Z})
\big(
\sum_{j=2}^{|\mathcal{Y}|} \epsilon_{y_j|z_1}\! \big(\!  -\! \frac{ \int_{\mathcal{X}}\!P_{X}(x)\, p_{y_j|x}\big(1 -\frac{p_{z_1|x}}{1-\epsilon_{Z}}\big)dx}{\epsilon_{y_j}}   
\\
& \quad + o(\epsilon_Y)\big)
\big) +o(\epsilon_Y\epsilon_Z) \\
&= o(\epsilon_Y\epsilon_Z),
\end{align*}}
where the last asymptotic equality follows by recognizing that $\lim_{\epsilon_Z\to 0}\big(1 -\frac{p_{z_1|x_i}}{1-\epsilon_{Z}}\big) = o(\epsilon_Z)$.
\end{IEEEproof}
\vspace{5mm}




\vspace{5mm}


\begin{IEEEproof}[Proof of Lemma~\ref{theo:condmi}
Conditional I-MMSE relation]\label{proof:condmi}
We provide the complete proof, which fixes some typos in the proof provided in \cite{ostergaard:2011}. 
We will make use of a similar proof technique as that used in~\cite{guoshamai:2005}.
{\allowdisplaybreaks
\begin{align} \notag
&I(X;Y|Z) = \mathbb{E}_Z [ \int P_{X,Y|Z} \log \frac{ P_{X,Y|Z} }{P_{X|Z} P_{Y|Z} } ] \\  \notag
&= \mathbb{E}_Z [  \int P_{X|Z} P_{Y|Z,X} \log \frac{ P_{Y|Z,X} }{ P_{Y|Z} }  ]\\  \notag
&= \mathbb{E}_Z [ \!\int\!\! P_{X|Z} P_{Y|Z,X} \log \frac{ P_{Y|Z,X} }{ P_{Y'|Z'} }  + \int\!\! P_{X|Z} P_{Y|Z,X} \log \frac{ P_{Y'|Z'} }{ P_{Y|Z} } ]\\  \notag
&= \mathbb{E}_Z [ \int P_{X|Z} P_{Y|Z,X} \log \frac{ P_{Y|Z,X} }{ P_{Y'|Z'} }  + \int P_{Y,X|Z} \log \frac{ P_{Y'|Z'} }{ P_{Y|Z} } ]\\  \notag
&= \mathbb{E}_Z [ \int P_{X|Z} P_{Y|Z,X} \log \frac{ P_{Y|Z,X} }{ P_{Y'|Z'} }  + \int P_{Y|Z} \log \frac{ P_{Y'|Z'} }{ P_{Y|Z} } ]\\ \label{eq:div1}
&= \mathbb{E}_Z [ \mathbb{E}_{X|Z} [D( P_{Y|Z,X} \| P_{Y'|Z'} )]  - D(  P_{Y|Z} \| P_{Y'|Z'} ) ].
\end{align}
We can freely choose the distribution of $Y'|Z'$ as long as it has the same support as $Y|Z$. If we choose it to be Gaussian, then the first term in \eqref{eq:div1} is the divergence between two Gaussian distributions, which can be determined as follows~\cite{guoshamai:2005}:
\begin{align}\notag
&D(\mathcal{N}(m_1,\sigma_1)\| \mathcal{N}(m_0,\sigma_0)) \\ \label{eq:div2gauss}
=&\frac{1}{2}\log\bigg(\frac{\sigma_0^2}{\sigma_1^2}\bigg) +\frac{1}{2}\bigg( \frac{ (m_1-m_0)^2}{\sigma_0^2} + \frac{\sigma_1^2}{\sigma_0^2} - 1\bigg)\log e.
\end{align}
Since $\mathbb{E}[Y|X,Z] = \mathbb{E}[Y|X] = \sqrt{\gamma}X$ and $\mathbb{E}[\mathrm{var}(Y|X,Z)]=\sigma_N^2=1$, it follows that $Y|X,Z \sim \mathcal{N}(\sqrt{\gamma}X, 1)$.  We now make the choice of $Y'|Z' \sim  \mathcal{N} (\sqrt{\gamma}  \mathbb{E}[X|Z], 1 + \gamma\mathbb{E}[\mathrm{var}(X|Z))]$, which when inserted into \eqref{eq:div2gauss} leads to:
\begin{align} \notag
&\mathbb{E}_{X|Z} [D( P_{Y|X} \| P_{Y'|Z'} )] = 
\frac{1}{2}\log(1 + \gamma \mathbb{E}[\mathrm{var}(X|Z)]) \\ \notag
&\quad - \frac{1}{2}\log(e) +\frac{1}{2}\log(e)\frac{\mathbb{E}_{X|Z}[(\sqrt{\gamma}X - \sqrt{\gamma}\mathbb{E}[X|Z])^2]}{1+\gamma\mathbb{E}[\mathrm{var}(X|Z)]} \\
&\quad+ \frac{1}{2}\log(e)\frac{1}{1+\gamma\mathbb{E}[\mathrm{var}(X|Z)]} \\ \notag 
&=\frac{1}{2}\log(1 + \gamma\mathbb{E}[ \mathrm{var}(X|Z)]) \\ \notag
& \ +\frac{1}{2}\big(\frac{ \gamma\mathbb{E}[\mathrm{var}(X|Z)]}{1+\gamma\mathbb{E}[\mathrm{var}(X|Z)]} + \frac{1}{1+\gamma\mathbb{E}[\mathrm{var}(X|Z)]} -1\big)\log(e)\\  \label{eq:div_simple}
&= \frac{1}{2}\log(1 + \gamma \mathbb{E}[\mathrm{var}(X|Z)]).
\end{align}
When taking the limit $\gamma\to 0$, it follows from \eqref{eq:div_simple} that:
\begin{equation} \label{eq:1stterm}
\lim_{\gamma\to 0}  \frac{1}{\gamma}\mathbb{E}_Z \big[ \mathbb{E}_{X|Z} [D( P_{Y|X} \| P_{Y'|Z'} )]  = \frac{1}{2} \mathbb{E}[\mathrm{var}(X|Z)],
\end{equation}
since $\lim_{\gamma\to 0} \frac{1}{\gamma}\log (1 + \gamma c) = c$. 

We now look at the second equation in \eqref{eq:div1} and use that $P_{Y|Z} =  \mathbb{E}_{X|Z} P_{Y|Z,X}$. Following along similar lines as in~\cite{guoshamai:2005}, we can write:
\begin{align} \notag
&\log\big(\frac{P_{Y|Z}(y|z)}{P_{Y'|Z'}(y|z)}\big) = \log\big(\frac{
\mathbb{E}_{X|Z}[P_{Y|Z,X}(y|z,x)]}{P_{Y'|Z'}(y|z)}\big) \\ \notag
&=
\log\big( \frac{\sqrt{2\pi(1+\gamma\mathrm{var}(X|z))}}{\sqrt{2\pi \sigma_N^2}} \big) \\ \notag
&\quad + \log\big( 
\frac{\mathbb{E}_{X|Z}[ \exp( -\frac{1}{2\sigma_N^2}(y-\sqrt{\gamma}X)^2)] } 
{\exp(-\frac{1}{2(1+\gamma\mathrm{var}(X|z))}(y-\sqrt{\gamma}\mathbb{E}[X|z])^2}
 \big) \\ \notag
&= \frac{1}{2}\log\big(1+\gamma\mathrm{var}(X|z) \big) \\ \notag
&\quad + \log\bigg( 
\frac{\mathbb{E}_{X|Z}[ \exp( -\frac{1}{2\sigma_N^2}(y-\sqrt{\gamma}X)^2)] } 
{\exp(-\frac{1}{2(1+\gamma\mathrm{var}(X|z))}(y-\sqrt{\gamma}\mathbb{E}[X|z])^2}
 \bigg) \\ \notag
&= \frac{1}{2}\log\big(1+\gamma\mathrm{var}(X|z) \big) \\  \notag
&\quad+ \log\bigg( 
\mathbb{E}_{X|Z}[ \exp( -\frac{1}{2}(y-\sqrt{\gamma}X)^2  \\  \notag
&\quad +\frac{1}{2(1+\gamma\mathrm{var}(X|z))}(y-\sqrt{\gamma}\mathbb{E}[X|z])^2) ]
 \bigg) \\ \notag
&\overset{(a)}{=} \frac{1}{2}\log\big(1+\gamma\mathrm{var}(X|z) \big) \\ \notag
&\quad+ \log\bigg( 
\mathbb{E}_{X|Z}\bigg[ (X-\mathbb{E}[X|z])y\sqrt{\gamma} 
+ \frac{1}{2}y^2\gamma(\mathbb{E}[X|z] - X)^2  \\ \notag
&\quad -\frac{1}{2}X^2 \gamma \bigg] +1 -\frac{1}{2}y^2\gamma\mathrm{var}(X|z) + \frac{1}{2}\gamma\mathbb{E}[X|z] \\
&\quad + d(y,z)\gamma^{3/2} + o(\gamma^{3/2})
 \bigg) \\ \notag
&=\frac{1}{2}\log\big(1+\gamma\mathrm{var}(X|z) \big) +  \log\big( 1  + \frac{1}{2}\gamma(\mathbb{E}[X|z]^2 - \mathbb{E}_{X|Z}X^2) \\ \notag
&\quad+ d(y,z)\gamma^{3/2} + o(\gamma^{3/2}) \big)  \\  \notag
&=\frac{1}{2}\log\big(1+\gamma\mathrm{var}(X|z) \big) \\ \label{eq:2ndterm}
&\quad+  \log\big( 1  - \frac{1}{2}\gamma\mathrm{var}(X|z)  +  d(y,z)\gamma^{3/2} + o(\gamma^{3/2}) \big) 
\end{align}
where $(a)$ follows by the series expansion of $\exp(\cdot)$ in terms of $\gamma$, and $d(y,z)$ denotes a bounded function, which depends upon $y$ and $z$ but not directly on $\gamma$. 
}
Clearly, if we divide  \eqref{eq:2ndterm} by $\gamma$, and let $\gamma\to 0$ the result will be zero, i.e., 
\begin{align*} 
&\lim_{\gamma\to 0} \frac{1}{\gamma} \bigg( \frac{1}{2}\log\big(1+\gamma\mathrm{var}(X|z) \big) \\  
&\quad+  \log\big( 1  - \frac{1}{2}\gamma\mathrm{var}(X|z)  +  d(y,z)\gamma^{3/2} + o(\gamma^{3/2}) \big)\bigg) \\
&= 0.
\end{align*}
We are allowed to move the limit operator inside the expectation over $Z$, since both $\mathrm{var}(X|z)$ and $\mathbb{E}[X|z]$ are bounded functions in $z$, and the limit when $\gamma\to 0$ is finite for all $z$. Moreover, the expression in \eqref{eq:2ndterm} is bounded for all finite $\gamma$.
Since \eqref{eq:1stterm} does not tend to zero, it follows that the left term in \eqref{eq:div1} dominates the mutual information for small $\gamma$.

For finite $\gamma$ it is insightful to subtract  the expected value of \eqref{eq:2ndterm} from \eqref{eq:div_simple} and keep the low order (in $\gamma$) terms in the series expansion of this difference, i.e.,
\begin{align} \notag
&\frac{1}{2}\log(1 + \gamma \mathbb{E}[\mathrm{var}(X|Z)]) - \frac{1}{2}\mathbb{E}[\log\big(1+\gamma\mathbb{E}[\mathrm{var}(X|Z)]] \big) \\ \notag
&\quad-  \mathbb{E}[\log\big( 1  - \frac{1}{2}\gamma\mathbb{E}[\mathrm{var}(X|Z)]  +  d(Y,Z)\gamma^{3/2} + o(\gamma^{3/2}) \big)]  \\ \notag
&= \mathbb{E}[\log(1+ d(Y,Z)\gamma^{3/2})] + \frac{1}{2}\gamma \mathbb{E}[\mathrm{var}(X|Z)] \\  \notag
&\quad - \frac{1}{4} \gamma^2 \mathbb{E}[\mathrm{var}(X|Z)^2]  + o(\gamma^2) \\ \notag 
&=  \frac{1}{2}\gamma \mathbb{E}[\mathrm{var}(X|Z)]  + \mathbb{E}[d(Y,Z)]\gamma^{3/2}  - \frac{1}{4} \gamma^2 \mathbb{E}[\mathrm{var}(X|Z)^2] \\  \label{eq:dif1}
&\quad + o(\gamma^2).
\end{align}
We will now show that $ \mathbb{E}[d(Y,Z)]\gamma^{3/2} = o(\gamma^2)$. First we note that:
\begin{equation}\label{eq:y}
\mathbb{E}_{X|Z}\mathbb{E}[Y|Z] = \sqrt{\gamma}\mathbb{E}_{X|Z}[X]= \sqrt{\gamma} \mathbb{E}[X|Z],
\end{equation}
and that 
\begin{align}\notag 
&\mathbb{E}_{X|Z}\mathbb{E}[Y^3|Z] =\mathbb{E}_{X|Z} \mathbb{E}[(\sqrt{\gamma}X+N)^3|Z] \\ \notag
&= \mathbb{E}_{X|Z} \mathbb{E}[(\sqrt{\gamma}^3X^3+N^3 + 3N^2\sqrt{\gamma}X + 3\gamma X^2N)|Z] \\ \notag
&\overset{(a)}{=}  \mathbb{E}_{X|Z} \mathbb{E}[(\sqrt{\gamma}^3X^3 + 3N^2\sqrt{\gamma}X + 3\gamma X^2N)|Z] \\ \notag
&=  3\sqrt{\gamma}\mathbb{E}[N^2] \mathbb{E}[X|Z] + o(\sqrt{\gamma}) \\ \label{eq:y3}
&= 3\sqrt{\gamma}\mathbb{E}[X|Z] + o(\sqrt{\gamma}),
\end{align}
where $(a)$ follows since $ \mathbb{E}[N^3|Z] =  \mathbb{E}[N^3] = 0$, due to $N$ being zero-mean Gaussian. 
At this point, we write out $d(y,z)$ after taking the expectation w.r.t.\ $X|Z$ and simplifying terms: 
\begin{align}\notag 
\mathbb{E}_{X|Z} d(y,Z) &=
\frac{y}{2}\big( 2\mathbb{E}[X|Z]\mathbb{E}[\mathrm{var}(X|Z)]  \\ \notag 
&\quad +  \mathbb{E}[X|Z] \mathbb{E}_{X|Z}[X^2] - \mathbb{E}_{X|Z}[X^3]\big) \\ \notag 
&\quad + \frac{y^3}{6}\big( \mathbb{E}[X|Z]^3 - 3 \mathbb{E}[X|Z]  \mathbb{E}_{X|Z}[X^2] \\ \label{eq:X3}
&\quad + \mathbb{E}_{X|Z}[X^3] \big)
\end{align}
Taking the expectation of \eqref{eq:X3} wrt.\ $Y|Z$ and inserting \eqref{eq:y} - \eqref{eq:y3} leads to:
\begin{align*}\notag
\mathbb{E}_{Y|Z}\mathbb{E}_{X|Z} d(Y,Z) &= \sqrt{\lambda}\mathbb{E}[X|Z]^2\big(
\mathbb{E}[\mathrm{var}(X|Z)]  \\
& \quad -  \mathbb{E}_{X|Z}[X^2] 
+ \mathbb{E}[X|Z]^2)  + o(\sqrt{\gamma}) \\
 &= o(\sqrt{\gamma}),
\end{align*}
since 
\begin{equation*}
- \mathbb{E}_{X|Z}[X^2] 
+ \mathbb{E}[X|Z]^2 = -\mathbb{E}[\mathrm{var}(X|Z)].
\end{equation*}
Thus, we have shown that  $\mathbb{E}[d(Y,Z)]\gamma^{3/2} = o(\gamma^2)$, which when used in \eqref{eq:dif1} implies that 
\begin{equation*}
I(X;Y|Z) = \frac{1}{2}\gamma \mathbb{E}[\mathrm{var}(X|Z)]  - \frac{1}{4} \gamma^2 \mathbb{E}[\mathrm{var}(X|Z)^2] + o(\gamma^2).
\end{equation*}
\end{IEEEproof}




%


\bibliographystyle{IEEEtran}
\bibliography{mdfb}



%


\begin{IEEEbiographynophoto}{Jan {\O}stergaard}
(S'98--M'99--SM'11) received the M.Sc.E.E.\ degree from Aalborg University, Aalborg, Denmark, in 1999 and the PhD.E.E.\ degree (with cum laude) from Delft University of Technology, Delft, The Netherlands, in 2007. From 1999 to 2002, he worked as an R{\&}D Engineer at ETI A/S, Aalborg, and from 2002 to 2003, he was an R{\&}D Engineer with ETI Inc., VA, USA. Between September 2007 and June 2008, he was a Postdoctoral Researcher at The University of Newcastle, NSW, Australia. He has been a Visiting Researcher at Tel Aviv University, Israel, and at Universidad Técnica Federico Santa María, Valparaíso, Chile. Dr. Østergaard is currently a Full Professor in Information Theory and Signal Processing, Head of the Section on AI and Sound, and Head of the Centre on Acoustic Signal Processing Research (CASPR), at Aalborg University. He has received a Danish Independent Research Council’s Young Researcher’s Award, a Best PhD Thesis award by the European Association for Signal Processing (EURASIP), and fellowships from the Danish Independent Research Council and the Villum Foundations Young Investigator Programme. 
His research interests are in the areas of statistical signal processing, information theory, source coding, joint source-channel coding, networked control theory, and acoustic signal processing.
He is an Associate Editor for Signal Processing and Source Coding for the IEEE TRANSACTIONS ON INFORMATION THEORY. 
\end{IEEEbiographynophoto}


\begin{IEEEbiographynophoto}{Uri Erez}
(Member, IEEE) was born in Tel-Aviv, Israel, in 1971. He received the B.Sc. degree in mathematics and physics and the M.Sc. and Ph.D. degrees in electrical engineering from Tel-Aviv University in 1996, 1999, and 2003, respectively. From 2003 to 2004, he was a Post-Doctoral Associate with the Signals, Information and Algorithms Laboratory, Massachusetts Institute of Technology (MIT), Cambridge, MA, USA. Since 2005, he has been with the Department of Electrical Engineering-Systems, Tel-Aviv University. His research interests are in the general areas of information theory and digital communication. From 2009 to 2011, he served as an Associate Editor for Coding Techniques for the IEEE TRANSACTIONS ON INFORMATION THEORY.
\end{IEEEbiographynophoto}

\begin{IEEEbiographynophoto}{Ram Zamir}(Fellow, IEEE) was born in Ramat Gan, Israel, in 1961. He received the B.Sc., M.Sc. (summa cum laude), and D.Sc. (Hons.) degrees in electrical engineering from Tel-Aviv University, Israel, in 1983, 1991, and 1994, respectively. From 1994 to 1996, he was a Post-Doctoral Student at Cornell University, Ithaca, NY, USA, and the University of California at Santa Barbara. In 2002, he spent a sabbatical year at MIT. In 2008 and 2009, he spent short sabbaticals at ETH and MIT, respectively. Since 1996, he has been with the Department of Electrical Engineering–Systems, Tel Aviv University. He has also been consulting in the areas of radar and communications (DSL and WiFi), where he was involved with companies, like Orckit and Actelis. From 2005 to 2014, he was the Chief Scientist at Celeno Communications. He has been teaching information theory, data compression, random processes, communications systems, and communications circuits at Tel Aviv University. His book Lattice Coding for Signals and Networks was published in 2014. His research interests include information theory, including lattice codes for multi-terminal problems, source coding, communications, and statistical signal processing. From 2013 to 2015, he was a member of the BOG of the Society. From 2000 to 2005, he headed the Information Theory Chapter of the Israeli IEEE Society. From 2001 to 2003, he served as an Associate Editor for source coding in the IEEE TRANSACTIONS ON INFORMATION THEORY.
\end{IEEEbiographynophoto}

\end{document}